\documentclass[12pt]{iopart}

\usepackage{graphicx}
\usepackage[section]{placeins}

\expandafter\let\csname equation*\endcsname\relax 
\expandafter\let\csname endequation*\endcsname\relax
\usepackage{amsmath} 

\begin{document}

\title{An Ultra-Compact X-Ray Free-Electron Laser}

\author{J. B. Rosenzweig$^1$, N. Majernik$^1$, R. R. Robles$^1$, G. Andonian$^1$, O. Camacho$^1$, A. Fukasawa$^1$, A. Kogar$^1$, G. Lawler$^1$, Jianwei Miao$^1$, P. Musumeci$^1$, B. Naranjo$^1$, Y. Sakai$^1$,  R. Candler$^2$, B. Pound$^2$, C. Pellegrini$^{1,3}$, C. Emma$^3$,  A. Halavanau$^3$,  J. Hastings$^3$, Z. Li$^3$, M. Nasr$^3$, S. Tantawi$^3$, P. Anisimov$^4$, B. Carlsten$^4$, F. Krawczyk$^4$, E. Simakov$^4$,    L. Faillace$^5$, M. Ferrario$^5$, B. Spataro$^5$,  S. Karkare$^6$, J. Maxson$^7$, Y. Ma$^8$,  J. Wurtele$^9$, A. Murokh$^{10}$, A. Zholents$^{11}$, A. Cianchi$^{12}$, D. Cocco$^{13}$, S. B. van der Geer$^{14}$}

\address{$^1$ Department of Physics and Astronomy,  University of California, Los Angeles, 405 Hilgard Ave., Los Angeles, CA 90095, USA}
\address{$^2$ Department of Electrical Engineering,  University of California, Los Angeles, 405 Hilgard Ave., Los Angeles, CA 90095, USA}
\address{$^3$ SLAC National Accelerator Laboratory, 2575 Sand Hill Rd, Menlo Park, CA 94025, USA}
\address{$^4$ Los Alamos National Laboratory, Los Alamos, NM 87545, USA}
\address{$^5$ Laboratori Nazionali di Frascati, INFN, Via E. Fermi, 00044 Frascati RM, Italy}
\address{$^6$ Department of Physics, Arizona State University, Tempe, AZ 85287, USA}
\address{$^7$ Department of Physics, Cornell University, Ithaca, New York, USA}
\address{$^8$ School of Engineering, University of California, Merced, 5200 North Lake Rd. Merced, CA 95343, USA}
\address{$^9$ Department of Physics, University of California, Berkeley, CA 94720, USA}
\address{$^{10}$ RadiaBeam Technologies, Santa Monica, CA 90404, USA}
\address{$^{11}$ Argonne National Laboratory, 9700 S. Cass Ave., Lemont, IL 60439, USA}
\address{$^{12}$ Dipartimento di Fisica, Università degli Studi di Roma ``Tor Vergata", Via della Ricerca Scientifica 1,  00133 Roma RM, Italy}
\address{$^{13}$ Lawrence Berkeley National Laboratory, 1 Cyclotron Rd., Berkeley, CA, 94720, USA}
\address{$^{14}$ Pulsar Physics, Burghstraat 47, 5614 BC Eindhoven, The Netherlands}

\ead{rosen@physics.ucla.edu}

\begin{indented}
\item[]March 2020
\end{indented}

\begin{abstract}
In the field of beam physics, two frontier topics have taken center stage due to their potential to enable new approaches to discovery in a wide swath of science. These areas are: advanced, high gradient acceleration techniques, and x-ray free electron lasers (XFELs). Further, there is intense interest in the marriage of these two fields, with the goal of producing a very compact XFEL. In this context, recent advances in high gradient radio-frequency cryogenic copper structure research have opened the door to the use of surface electric fields between 250 and 500 MV/m. Such an approach is foreseen to enable a new generation of photoinjectors with six-dimensional beam brightness beyond the current state-of-the-art by well over an order of magnitude. This advance is an essential ingredient enabling an ultra-compact XFEL (UC-XFEL). In addition, one may accelerate these bright beams to GeV scale in less than 10 meters. Such an injector, when combined with inverse free electron laser-based bunching techniques can produce multi-kA beams with unprecedented beam quality, quantified by ~50 nm-rad normalized emittances. The emittance, we note is the effective area in transverse phase space ($(x,p_x/m_ec)$ or $(y,p_y/m_ec)$ occupied by the beam distribution, and it is relevant to achievable beam sizes as well as setting a limit on FEL wavelength, as discussed below. These beams, when injected into innovative, short-period (1-10 mm) undulators uniquely enable UC-XFELs having footprints consistent with university-scale laboratories. We describe the architecture and predicted performance of this novel light source, which promises photon production per pulse of a few percent of existing XFEL sources. We review implementation issues including collective beam effects, compact x-ray optics systems, and other relevant technical challenges. To illustrate the potential of such a light source to fundamentally change the current paradigm of XFELs with their limited access, we examine possible applications in biology, chemistry, materials, atomic physics, industry, and medicine -- including the imaging of virus particles -- which may profit from this new model of performing XFEL science.
\end{abstract}

\vspace{2pc}
\noindent{\it Keywords}: free-electron laser, high accelerating gradient, cryogenic RF, IFEL, high brightness beams 

%
%
%
%

\section{Introduction}

The x-ray free electron laser (XFEL) is a transformative instrument, producing coherent x-ray pulses with peak brightness 10 orders of magnitude greater than preceding approaches \cite{PellegriniRMP}. The existence of a coherent {\AA}-wavelength source with femtosecond pulses has changed the landscape of science. After only a decade of operation of the first hard x-ray FEL, the LCLS at SLAC, its unprecedented source parameters and associated instruments have combined to form an invaluable tool for research in chemistry, biology, materials science, medicine, and physics \cite{Bostedt2016}. As such, XFELs are laboratories with an inherent and deep multi-disciplinary flavor. The LCLS and the other hard x-ray FELs  \cite{Emma2010,Geloni2017} worldwide are based on two enabling technologies: conventional or superconducting RF accelerating structures and magnetic undulators with periods of at least 1.5 cm. These each contribute to the footprint (~km scale, with the smallest instruments near 0.7 km in length) and cost of recent generation XFELs. This combination of unique research significance and high cost means user demand significantly outstrips supply. As a concrete example there exists only a single XFEL facility in the US, the LCLS, which is able to satisfy the beam-time requests of less than 20 percent of the proposed experiments. The types of science that can be engaged in this constrained model are limited – there can be little cross-checking and iteration based on empirical feedback. There are also few opportunities for translational research in industry, medicine and other applied fields. Smaller, less expensive XFELs have been built with more constrained X-ray production. To illustrate this point, we look to two  examples, SwissFEL and SACLA, with their shared emphasis on high brightness electron beams and high gradient acceleration techniques. They have footprints of 550 m and 600 m in total length, and comprehensive total costs of approximately 400 million dollars. 

The current generation of XFELs has greatly exceeded performance expectations.  The progressive successes of  XFELs has made a case that next-generation XFELs are essential. This interest in expansion is manifested in the soft x-ray regime by LCLS-II \cite{LCLS-II-CDR}, a billion-dollar-class facility aimed at exploitation of coherent, ultra-fast photons at longer wavelength, where new opportunities in spectroscopy await. There is also compelling interest in harder x-rays, above 40 keV, to perform imaging in dense, high-Z, mesoscale systems, such as the MaRIE XFEL proposed at LANL \cite{Sheffield2017,Russell2015,Lewellen2015}. 

While x-ray free-electron lasers have attracted multiple billions of dollars in cumulative investment, they still number only a handful worldwide \cite{Pile2011, Kang2017, European-XFEL-TDR, SwissFEL-CDR, FERMI-CDR, Schreiber2015}. Despite the scientific success and tremendous demand from the user community, the XFEL in its present range of configurations, due to the expense and user limitations involved, has stimulated competitive, alternative approaches to appear.  These seek to preserve the extraordinary flexibility and temporal advantages of an XFEL while dramatically reducing the size and cost of the instrument.  An XFEL that is both highly capable and miniaturized has the power to resolve the issue of access that is currently constraining scientific discovery. To develop an ultra-compact XFEL, however, one must reinvent the approach to creating x-rays using the free-electron laser mechanism. Given this history and current status, it is notable that the birth of the XFEL was based on proof-of-principle physics experiments in previous decades that were carried out by small research teams working in an innovative, exploratory fashion \cite{Pellegrini2016a}. To reinvent the XFEL we seek to profit from the same approach. 

Previous proposals on realizing very compact XFELs have concentrated on the use of advanced accelerator techniques \cite{RMPLPA, DLARMP} to minimize the length of the accelerator, thus achieving 10 GeV-level beams in a length in the 10’s of meter range or below. This class of instrument is seen as a step beyond the XFEL, which is termed a 4\textsuperscript{th} generation light source -- thus the miniaturization of the XFEL with an advanced accelerator is termed a 5\textsuperscript{th} generation light source.  These proposals, at present emphasizing a range of techniques from existing high gradient radio-frequency linear accelerators (linacs), to plasma-based acceleration, have not profited from potentially transformative changes in magnetic undulator design. As we shall see in the following, this is due, particularly in the case of plasma acceleration, to the lack of advances in beam quality needed to use accelerators with lower energy, employed in tandem with advanced, short-period undulator methods. 

In this paper, we will present the details of a new approach that exploits multiple scientific advances to realize the design of an ultra-compact XFEL (UC-XFEL). We emphasize first the fundamental performance aspects of the XFEL, to ascertain the approach to an optimized design.  Instead of adapting the FEL to the acceleration technique used, we employ a comprehensive design philosophy that is derived from simultaneously examining multiple aspects of the FEL system.  Given the decade of dedicated research relevant to components of the 5\textsuperscript{th} generation light source, this integrated approach can now be crystallized; the  concept for an ultra-compact x-ray FEL presented here utilizes a recipe based on dramatic advances in the critical component ingredients of the FEL, with the key aspect being the use of an electron beam with unprecedentedly high six-dimensional brightness. 

On this basis, we propose here a new class of UC-XFEL sources that can be implemented at the university level, at size and cost diminished more than ten-fold. Despite this down-scaling, we project that this type of UC-XFEL may produce photon fluxes per pulse that yield several percent of existing full-scale XFELs, in both the soft and hard x-ray regions. This model of affordable, distributed x-ray lasers will make intense, coherent, ultra-fast sources available to the broad scientific community in a ubiquitous way, similar to the optical lasers now employed at university labs in many diverse fields.  Present XFEL experiments, by their nature, have a certain metric for success, which is found in producing one-of-a-kind results in a short period of running – on the scale of days. Relaxing this constraint will inevitably lead to new scientific results, much like the explosion of scientific activity in the early days of synchrotron radiation sources, with the subsequent, rapid development of hundreds of beamlines worldwide \cite{SRhistory}. It is similarly envisioned that the UC-XFEL, which has a predicted cost that is similarly less than five percent  of existing XFELs, will permit wide availability of \textit{coherent} x-rays to a significantly broadened user community.

\begin{figure}[h!]
    \centering
    \includegraphics[width=0.95\linewidth]{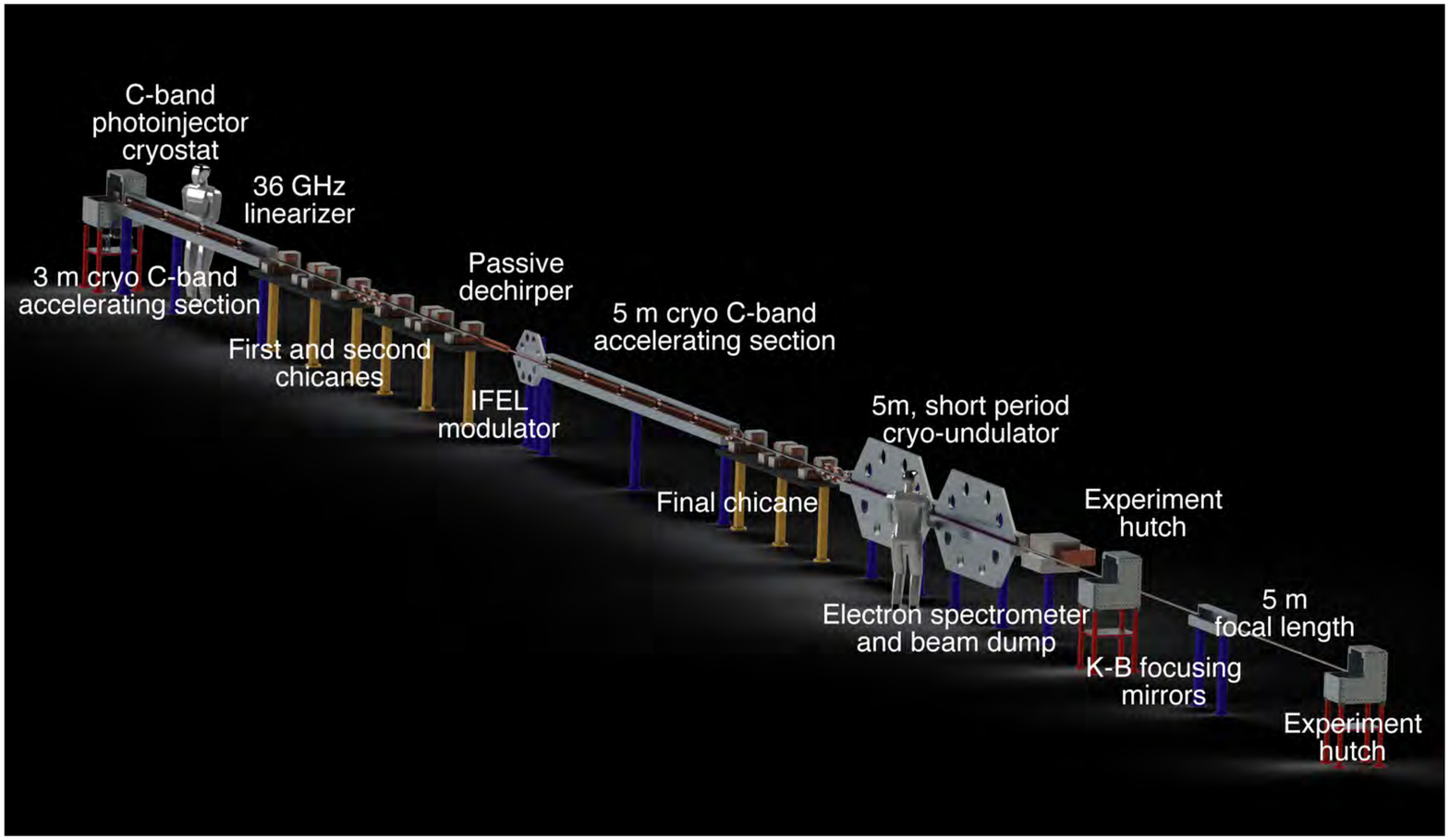}
    \caption{To-scale conceptual layout of the soft x-ray (SXR) UC-XFEL, marking major component systems, with an end-to-end length of approximately 40 meters. Human figures shown for size comparison.}
    \label{fig:overviewImage}
\end{figure}

We discuss below a recipe for taking the km-scale XFEL and miniaturizing it to fit inside the footprint of a university building. To inform the discussion that follows, and to aid in visualization, we show in Figure \ref{fig:overviewImage} a conceptual layout of the UC-XFEL, including its major components at scale. It can be appreciated that the total longitudinal footprint of the instrument is below 50 m. The rationale for the choices made in the UC-XFEL are motivated in what follows. The discussion in this paper is quite detailed, examining many advanced aspects of the UC-XFEL's physical and technological properties. The depth of this discussion is demanded by the challenging nature of the cutting edge techniques proposed for use, and by the intricate physical coupling of various subsystems in the UC-XFEL.  Critically, this detailed discussion shows how it is possible to balance the myriad of advantages and limitations attendant with the suite of advanced methods employed, and mesh them together into a high-performing, functioning whole.  Through this discussion it is shown that a highly credible path to realizing this paradigm-changing instrument, the ultra-compact x-ray free-electron laser, exists. 

\section{A Recipe for an Ultra-Compact XFEL}

As noted above, there has been considerable existing effort aimed at imagining, in detail, an ultra-compact XFEL, without definitive result. To shed light on this challenge, we review a number of relations that one should introduce to quantify the task of realizing the UC-XFEL. The most important gives the dependence of the resonant, on-axis lasing wavelength, $\lambda_r$, on the beam and magnetic undulator parameters: 

\begin{equation}
    \lambda_r = \frac{ \lambda_u}{2\gamma^2}\left(1+\frac{1}{2}K^2_u\right).
\end{equation}

\noindent Here $\gamma=U_e/m_ec^2$ is the beam energy, $U_e$, normalized to the rest energy, $m_ec^2$, and the planar undulator parameter

\begin{equation}
    K_u = \frac{ eB_0}{k_um_ec}, 
\end{equation}

\noindent where the magnetic field in the undulator's symmetry plane is taken to be of the form

\begin{equation}
   B_y = B_0\cos(k_uz).
\end{equation}

\noindent For present-day undulators, with period $\lambda_u=2\pi/k_u$ of a few cm, the parameter $K_u$ slightly exceeds unity in most devices. The physical significance of $K_u$ is that it indicates the vector potential amplitude, normalized to the measure of relativistic momentum, $m_ec$. Further, one may note that it  measures the coupling of the electromagnetic wave in the FEL to the electron's undulating motion, as the maximum angle found in the oscillating design trajectory is $\theta_{\max}=K_u/\gamma$. This angle reflects the degree with which the electrons can exchange energy with the FEL light.

The present authors have examined in some detail the mitigation of beam energy demands in the XFEL through a significant shortening of the undulator period from $\lambda_{u,>}$ to $\lambda_{u,<}$, to the millimeter level or below \cite{Harrison2012}. This period-shortening has the effect of reducing the beam energy demanded, $\gamma_<$, from a reference design, $\gamma_>$, by a factor 

\begin{equation}
  R_\gamma =\frac{\gamma_<}{\gamma_>} = \sqrt{\left(\frac {\lambda_{u,<}}{\lambda_{u,>}}\right)\frac{1+\frac{1}{2} K^2_{u,<}}{1+\frac{1}{2} K^2_{u,>}}}.
\end{equation}

\noindent With peak fields $B_0$ near the 1 T level, scaling to mm-period reduces $K_{u,<}$ to below unity. This reduction factor in the beam energy demanded is significant; in the limit of small (relative to unity)  $K_{u,<}$ and large $K_{u,>}$ it is approximately $ \sqrt{\frac {\lambda_{u,>}}{2\lambda_{u,<}}}K_{u,>}$. It is clear from this relation that one may decrease the energy needs of the FEL in this manner by as much as an order of magnitude, dependent on the details of the undulator parameters chosen. We will in this paper discuss various options for employing short-period undulators, and progress in their realization. The examples chosen will have energies in the 1 to 1.6 GeV range, to be compared with 4.3 GeV and 5.8 GeV in the cases of LCLS-II and SwissFEL \cite{SwissFEL-CDR}, respectively. 

We note that the use of shorter undulator period also inherently shortens the undulator length required to achieve FEL saturation. This is explicitly seen through the expression for the exponential gain length of the FEL instability \cite{Bonifacio1984},

\begin{equation}
   L_\mathrm{g} =\frac{\lambda_u}{4\pi\sqrt{3}\rho}.
\end{equation}

\noindent In a single pass, unseeded XFEL which relies self-amplification of spontaneous emission (SASE), this instability typically proceeds to saturation within 20$L_\mathrm{g}$, indicating that the undulator length is shortened proportionally to $\lambda_u$, without consideration  of the variation of other parameters.  This expression also introduces the Pierce (dimensionless gain) parameter, given by 

\begin{equation}
    \rho^3 = \frac{I}{4I_A}\frac{\gamma K_u^2 [\mathrm{JJ}]^2(K_u^2)}{(1+K_u^2/2)^2(k_r\sigma_x)^2}, 
    \label{eq:rhodefn}
\end{equation}

\noindent where $[\mathrm{JJ}]^2(K_u^2)$ is a Bessel-function dependent parameter that is near unity for $K_u^2<1$, $I$ is the beam current, $I_A$ is the Alfv\'{e}n current, $I_A=ec/r_e \approx$ 17.045 kA, where $r_e$  is the classical electron radius, $k_r=2\pi/\lambda_r$ and the transverse beam rms spot size is $\sigma_x$. It can be seen that $\rho^3\propto I/\sigma_x^2$, the beam current density. Its maximization, or alternatively the minimization of the gain length, is accomplished through high current (kA-level), small emittance, $\epsilon_n$ (\emph{i.e.} the rms size of the beam distribution in transverse phase space), and tight focusing of the beam. It is often stated that the $\rho^3$ is proportional to the five-dimensional beam electron brightness, $B_\mathrm{5D} \equiv 2I/\epsilon_n^2$. We will, in the following section, give a more direct, quantified analysis of the dependence of the gain parameter on beam quality, using the six-dimensional beam brightness

\begin{equation}
  B_\mathrm{6D} \equiv \frac{2I}{\epsilon_n^2\sigma_{\gamma}},
\end{equation}

\noindent where $\sigma_\gamma =\sigma_{U_e}/m_ec^2$ is the electron beam's normalized rms energy spread. We note that this energy spread also is directly related to $\rho$, in that we require $\sigma_{\gamma}/\gamma <\rho$ in order achieve lasing -- larger energy spreads cause Landau damping to extinguish the FEL instability. Further, one may relate the efficiency of beam energy extraction to FEL radiation with the Pierce parameter, $\eta_\mathrm{FEL}=N_\gamma \hbar \omega_r/N_b U_e\simeq \rho$.

One may also note that the gain parameter is dependent on the efficacy of the focusing applied to the beam. Innovative approaches to this focusing, using advanced high field quadrupoles, may be needed to provide an optimal spot size $\sigma_x$; these methods are discussed below. This spot size is limited from below by betatron motion induced slippage and diffraction effects, which require that the radiation Rayleigh range as approximated by $Z_r=4\pi\sigma_x^2/\lambda_r$ is notably larger than $L_g$. For XFELs, this effect is often ignorable due to the short wavelength of the lasing photons. Finally, the emittance should also be small to obey the coherence requirement termed the Pellegrini criterion, $\epsilon_n<\gamma\lambda_r/4\pi$, which guarantees the overlap of the radiation of individual electrons in the beam with each other to coherently add and create the lasing mode \cite{KIM198671}. Quantitatively, as we are attempting to strongly lower the beam energy by the use of short-period undulators, a very low normalized emittance is demanded. 

Much recent progress has been made in understanding how to improve the emittance and attendant brightness of electron beams. The introduction of the high field RF photoinjector approximately 30 years ago \cite{Fraser1985} was a critical step forward in this regard, an advance that yielded an order of magnitude increase in beam brightness. This improvement, which was based on the use of large accelerating fields and optimized beam optics (emittance compensation) techniques was a key element in the realization of the SASE FEL \cite{Akre2008}.  Recently, it has been shown by a SLAC-UCLA collaboration that one may strongly increase the peak operating surface field in copper RF cavities from the nominal current value of $E_0=$120 MV/m by a factor of up to four. This is accomplished by cryogenically cooling the copper, to enter into the anomalous skin effect (ASE) regime \cite{ChambersASE}. The combination of resulting lower dissipation due to diminished surface resistivity with increased material yield strength and mitigation of thermal expansion are the physical effects underpinning this remarkable advance. Applying increased fields in the photoinjector should have profound implications for beam brightness, which stands to be increased 50-fold over the original LCLS design \cite{Rosenzweig2018,Rosenzweig2019,LCLSCDR}, and similarly advance the recent state-of-the-art \cite{Prat2019}. We discuss the expected performance of such a high field photoinjector, operated at $E_0=240$ MV/m below, and deepen previous discussions to examine the implications of more advanced RF designs. 

This new approach to high field RF acceleration also permits a dramatic reduction in length of the accelerator needed for the UC-XFEL. A cryogenically-cooled C-band linear accelerator structure is now being developed for linear collider applications at SLAC and UCLA \cite{Rosenzweig2018, Rosenzweig2019}, with operation at an average accelerating gradient of $eE_\mathrm{acc}=125$ MeV/m; this entails using a peak surface field of $250$ MV/m, nearly identical to that found in the photoinjector. To put this gradient in perspective, it is over a factor of six larger than that employed at LCLS \cite{Emma2010} and LCLS-II. To reach $U_e=$1 GeV (for our soft x-ray example) one would need eight meters of active length with this approach. Between the reduction in energy needed and enhanced gradient employed, the accelerator is shortened by a factor of over 25. The proposed accelerator sections are of an innovative design where the coupling is achieved independently through a wave-guide manifold; there is negligible cell-to-cell coupling in this standing wave design. As such, the accelerating structure may be optimized to have a very high power efficiency \cite{tantawi2018distributed}, as measured by the shunt impedance.

The approach described above yields, in simulation, a beam from the photoinjector having 20 A peak current and $\epsilon_n\simeq 50$ nm-rad. This extremely low emittance must be preserved during both high field acceleration and pulse compression, which entails enhancing the current to several kA to achieve strong XFEL gain. We explore this process in detail, identifying a compact (total length $<$10 m) two-stage compression scheme: the first is a compact chicane that yields $I=400$ A as explored in preliminary investigations in \cite{Robles2019}; the second is an optical micro-bunching technique that utilizes the inverse free-electron laser (IFEL) mechanism \cite{Zholents2005, SashaTrieste}. The advantages of such an approach, which have been outlined in several recent works \cite{Robles2019,Carlsten2019}, are numerous. If one does not attempt to compress the beam into a single pulse, but instead organizes the beam into a pulse train on the scale of a laser wavelength $\lambda_L$, then the bending and accompanying longitudinal motion is limited. This approach suppresses collective effects during compression which can increase the emittance. 

The IFEL produced micro-bunch train, may produce a laser-synchronized train of x-ray FEL pulses \cite{Zholents2005}. To permit robust lasing performance in this system, the micro-bunches cannot be too short, as the radiation slippage over a gain length $z_s =L_\mathrm{g}(\lambda_r/\lambda_u)$ needs to be limited to a fraction of a micro-bunch length \cite{Bonifacio1984, Zholents2005}. Operation at short $\lambda_r$ aids in satisfying this condition, but one must also have robust gain, which in turn is a function of beam brightness, as discussed below. Assuming micro-bunch lasing is achieved, the x-ray pulse train obtained presents distinct advantages in pump-probe experimental applications, as has been noted in other laser-synchronized systems such as high harmonic generation \cite{Popmintchev2012} and Compton scattering sources based on IFEL \cite{Gadjev2019}.  

Finally, a micro-bunched format for achieving the needed peak current serves to ameliorate a problem that particularly afflicts short period undulators. As the gap used in the undulator scales with $\lambda_u$, the material boundaries in mm-period undulators are only 100's of $\mu$m away, and serious issues may arise from resistive wall wakefields. Use of micro-bunches can strongly mitigate this issue \cite{Zholents2004,Carlsten2019,SashaTrieste}, and this aspect of XFEL design is therefore attractive in our case.

With the above-listed ingredients, a beam capable of using 1 to 10 mm period undulators, reaching hard and soft x-ray FEL operation, may be envisioned. In this context, we discuss the present state-of-the-art in sub-cm period undulators, and identify paths to extend current designs (existing down to $\lambda_u=7$ mm) to the 1 mm level. We show through start-to-end simulation analysis, that the saturation length for producing multi-10's of gigawatt XFEL power is 4 m at 1 nm, and 6 m at 1.5 {\AA} operation, respectively. The per-pulse photon flux predicted is nearly five percent of current LCLS operations, in an instrument with a total footprint below 30 m in length. 

Such a compact system demands, for consistency, x-ray optics and experimental end stations which are scaled down in size in a similar fashion, to less than 10 m in length. We discuss approaches to such optics which take advantage of several notable differences between the UC-XFEL and current generation full-scale XFELs: the reduction in peak power and integrated fluence, and increased divergence in the radiation. This analysis is informed by a discussion of a variety of experimental opportunities opened by the realization of an ultra-compact XFEL based on these emergent technologies discussed above, and associated advances in physics design principles. 

\section{Six-dimensional Brightness Scaling in SASE FELs}

While the recipe introduced above explains the conceptual dependencies connecting the suite of ideas introduced that enable the UC-XFEL, it is more direct and intuitive to describe the XFEL in the order in which the electron beam encounters the component systems. This begins with the electron source, which plays a central role in enabling the needed performance of the UC-XFEL.  

Arguments in favor of the advantages of high brightness beams for driving SASE FELs have traditionally been based on the 5D brightness, $B_\mathrm{5D}$. This viewpoint is problematic, however, as $B_\mathrm{5D}$ is not a conserved quantity.  On the other hand, the Liouville theorem indicates that the six-dimensional brightness $B_\mathrm{6D}$ is conserved in a local sense.  We thus present here an analysis of the Pierce gain parameter, $\rho$, that seeks to reveal its dependence on $B_\mathrm{6D}$. Here we assume that there are no correlations in the 6D phase space after the beam is prepared for lasing, and that dilution of $B_\mathrm{6D}$ (a measure based on rms quantities) may be ignored. 

We begin by noting that for values of the undulator strength parameter $K_u<1$ as encountered in the UC-XFEL, the expression for the Pierce parameter can be simplified to read approximately 

\begin{equation}
    \rho^3 \simeq \frac{I}{4I_A}\frac{\gamma K_u^2}{k_r^2\sigma_x^2}.
\end{equation}

\noindent We have written $\rho$ in terms of the radiation wave-number, as we will aim to reveal scaling of parameters while holding the FEL wavelength fixed. 

To relate the gain to the beam brightness, we first write the emittance in terms of the Pellegrini criterion $\epsilon_x \simeq \epsilon_n/\gamma=\eta_\epsilon/2k_r$, {\it i.e.}, where $\eta_\epsilon$ is less than or equal to unity. When it takes the value 1, the electron and photon beam emittances are equal. Additionally, we will assume the peak undulator fields to be fixed by design limits at $B_0\simeq 1$ T and thereby rewrite the undulator parameter in the following way,

\begin{equation}
    K_u = \frac{eB_0}{k_um_ec}\equiv \frac{1}{k_ur_B}=\frac{2\gamma^2}{r_Bk_r} ,
\end{equation}

\noindent where the parameter $r_B=1.7$ mm can be understood as the radius of curvature of the trajectory an electron having momentum $m_ec$ follows in a uniform field of strength 1 T. In the last equality we have approximated the FEL resonance condition valid to lowest-order in $K_u$, $k_u=k_r/2\gamma^2$. 

To optimize the focusing of the electron beam,  we additionally take the average Twiss $\beta$-function of the electron beam to be near to the FEL gain length,

\begin{equation}
    \beta_x=\frac{L_\mathrm{g}}{\eta_\beta}=\frac{1}{2\eta_\beta k_u \sqrt{3}\rho}
\end{equation}

\noindent where we have introduced another factor, $\eta_\beta$. This tuning factor is near to, but generally slightly smaller than unity, depending on final FEL optimization.  This scaling is due to the need to avoid excessive transverse angles in the  beam. We insert these scaling relations into the expression for $\rho^3$, thus eliminating one power of $\rho$,

\begin{equation}
    \rho^2=\frac{\sqrt{3}}{4I_A}\frac{2I}{\epsilon_{n}^2}\frac{\eta_\epsilon\eta_\beta \gamma^5}{ r_B^2k_r^4}=\frac{\sqrt{3}}{4I_A}B_\mathrm{5D}\frac{\eta_\epsilon\eta_\beta\gamma^5}{r_B^2k_r^4} .
\end{equation}

\noindent In this last equality the 5D brightness is used. We now pass to a description of the scaling which employs the six-dimensional rms brightness which, in the absence of transverse and longitudinal emittance growth or beam loss, may be conserved. We relate the 5D and 6D brightness by introducing the normalized rms energy spread $\sigma_\gamma$, and indicate the familiar energy spread requirement as 

\begin{equation}
   \frac{\sigma_\gamma}{\gamma}=\eta_\gamma\rho, 
\end{equation}

\noindent where $\eta_\gamma$ is another optimization factor, similarly limited above by unity. In this case we can rewrite the expression for $\rho$ utilize the 6D beam brightness,

\begin{equation}
   B_\mathrm{6D}=B_\mathrm{5D}/\sigma_\gamma
\end{equation} to obtain
\begin{equation}
    \rho = \frac{\sqrt{3}}{4I_A}\frac{\eta_\epsilon\eta_\beta\eta_\gamma \gamma^6}{r_B^2k_r^4}B_\mathrm{6D}.
    \label{eq:b6dscaling}
\end{equation}

\noindent From this analysis we see that the Pierce parameter, which dictates the gain length and efficiency characteristics of the XFEL, has linear scaling with the six-dimensional electron beam brightness. This is much more striking than the commonly quoted dependence on 5D brightness $\rho \propto B_\mathrm{5D}^{1/3}$. The scaling with energy also appears as quite strong, depending as $\gamma^6$. Conversely, if one includes the implicit variation of $k_r(\gamma)$, holding the undulator period constant (as is more often the case in FEL design strategies), this reverses, and $\rho \propto \gamma^{-2}$.  However, in the UC-XFEL, we are concerned with achieving a certain $k_r=2\pi/\lambda_r$ through increasing $k_u$ and thus using smaller $\gamma$. This strong energy dependence indicates that we must, in order to operate at smaller $\gamma$, dramatically increase the electron beam 6D brightness. The approach to the electron source yielding this critical advance is discussed in the next section. 

Before introducing the approach to the initial production of high beam brightness, we  note that the  factors $\eta_\epsilon$, $\eta_\beta$,and $\eta_\gamma$ can be analyzed with sophisticated optimization procedures. These procedures are aided by theoretical work that has been based, as is presented here, on an analysis based on 6D brightness \cite{LBNLbrightness}. Similarly, design trade-offs are necessary to include the effects of non-one-dimensional phenomena such as diffraction. This type of optimization has been studied in the context of the well-known Xie analysis, found in Refs. \cite{Xie1995,Xie2000}, and extended to include space-charge in Ref. \cite{Marcus2011}. 

\section{High Gradient Cryogenic Photoinjector}

\begin{figure}[h!]
    \centering
    \includegraphics[width=0.7\linewidth]{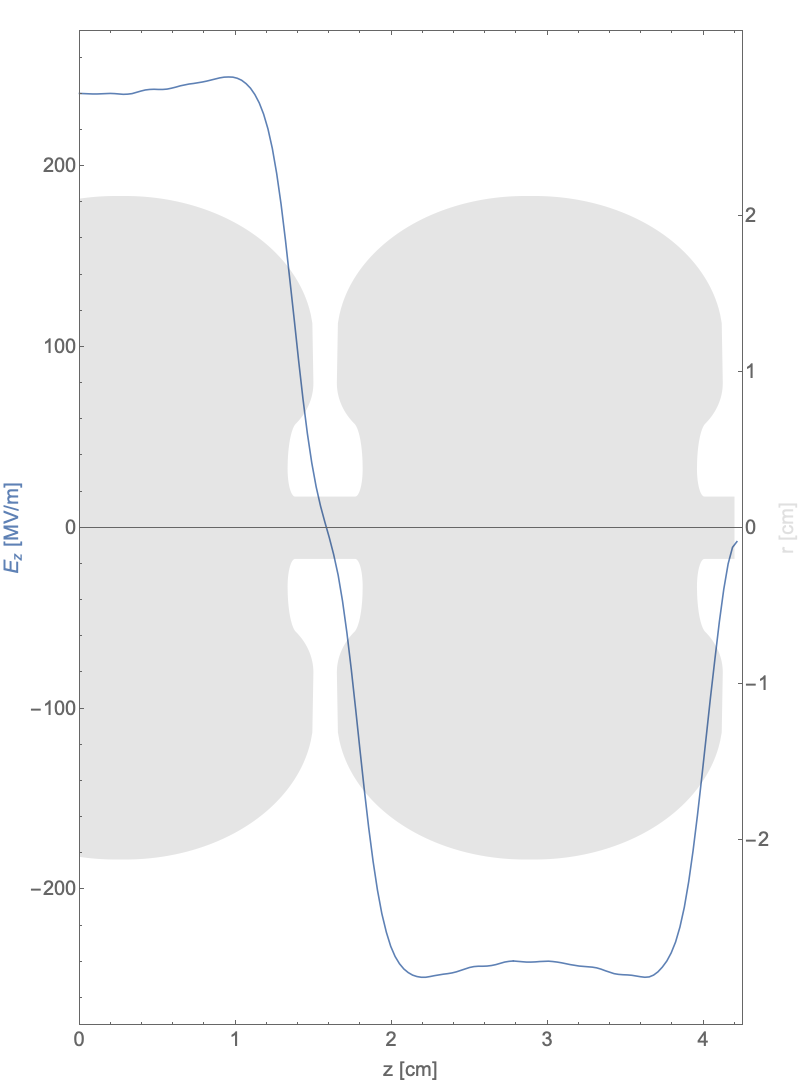}
    \caption{Geometry of 1.6 cell C-band cryogenic photoinjector, without coupling. Overlaid with the longitudinal electric field, illustrating significant higher spatial harmonics.}
    \label{fig:ezPlotCombined}
\end{figure}

The approach to photoinjector for UC-XFEL has been chosen by extending several previous studies of the potential use of cryogenic cooling of the RF cavities \cite{Rosenzweig2018,Rosenzweig2019}. These studies, along with experimental investigations of the breakdown and dark-current emission performance of cryogenic cavities \cite{Cahill2018b}, have yielded in  previous work an optimized surface electric field on the photocathode $E_0=240$ MV/m. This field, obtained in a high shunt-impedance RF structure geometry, is a factor of two below the measured breakdown limit in cryogenic copper \cite{Cahill2018}, and also well below the threshold (300 MV/m) of large dark current emission. The novel shape of the structure, inspired by the optimized linear accelerator structure discussed in the next section, is indicated in Figure \ref{fig:ezPlotCombined}. Even with derating of the maximum field, the brightness obtained by this new generation of RF photoinjector should significantly increase over current values. This is due to, above all, the increased beam density at emission. As discussed in Ref. \cite{Rosenzweig2019}, the five-dimensional brightness is expected to scale as $B_\mathrm{5D}\propto E_i^n$, where $E_i$ is the amplitude of the initial launch field, and the exponent $n$ is between 1.5 and 2, depending on the beam shape. In the regime where we intend to operate, $n\simeq 1.5$. The six-dimensional brightness, including the effect of space-charge on slice energy spread ($\propto E_i^{-0.5}$) \cite{jrecscaling, KIM1989201,ZHuang2005}, the scales as $B_\mathrm{6D}\propto E_i^2$. The proposed high field cryogenic C-band RF photoinjector source takes advantage of this scaling. We will see that this step forward must also be accompanied by innovative approaches to brightness preservation during beam manipulations after the injector. 


\begin{table}[h!]
    \centering
    \begin{tabular}{|c|c|c|}
        \hline
        Parameter & Units & Value \\
        \hline 
       $\pi$-mode cells & $-$ & 1.6 \\
        Peak axial field, $E_0$ & MV/m & 240 \\
        Charge per pulse  & pC & 100  \\
        Peak solenoid field, $B_\mathrm{peak}$ & T & 0.55 \\
        Spot size on cathode, $\sigma_x$  & $\mu$m & 75 \\
        Thermal emittance at cathode  & nm-rad &  38 \\
        Pulse length  & psec & 5 \\
        Energy after acceleration & MeV & 153 \\
        Final normalized emittance, $\epsilon_n$ & nm-rad & 55 \\
        \hline
    \end{tabular}
    \caption{Parameters of the ultra-high brightness RF photoinjector for UC-XFEL used in GPT simulation.}
    \label{tab:RFgun}
\end{table}

As illustrated, the proposed photoinjector RF structure is a 1.6 cell C-band gun with an optimized shape that minimizes the magnetic field on the surfaces. This is a new geometry, with many different features from previously studied photoinjectors \cite{Rosenzweig2018}. The chosen form of the cavity, having pronounced re-entrant irises, permits lower input power and mitigates dissipation at cryogenic temperatures. With an expected overall repetition rate of 100 Hz, and nominal 300 nsec RF pulses, at the foreseen operating temperature of 27 K this dissipation is 11 W, requiring over 0.5 kW cooling power.  This operating point is chosen, in part, to take advantage of the faster response times and associated mitigation of power considerations in the UC-XFEL, as well as its possible utility for application in the MaRIE XFEL. 


This optimized RF structure has several potentially important impacts on the performance of the photo-emitted beam. The significant higher spatial harmonic content (see Figure \ref{fig:ezPlotCombined}) in the RF profile arising from optimizing the shunt impedance (to over 400 M$\Omega$/m) through re-entrant features provides a beneficial effect: enhancement of the second-order RF transverse focusing experienced by the electrons \cite{Rosenzweig1994}. These spatial harmonics may also introduce a nonlinear radial dependence of the focusing and defocusing fields in the gun cavity which could, for beams with non-trivial radial extent, lead to rms emittance growth. This has been found to not be the case for the parameters of the UC-XFEL photoinjector.  This has been shown through simulations in which the full, nonlinear, 3D form of the transverse fields are used. They are observed to differ from those where the fields are are linearly extrapolated off-axis from the on-axis field behavior in producing emittances which increase by no more than one nanometer-radian. In this regard, we note that beam dynamics simulations  of the injector using the code GPT  \cite{GPTSite} include modeling of the following physical effects: cathode emission properties; 3D electromagnetic fields; 2D solenoid fields; and space-charge forces (including image-derived) obtained from placing the charge and current on a 3D grid. The effects of intra-beam scattering (IBS) are included in an analytical estimate discussed below.

\begin{figure}[h!]
    \centering
    \includegraphics[width=0.8\linewidth]{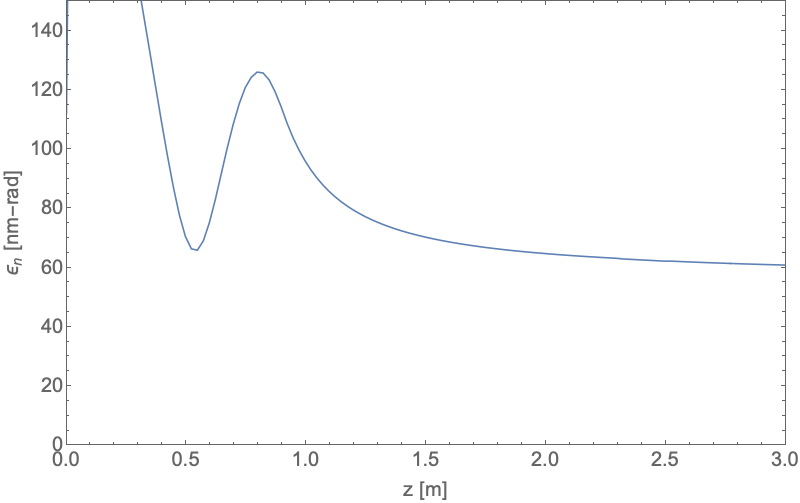}
    \caption{Emittance evolution from a C-band cryogenic RF photoinjector, with final value reaching $\epsilon_n=55$ nm-rad.}
    \label{fig:emitcompensation}
\end{figure}

These effects have thus been studied along with the overall performance of the photoinjector, and found to yield  beam brightness further enhanced over that reported in Ref. \cite{Rosenzweig2018}. This study includes solenoid focusing similar to that of a scaled LCLS gun with a 1.05 meter drift to the first post-acceleration linac (described in the next section). Designs of a newly-conceived compact, cryogenic high field ($B_\mathrm{peak}=$0.55 T) solenoid have been employed in the beam dynamics simulations. Results from these studies include the evolution of the rms emittance and beam size during the focusing and accelerating process, yielding emittance compensation, as shown in Figure \ref{fig:emitcompensation}. The final emittance after acceleration to $U_e=$ 153 MeV is excellent: $\epsilon_n = 55$ nm-rad (at $z=4.4$ m).  This is achieved with $Q_b=$100 pC charge, and 5 psec FWHM pulse length, with peak current $I_p =20$ A. As we shall see, this value of emittance is near to the value which can be preserved in accelerating, transporting, and compressing the beam to its final current of ~4 kA. As we are interested in the 6D brightness of the source, which is at best preserved in acceleration and transport, it is important to evaluate the longitudinal slice energy spread which, after including IBS, is near 1.6 keV. In this case, the achieved 6D brightness is $B_\mathrm{6D}=5\times 10^{18}$ A/(m-rad)$^2$. This can be compared to the value associated with the LCLS design, at $B_\mathrm{6D}=5.1\times 10^{16}$ A/(m-rad)$^2$, which is two orders of magnitude smaller. Advances in 6D brightness have been noted since the LCLS design was introduced; recent experiments in generating very low emittance beams at SwissFEL have given a 6D source brightness of $5.5\times 10^{17}$ A/(m-rad)$^2$ \cite{Prat2019}. Thus there is an order of magnitude improvement of the high field cryogenic source in this parameterization of beam quality over a current state-of-the-art value.

\begin{figure}[h!]
    \centering
    \includegraphics[scale=0.38]{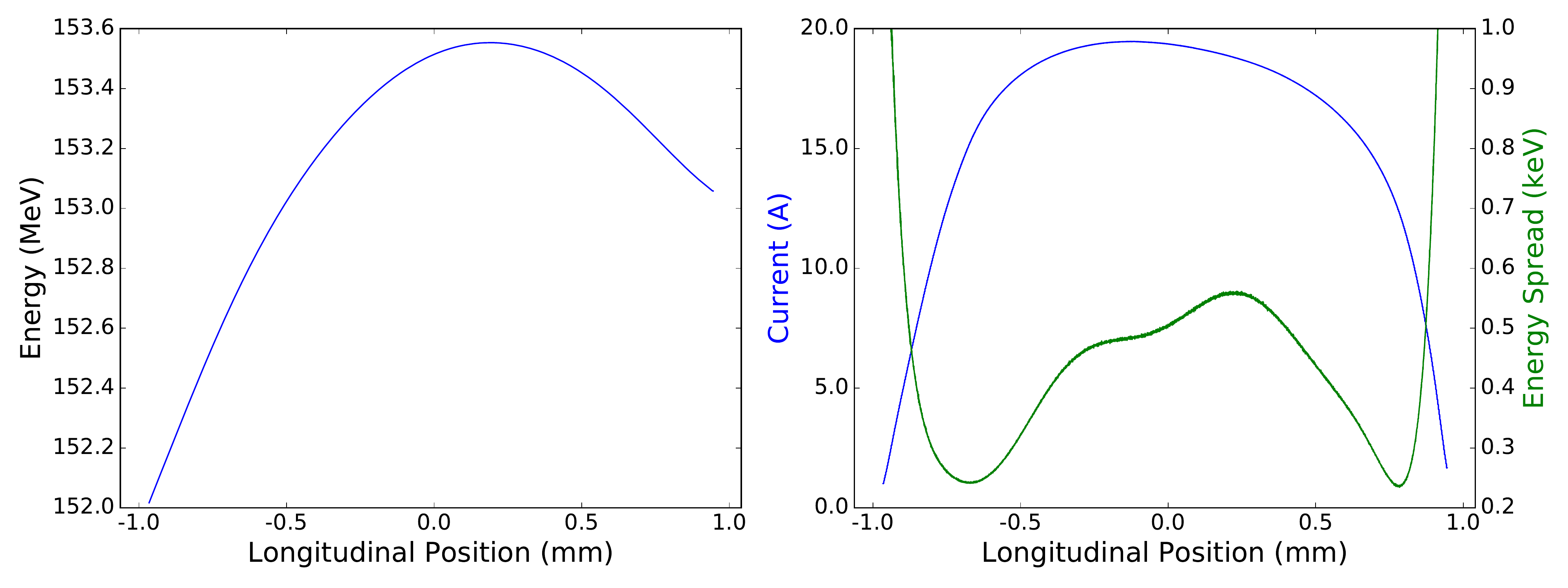}
    \caption{(Left) Longitudinal phase space at the exit of the photoinjector and (right) plots of the current profile and slice energy spread. The beam head is to the right.}
    \label{fig:TOPGUNInput}
\end{figure}

The characteristics of the beam longitudinal phase space at the injector exit are shown in the GPT simulation results displayed in Figure \ref{fig:TOPGUNInput}. Use of a stronger solenoid (0.64 T) can result in a beam with lower emittance, ($\epsilon_n = 50$ nm-rad) through control of the transverse beam oscillation amplitudes.  The decrease in beam size, however, yields some pulse length expansion from longitudinal space-charge effects, to 6.2 psec. For the remainder of this paper, we utilize the shorter beam option, as it stresses the sensitive downstream compression processes less. 

We note that the beam dynamics and emittance compensation process have been optimized in this case considering a copper photocathode with nominal ``thermal" rms normalized emittance \cite{Vecchione2013} given by 0.5 mm-mrad/mm \cite{dowellqemete}. This can be improved upon in principle by using advanced methods involving novel cathode materials, or by taking advantage of the cryogenic state of the emitting surface \cite{cryoemission}. This approach permits the maintenance of a small emittance at larger emitting spot sizes, while giving shorter pulses and higher peak current. Studies of this optimization path in the zero-thermal-emittance limit have shown notably higher brightness \cite{pierce2020role}. While this potentially advantageous approach remains under study (in simulation \cite{pierce2020role} and in current experiments at UCLA), the predicted performance of the UC-XFEL as presented here does not depend on its implementation. In this regard, it should be stated that introduction of novel photo-cathodes would also have an effect on the challenge of operating in a large field-emission (dark current) environment \cite{Cahill2018b}, which remains a key issue in the experimental realization of this high field, high brightness photoinjector.  Similarly, potential problems in managing emission characteristics are found in a large laser fluence environment \cite{Bae2018}, as may be necessary in the high-field photoinjector, with its emphasis on small emitting areas. 

The photoinjector system functionally consists of the very high field RF gun, focusing optics, and post-acceleration cryogenic linacs operated at an average acceleration gradient of $125$ MeV/m. A first version of both the cryogenic C-band RF gun and linear accelerator (linac) structures are now under development by a UCLA-Stanford-LANL-INFN collaboration, with an initial emphasis on linear collider applications \cite{bane2018advanced}. In the experimental work relevant to linear colliders, the RF gun beam dynamics concentrate on the case of a magnetized photocathode, which is used with skew quadrupoles after post-acceleration in the linac to produce highly asymmetric emittances after removal of the angular momentum in the beam \cite{Brinkmann2001}.  The symmetric, unmagnetized beam case relevant to UC-XFEL is also to be examined using the same experimental infrastructure. The design of the linac structure used in these experiments, and in the UC-XFEL, are described in the next section.

\section{Linear Accelerating Structure}

\begin{figure}[h!]
    \centering
    \includegraphics[scale=0.465]{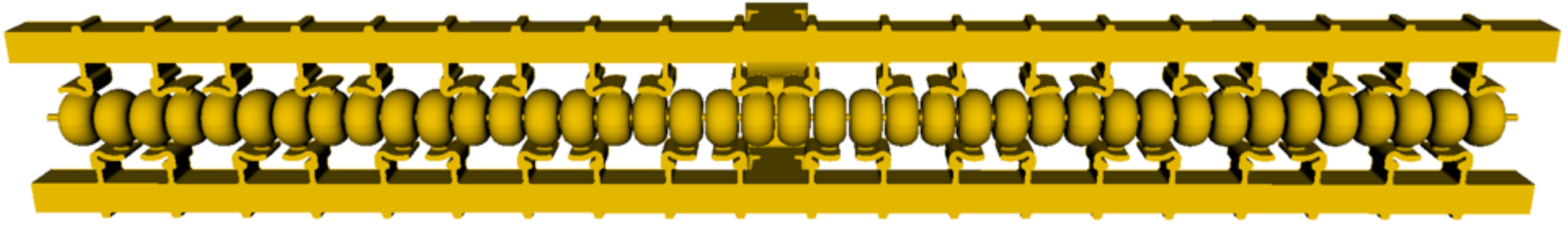}
    \caption{Rendering of 40-cellcryogenic copper linear accelerator structure, with individual cell coupling from wave-guide manifold.}
    \label{fig:Cryolinac}
\end{figure}

The linac structures proposed for the UC-XFEL, as noted, also rely on use of cryogenic C-band copper cavities.  Their design is based on a similar philosophy as the RF gun cavities, with a highly optimized shape. At cryogenic temperatures, this optimized structure yields a shunt impedance of $460$ M$\Omega/$m. This quite high value is obtained by eliminating cell-to-cell coupling through the irises, in favor of a re-entrant geometry. In applications such as the linear collider or XFEL, this geometry has notable advantages over previous generations of X-band linacs, where such a re-entrant structure would provoke transverse wakefields that are very challenging to manage. As the re-entrant design isolates each cavity, the linac section design thus exploits a new coupling scheme, with independent coupling from a wave-guide manifold to each individual cell \cite{tantawi2018distributed}. The resulting form of the linac structure is shown in Figure \ref{fig:Cryolinac}.

These structures are now under development and initial testing at SLAC and UCLA. The initial goal of these studies is the investigation, introduced above, into the production of linear collider-quality, asymmetric emittance beams through the use of a magnetized photocathode \cite{Brinkmann2001}, and the demonstration of subsequent high gradient acceleration, at both 27 K and 77 K operating temperatures. While the  design and innovative realization of these cryogenic structures is rapidly maturing, there are compelling issues left to address. Prominent among these topics is the control of transverse field effects, including multi-bunch beam break-up (BBU), through the damping of higher order modes (HOM) \cite{Shintake_1992, MOSNIER198781}. Simulation and optimization studies are now underway to consider the addition of mode damping and frequency detuning to address BBU in this novel geometry.  Experimentally, peak accelerating gradients of up to 140 MeV/m, consistent with the design parameters demanded here, have been demonstrated in X-band linac prototypes based on this design \cite{tantawi2018distributed}. It should also be noted that multi-cell cryogenic copper accelerating sections are now being fabricated by creating two pieces split along the midplane, and then joining these pieces using innovative techniques such as electron-beam welding. In such a way, one may create intricately shaped cavities for optimizing the shunt impedance, as well as HOM damping, effectively and at lower cost. 

One of the essential advantages of using high gradient cryogenic RF linacs is to yield acceleration giving the beam energy needed for UC-XFEL on a ~10 m length scale. This serves the goal of making a compact system in its own right. The compactness of the linac (as well as the unique approach to final beam compression) also yields a secondary, but critically important advantage: it suppresses the micro-bunching instability (MBI). As this instability is particularly worrisome for high brightness beams, its mitigation, discussed further below, is essential for the UC-XFEL.

 In the longer term, research and development into cryogenic RF acceleration has other fundamental subjects to address. First, one should understand the dramatic increase in breakdown fields in more detail, to push demonstrated, useful performance of cryogenic RF cavities to $>$500 MV/m surface fields. To do so, it is essential to understand the dynamics of surface breakdown, and field emission currents, as well as their mitigation. These investigations have been initiated, and are enabled through a first-principles simulation effort using molecular dynamics codes developed at LANL \cite{perez_huang_voter_2018} for advanced materials investigations. These studies should permit an understanding of the potential role of copper alloys and other materials in extending breakdown towards the theoretical limit dictated by material stress.  Further, the use of coatings such as graphene and silicon oxynitride \cite{Theodore2006} is now under study for mitigation of field emission. These efforts, while not critical to the vision of the UC-XFEL presented here, are directed towards large long-term projects such as compact linear colliders \cite{bane2018advanced, bambade2019international} and next generation full-scale XFELs such as MaRIE \cite{Carlsten2019}.

\subsection{Alternative Approaches for High Brightness Beam Production}

The push towards an ultra-compact XFEL using advanced accelerators has taken place predominantly in the context of plasma accelerators. Particularly, laser-plasma accelerators (LPAs) that can produce, by injection of electrons from the background plasma, a beam that is accelerated up to multi-GeV \cite{Leemans2006} energies in a few cm distance, are very attractive to apply in the XFEL context. Indeed, due to this fortuitous energy reach, LPAs have been vigorously examined as candidates for ultra-compact FELs \cite{Fuchs2009}. Further, the beams produced by LPAs possess parameters that are competitive with certain classes of present day injectors. As such, much research has been focused recently on radiation production by LPA-produced beams, not only from FELs \cite{Labat2018, VanTilborg2017}, but also including hard photon emission from Compton scattering sources \cite{LPAICS}, and from betatron radiation \cite{Corde2013}. 

Progress, despite notable efforts, has been incremental in the field of LPA-based XFELs. As a concrete example of the level of beam quality presently achievable by LPAs, consider the parameters reported in Ref. \cite{Wang2016}, which explicitly cites the 6D brightness achieved. It is given, in the units of this current paper, by $B_\mathrm{6D} \simeq 1\times 10^{15}$ A/(m-rad)$^2$. This is a surprisingly small value, but one must keep in mind that, for similar charge, while the pulse length is notably shorter in the LPA, the beam emittance is an order of magnitude larger, and energy spread exceeds that of the photoinjector by two orders of magnitude. Thus one can see from the scaling given by Equation \ref{eq:b6dscaling}, that use of lower energy beams and/or short FEL wavelengths will be very challenging with such a source.  Stated another way, the approach adopted here gives a much more robust solution to the challenge of making a compact XFEL. 

There are two possible paths forward in plasma-based XFELs that arise from current research trends. The first is that one may utilize beams that are stretched either transversely, using a transverse gradient undulator \cite{HuangTGU}, or longitudinally through a chicane transformation \cite{Majernik19, Maier2012}, to mitigate the influence of energy spread on the gain of the FEL. These approaches produce only modest advantages in the gain performance of the FEL \cite{Barber2018}. A more definitive solution is found in changing the approach to particle injection, to emphasize the creation of a much higher 6D-brightness beam. For example, one may employ the plasma photocathode (or ``Trojan Horse") ionization scheme, to produce beams with much lower emittance, and improved energy spread \cite{Hidding2012}. Studies have shown \cite{Yunfeng2013} that the 6D brightness obtained from this scheme is at the level of $B_\mathrm{6D}\simeq 3.6\times 10^{18}$ A/(m-rad)$^2$. This value is superior to that found in the current case of a high field cryogenic photoinjector, and points to the critical role of the injection field; $E_0\simeq 10$ GV/m in this case, and very small source size. This promising approach to very high brightness beam production is in its experimental infancy, with beams still near in quality to the LPA-based injectors achieved in first tests \cite{Deng2019}. 

One may also examine the possibility of using a high brightness beam obtained from an RF photoinjector that is externally injected into a plasma accelerator. While this solution is not as compact as a plasma-based injection scheme, with an optimized approach to matching the beam transverse dynamics to the strong plasma focusing, the brightness needed for driving an FEL may be reached. This is one of the central goals of the EuPRAXIA project, which has reached the conceptual design phase \cite{FERRARIO2018}. This external injection approach gives up the advantage of creating very high current beams without compression. As we will see in the next section, this is a key challenge to be confronted in the present UC-XFEL vision. 

\section{Electron Beam Acceleration and Compression}

The electron beam is to be accelerated in the high gradient linac from 153 MeV (the photoinjector exit) to the full 1 GeV (soft x-ray, or SXR, FEL) or 1.6 GeV energy (hard x-ray, HXR, FEL) with two stages of compression taking the 20 A beams obtained from the photoinjector to $I_p =$ 4 kA.  The first stage uses a standard chicane, albeit with a higher harmonic RF linearization scheme \cite{smith1986intense, dowell1996boeing} operating in K$_\mathrm{a}$-band \cite{behtouei2020sw}. The need for such a high frequency device for linearization is widespread in the advanced FEL field, and the dedicated development work is now being carried out at INFN-LNF in the context of both this initiative and the XLS project \cite{behtouei2020sw}. The second stage of compression employs inverse free-electron laser bunching, in a scheme known as ESASE \cite{Zholents2005}. Throughout the beamline, both second-order RF focusing in the linacs \cite{Rosenzweig1994} and interspersed magnetic quadrupoles are used to control the beam envelope. We use 10 cm length quadrupoles placed after each linac with gradients restricted to no more than 20 T/m. We discuss the physics performance of each of these steps below. This discussion is guided by simulations of the beam performance using the beam dynamics code \texttt{elegant} \cite{Borland2000}, using the beam produced by the photoinjector as input for the acceleration and compression stages. The limitations of \texttt{elegant} in modelling 3D coherent synchrotron radiation effects are explored using GPT, and found to introduce small, benign changes in the beam dynamics.  The photoinjector simulations were performed using 1.3 million macroparticles. In moving this beam distribution to \texttt{elegant}, we employed the method presented in Ref. \cite{huang2005intrabeam} to estimate an increase in the uncorrelated energy spread of 1.5 keV in the photoinjector from IBS. This value adds in quadrature to the existing, roughly 500 eV, uncorrelated spread produced by the injector simulations. Since this value is an estimate and not the result of a rigorous simulation, we also discuss the general impact of fluctuations in the energy spread on the machine design in Subsection \ref{subsec:espread}. After numerically augmenting the energy spread to account for these effects, the number of particles is increased to 34 million using the SDDS program \texttt{smoothdist6s} to ensure proper modeling collective effects such as the microbunching instability. The subsequent \texttt{elegant} simulations include the effects of longitudinal space charge in all linac sections, a one-dimensional model of coherent synchrotron radiation (CSR) in the bunch compressors, incoherent synchrotron radiation in the bunch compressors, and intrabeam scattering (IBS) effects. To validate the use of the 1D CSR model, we present the results of additional simulations of the bunch compressors performed using the 3D CSR model employed in GPT \cite{Brynes_2018}. 

\subsection{Acceleration to the First Bunch Compressor}

The first compression occurs at 400 MeV, an energy chosen to balance the competing scaling of longitudinal space-charge and coherent synchrotron radiation (CSR). After the gun, the beam is accelerated in three linac sections up to 416 MeV then slightly decelerated in a sixth harmonic 34.272 GHz RF cavity \cite{Faillace2019} to 400 MeV to mitigate the second-order variation in the beam longitudinal phase space due to RF and wakefield effects. In this way, one may cancel the effects of the second-order momentum compaction in the chicane. The sixth harmonic was chosen over a more conventional 11.424 GHz X-band cavity due to RF voltage considerations; the required energy drop in the harmonic cavity scales with the inverse square of the harmonic number, resulting in a needed $>100$ MeV energy loss using 11.424 GHz. In addition to the increased length of an X-band cavity relative to a 34 GHz cavity, this would necessitate use of an additional meter long accelerating structure to maintain the 400 MeV working point. In Table \ref{tab:L1} we list the relevant parameters for the first linac section. This parameter set represents a functional solution that allows us to preserve the 6D beam brightness well in the subsequent compressor, but could potentially be further optimized to produce an even more robust design. During this first acceleration stage the beam accumulates approximately 200 eV of uncorrelated energy spread from IBS. This value has a small effect on the optimization of the second linac. The relatively benign effect of IBS is again owed to the limited interaction length implied by used of a compact, high gradient linac. 

\begin{table}[h!]
    \centering
    \begin{tabular}{|c|c|c|c|}
        \hline
        Parameter & Units & C-Band Linac & Harmonic Cavity  \\
        \hline
        Frequency & GHz & 5.712 & 34.272 \\
        Voltage change & MeV & 266.74 & 12.36 \\
        Phase & $^{\circ}$ & 75.91 & -98.20 \\
        \hline
    \end{tabular}
    \caption{Parameters for the first linac section and linearizing cavity.}
    \label{tab:L1}
\end{table}

\subsection{Dynamics in the Compressor}
In the interest of maintaining a compact design, only 5.5 m of the beamline is allocated to the first compressor, termed BC1. Additionally, to preserve the beam's 55 nm-rad emittance as well as possible, we make use of two consecutive smaller chicanes with a quadrupole triplet in the middle arrayed such that the second chicane mitigates the CSR effects induced by the first chicane \cite{DiMitri2013,Jing2013}. The relevant parameters for the two chicanes are reported in Table \ref{tab:bc1}. The parameter $R_{56}=\frac{\partial \zeta_f}{\partial (\delta p/p_0)}$ is the momentum compaction, \textit{i.e.} the final final longitudinal position $\zeta_f$  to the initial fraction momentum deviation from nominal $\delta p_i/p_0$.

\begin{table}[h!]
    \centering
    \begin{tabular}{|c|c|c|c|}
        \hline
        Parameter & Units & First Chicane & Second Chicane \\
        \hline 
        Magnet length & m & 0.2 & 0.2 \\
        Drift length & m & 1.29 & 0.21 \\
        Bend angle & $^{\circ}$ & 8.3 & 3.2 \\
        $R_{56}$ & mm & 59.85 & 2.15 \\
        Entrance $\beta_x$ & m & 16.25 & 5.5 \\
        Entrance $\alpha_x$ & & 4.1 & 3.1 \\
        \hline
    \end{tabular}
    \caption{Parameters for the chicanes of the first bunch compressor.}
    \label{tab:bc1}
\end{table}

The projected normalized emittance in \texttt{elegant} simulations grows to 80 nm-rad after the first chicane but is brought back down to the 65 nm-rad level by the end of the second. The projected emittance computed using only particles within the full-width at half-maximum of the current distribution is yet smaller, at $\epsilon_n =60$ nm-rad, indicating only 5 nm-rad effective projected emittance growth in the region of interest. As expected, the vertical emittance is largely unaffected by the compressor. The slice emittance in both planes is nearly unchanged by the first bunch compressor, with all growth resulting from misalignment of the $\textit{x'}$ centroids of the longitudinal slices of the beam.

\begin{figure}[h!]
    \centering
    \includegraphics[scale=0.4]{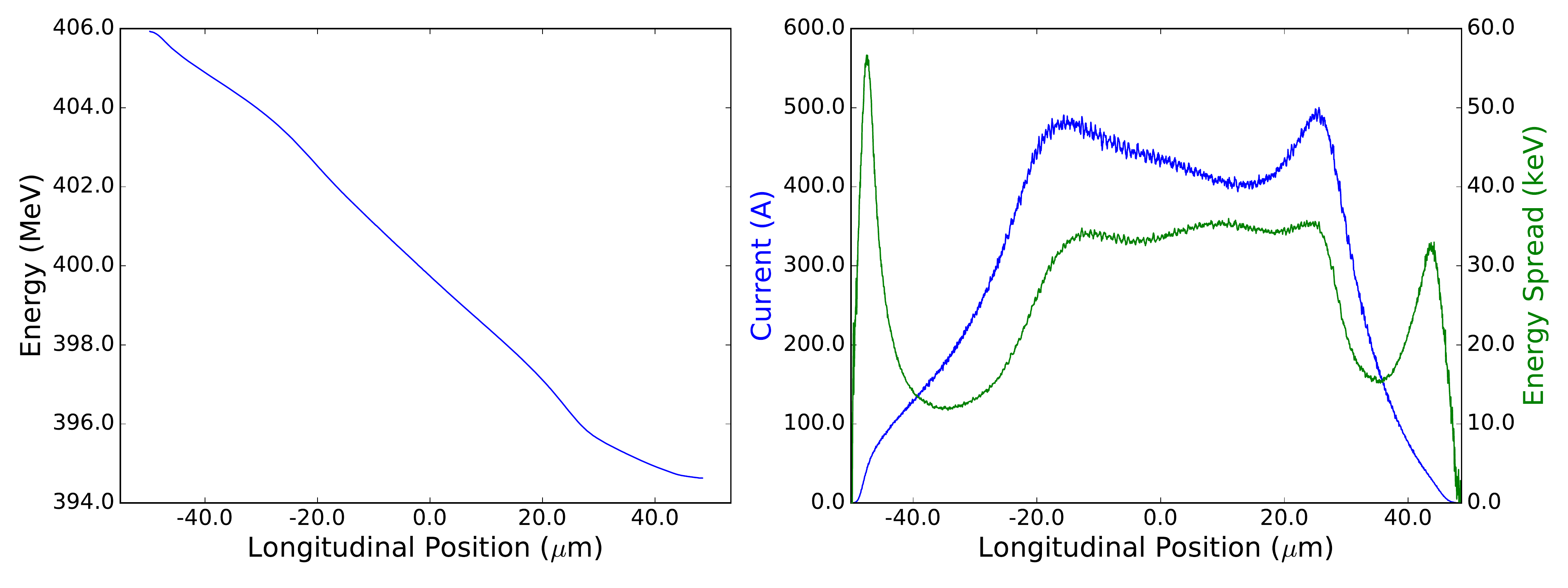}
    \caption{(Left) The longitudinal phase space of the beam after the first bunch compressor is shown alongside (right) the current profile and slice energy spread. The beam head is to the right. }
    \label{fig:AfterBC1}
\end{figure}

To conclude the description of the first compressor we plot in Figure \ref{fig:AfterBC1} the longitudinal phase space, current, and energy spread profile of the beam immediately following the chicane. There is a $\sim$45 $\mu$m longitudinal region in which the current lies between 400 and 500 A. The current profile has developed two small peaks at either end of the flat-top region which in a second bunch compressor might become excessively large, provoking enhanced collective effects. We will find below that the IFEL compressor is, however, insensitive to these peaks. The energy spread in the flat-top region is approximately 34 keV; very little dilution of $B_\mathrm{6D}$ is observed.

\section{Second Bunch Compressor}

\begin{figure}[h!]
    \centering
    \includegraphics[scale=0.4]{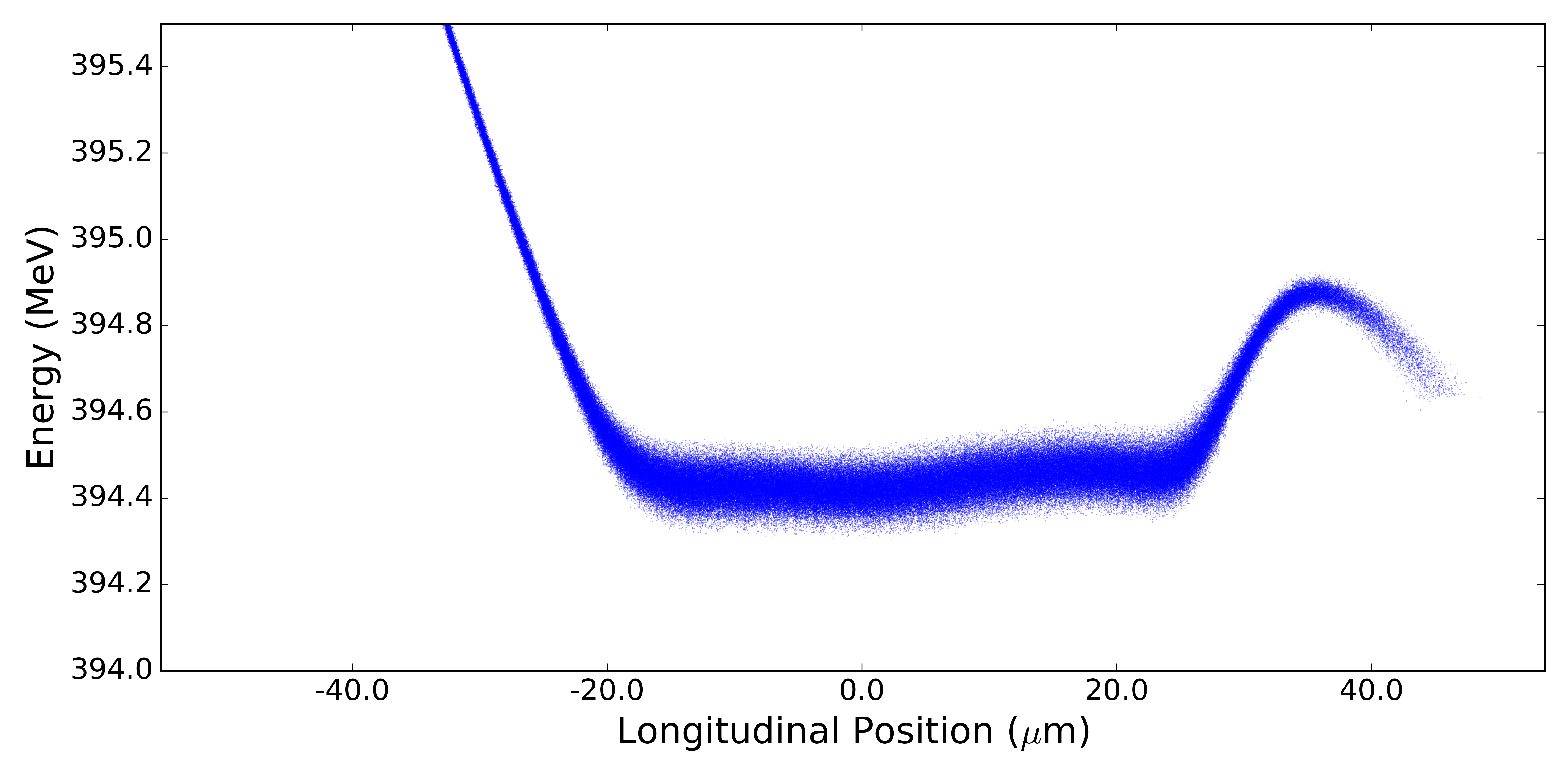}
    \caption{The longitudinal phase space of the beam after dechirping after the first compressor.}
    \label{fig:Dechirped}
\end{figure}

Immediately following the first bunch compressor a passive dechirping cavity \cite{BANE2016156} is employed to remove the remaining energy chirp from the beam. This device is a 1 m long corrugated pipe structure with inner diameter of 0.77 mm and wall periodicity 0.2 mm. This may be substituted for a Cartesian structure, but the effect of quadrupole field excitation on the emittance should then be evaluated. The dechirper is followed by a quadrupole triplet for transverse matching into the modulator. The longitudinal phase space following this dechirping cavity is shown in Figure \ref{fig:Dechirped}. 

\subsection{Overview of IFEL Compression}

In order to mitigate as much as possible deleterious collective effects while reaching a peak current of $I_p=4$ kA, the second compressor does not employ an RF and chicane-based transformation. Instead, it utilizes IFEL compression, introduced by Zholents and also known as the method of enhanced self-amplified spontaneous emission (ESASE) \cite{Zholents2005}. The conventional compressor approach would specifically introduce several complications: first, as we explore further in the subsequent sections, one of the primary advantages of using an ultra-high brightness beam is the ability to make use of several-mm-period, small gap undulators. These undulators naturally reduce the length of the system, but at the cost of enhanced resistive wall wakefields. Further, a standard chicane compressor would demand a quite large momentum compaction at the higher 1 GeV energy, which may lead to unmanageable slice emittance growth from CSR.  On the other hand, micro-bunched beams have been demonstrated to mitigate resistive wall wakefield effects \cite{Carlsten2019}, and we confirm (and extend to a new conductivity regime), and exploit this property in the sections that follow. Additionally, ESASE allows 4 kA current to be achieved with only a mm-scale momentum compaction, as we demonstrate below. This small momentum compaction entails notably less bending, and the attendant challenges in brightness preservation due to CSR are significantly reduced. 

For an ESASE compressor scheme, one first modulates the electron beam energy periodically in time using an infrared laser. The periodicity is imposed by the laser wavelength, chosen in our first example as $\lambda_L$ = 10 $\mu$m, which co-propagates with the electron beam inside of a short planar undulator. Once this modulation is compressed in a chicane, the beam is micro-bunched into a train of current spikes, each of which lases independently of the others in the final undulator. We describe the process using a sequence of two transformations. Let $p=(\gamma-\gamma_0)/\sigma_\gamma$ and $\theta=k_Ls$ where $\gamma_0$ is the mean beam energy, $\sigma_\gamma$ is the initial uncorrelated beam energy spread, and $k_L=2\pi/\lambda_L$. Then the ESASE process is described by 

\begin{equation}
    p_1 = p_0 + A\sin(\theta_0)
\end{equation}
\begin{equation}
    \theta_2 = \theta_1+Bp_1
\end{equation}

\noindent where $A=\Delta\gamma_L/\sigma_\gamma$ is the modulation amplitude normalized to the uncorrelated energy spread and $B=R_{56}k_L\sigma_\gamma/\gamma_0$ is the normalized compression strength. The final longitudinal particle coordinates are then related to the initial coordinates by 
\begin{equation}
    p_f = p_i+A\sin(\theta_i)    
\end{equation}
\begin{equation}
    \theta_f = \theta_i + Bp_i+AB\sin(\theta_i)
\end{equation}
The resulting periodic current profile can be written \cite{Zholents2005,Carlsten2019,Hemsing2014}
\begin{equation}
    I(\theta) = I_0\left(1+2\sum_{n=1}^\infty {J_n(-nAB)\mathrm{e}^{-n^2B^2/2}\cos(n\theta)}\right),
\end{equation}
where $J_n$ is the nth-order Bessel function of the first kind. 

\subsection{Choice of Operating Parameters} \label{sec:operatingParameters}

In the specific case that the ESASE IFEL system replaces a full bunch compressor, it is designed to achieve four goals: optimization of the peak current at a specified value, maximization of the micro-bunch full-width at half-maximum (FWHM) length, and minimization of both the slice energy spread and slice emittance. In general, for a given choice of modulation strength $A$ and design current $I$ there will be two values of the compressor strength $B$ which produce the design current at the center of the spikes: one which under-compresses and one which over-compresses. Over-compression can produce longer micro-bunches and subsequently mitigate slippage effects in the undulator \cite{Carlsten2019} at the cost of enhanced collective effects during compression. These deleterious effects are minimal for high-energy operation such as found in the proposed MaRIE XFEL (which would compress at 12 GeV), but they can prove to be quite harmful for the 1 GeV design presented here. For that reason we avoid over-compression, which also means that there is generally a unique value of $B$ which produces a desired current at a fixed modulation amplitude. Since we have restricted our discussion to a design current of 4 kA, there is thus only one independent parameter associated with the longitudinal dynamics.

\begin{figure}[h!]
    \centering
    \includegraphics[scale=0.6]{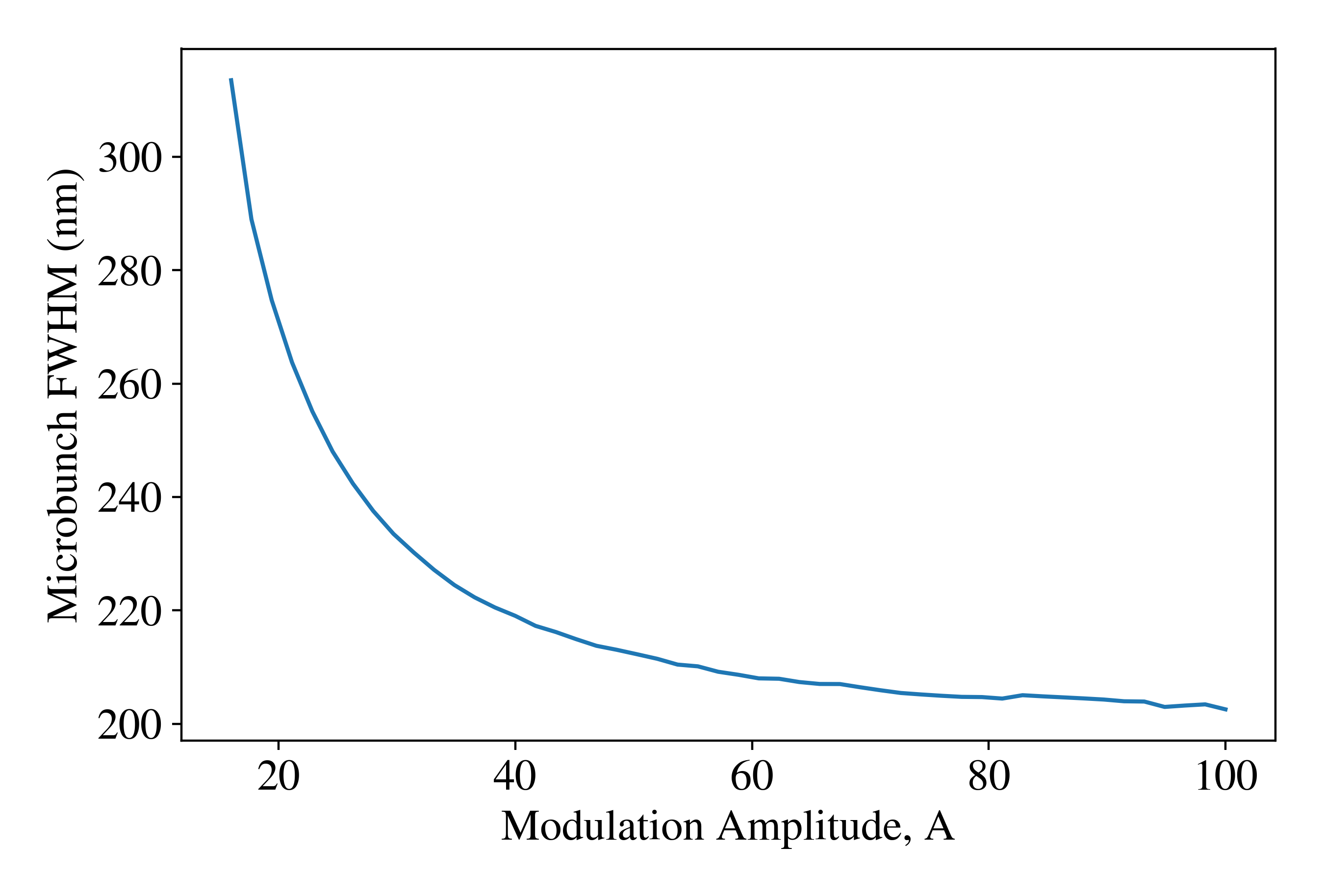}
    \caption{Full-width at half-maximum of 4 kA microbunches modulated by a 10 micron laser as a function of the normalized modulation strength, $A$.}
    \label{fig:FWHM}
\end{figure}

The determination of the modulation strength is motivated by a trade-off between the micro-bunch length and the slice energy spread. In the case at hand, the results of this trade-off become less clear as a result of collective effects. Here one is concerned primarily with CSR, which induces energy loss in the micro-bunches with a magnitude comparable to that of the applied energy modulation itself. In Figure \ref{fig:FWHM} we show the dependence of the micro-bunch FWHM on the normalized modulation amplitude for fixed 4 kA peak current and a 10 micron laser wavelength. This plot indicates a sharp drop-off in micro-bunch length for $A$ smaller than ~30 and a relatively gradual change for larger values. In our case, collective effects are too strong to preserve the slice emittance at a reasonable level for $A$ below 30; for the values of $A$ allowed by our system the micro-bunch FWHM is nearly independent of operating parameter variations. 

With these considerations in mind, we are left with two optimization goals: the simultaneous minimization of the energy spread and the emittance. Under these assumptions, these characteristics are tuned using three independent parameters: the modulation strength $A$ and the horizontal Courant-Snyder parameters at the entrance to the chicane $\alpha_{x}$ and $\beta_{x}$. We found that optimization of the beam envelope verified the common assumption that emittance growth is minimized when the beam reaches a small waist in the final bend magnet. This largely fixes $\alpha_x$ for a given $\beta_x$. In principle there is also the question of the optimal chicane geometry. We found a relatively simple dependence in this case: shorter bend magnets produced smaller emittance growth due up to a point where the bending strength is unfeasible. The minimum in our case was around 10 cm-long bend magnets, which we employ below. 

Once the above principles are followed we found a relatively weak dependence of the output parameters on the modulation amplitude as long as $A$ was not exceedingly small. We take $A=34$, which allows for a small compression strength $B=0.0256$, corresponding to an $R_{56}=1.14$ mm in the final chicane. This should be compared to the $R_{56}$ that would be required if we preserved the energy chirp from BC1 and compressed the full beam by a factor of 10, which would be roughly 6 mm, or roughly four times larger than the required ESASE chicane value. Furthermore, we chose $\beta_x=10$ m at the entrance of the chicane because this small value was relatively easy to achieve using our assumed quadrupole lattice. It should be noted that this set of parameters is not a unique global optimum -- it is simply a functional solution given the demands of the FEL. Other variations on the chosen parameter set did not produce notable improvements in the micro-bunching performance. 

\subsection{Dynamics in the Modulator and Second Linac}

\begin{figure}[h!]
    \centering
    \includegraphics[scale=0.4]{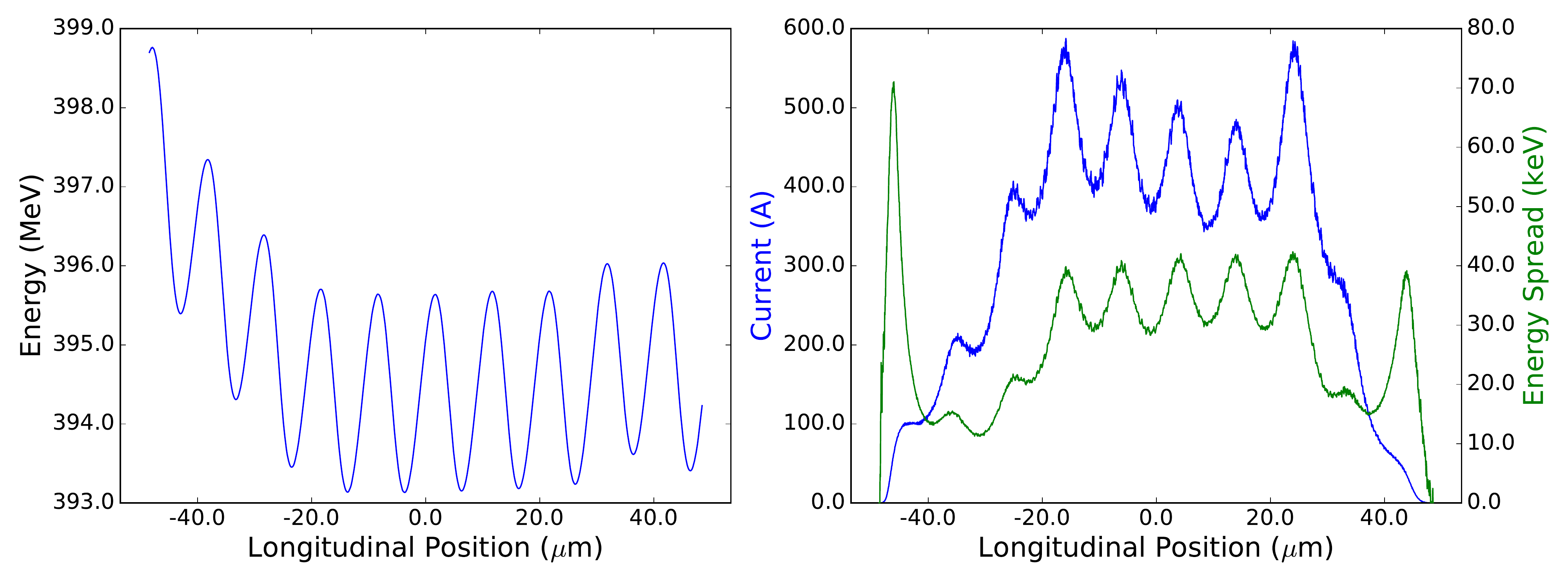}
    \caption{(Left) The longitudinal phase space and (right) current and energy spread profiles are plotted just after modulation. }
    \label{fig:Modulated}
\end{figure}

The electron beam must satisfy the resonance condition with the $\lambda_L=10$ $\mu$m laser inside of the modulator,
\begin{equation}
    \lambda_L = \frac{\lambda_\mathrm{mod}}{2\gamma_0^2}\left(1+\frac{K_\mathrm{mod}^2}{2}\right)
\end{equation}
where $\gamma_0$ is the average beam energy. The practicalities of dealing with the $1/\gamma_0^2$ scaling, causes us to choose to energy-modulate immediately after BC1. Further, we note that in order to wavelength-tune an XFEL with a low-$K_u$ undulator, it is necessary to change the final beam energy. As we are employing a resonant interaction in the IFEL modulator, it is convenient to use a fixed energy after tBC1. The parameters for the modulator are reported in Table \ref{tab:Modulator}, and the modulated phase space is plotted in Figure \ref{fig:Modulated}. We note that the momentum compaction in the modulator initiates the development of the current spikes. The modulated beam is then accelerated on-crest through five one-meter linac sections to its final energy, in the current case of interest, of 1 GeV.

\begin{table}[h!]
    \centering
    \begin{tabular}{|c|c|c|}
        \hline 
        Parameter & Units & Value  \\
        \hline 
        Undulator period, $\lambda_\mathrm{mod}$ & cm & 15\\
        Peak undulator field & T & 0.87\\
        Number of periods & & 10\\
        Laser wavelength, $\lambda_L$ & $\mu$m & 10\\
        Laser waist & mm & 0.926\\
        Laser peak power & MW & 145\\
        \hline 
    \end{tabular}
    \caption{Parameters of the laser modulator.}
    \label{tab:Modulator}
\end{table}

\subsection{Dynamics in the Second Compressor Chicane}

\begin{figure}[h!]
    \centering
    \includegraphics[scale=0.4]{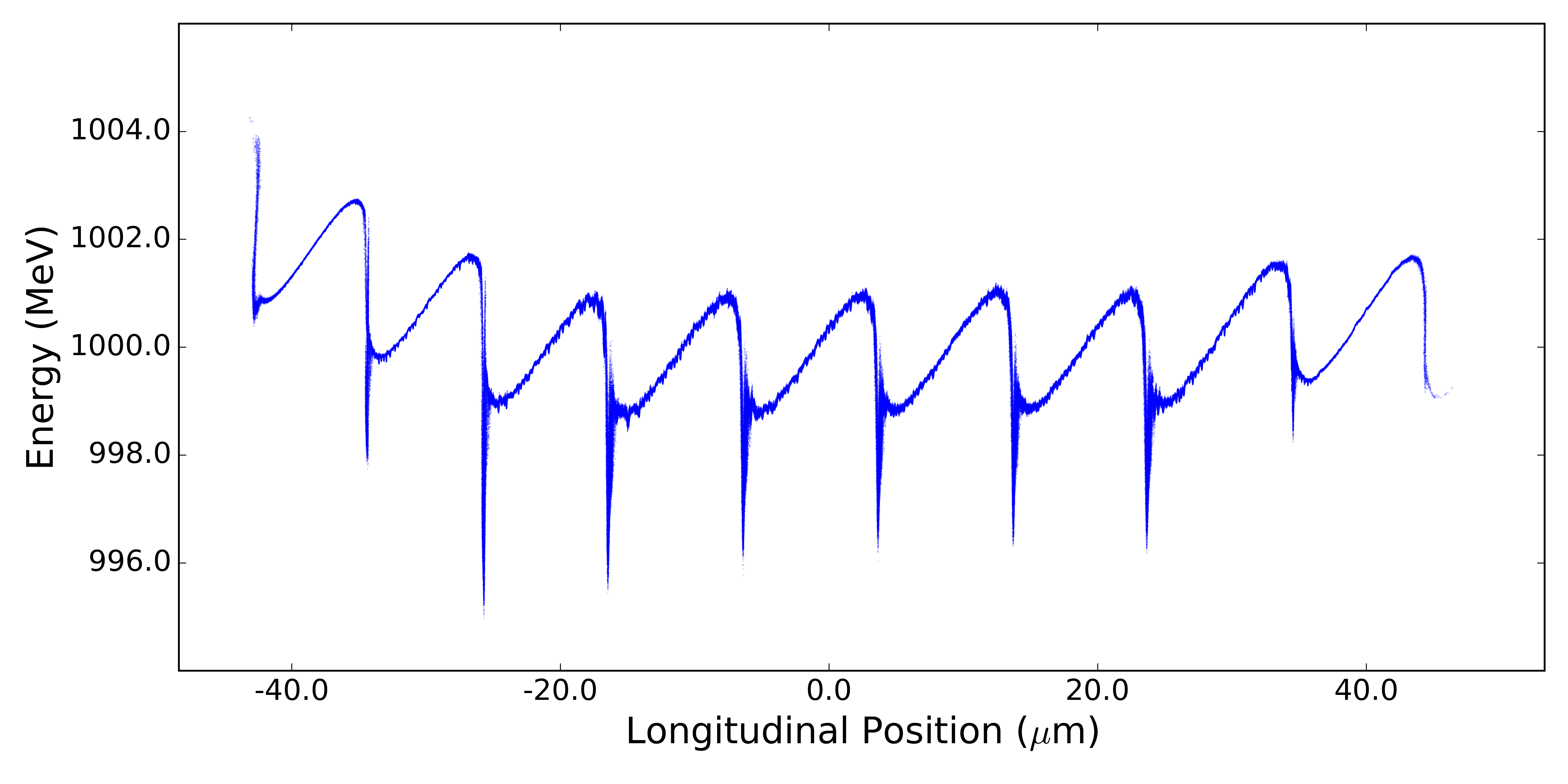}
    \caption{Full longitudinal phase space after the second compressor. The five micro-bunches with the most apparent energy loss, ranging from roughly $-20$ $\mu$m to $25$ $\mu$m, constitute the 4 kA current spikes. }
    \label{fig:Compressed}
\end{figure}

The second compressor (BC2) chicane is very compact, occupying a length of only 2.6 meters. The chicane design is specified in Table \ref{tab:BC2}, and an image of the longitudinal phase space after the chicane is shown in Figure \ref{fig:Compressed}. Although the ESASE method allows for smaller momentum compaction, 1D simulations have suggested that CSR can still present a non-negligible problem, as discussed briefly in \cite{Robles2019}. Due to the extremely short length scales present at relatively low energy, the energy loss induced by CSR effects in the compressor chicane is non-negligible. By shortening the magnet lengths, however, this loss may be diminished to the point where it does not disrupt the formation of the current spikes, incur excessive slice emittance growth, or otherwise strongly impact XFEL performance. 

\begin{table}[h!]
    \centering
    \begin{tabular}{|c|c|c|}
        \hline
        Parameter & Units & Value\\
        \hline 
        Magnet length & m & 0.1 \\
        Drift length & m & 1.0 \\
        Bend angle & $^{\circ}$ & 1.32 \\
        $R_{56}$ & mm & 1.14 \\
        Entrance $\beta_x$ & m & 10 \\
        Entrance $\alpha_x$ & & 4 \\
        \hline
    \end{tabular}
    \caption{Parameters of the final bunch compressor.}
    \label{tab:BC2}
\end{table}

\begin{figure}[h!]
    \centering
    \includegraphics[scale=0.4]{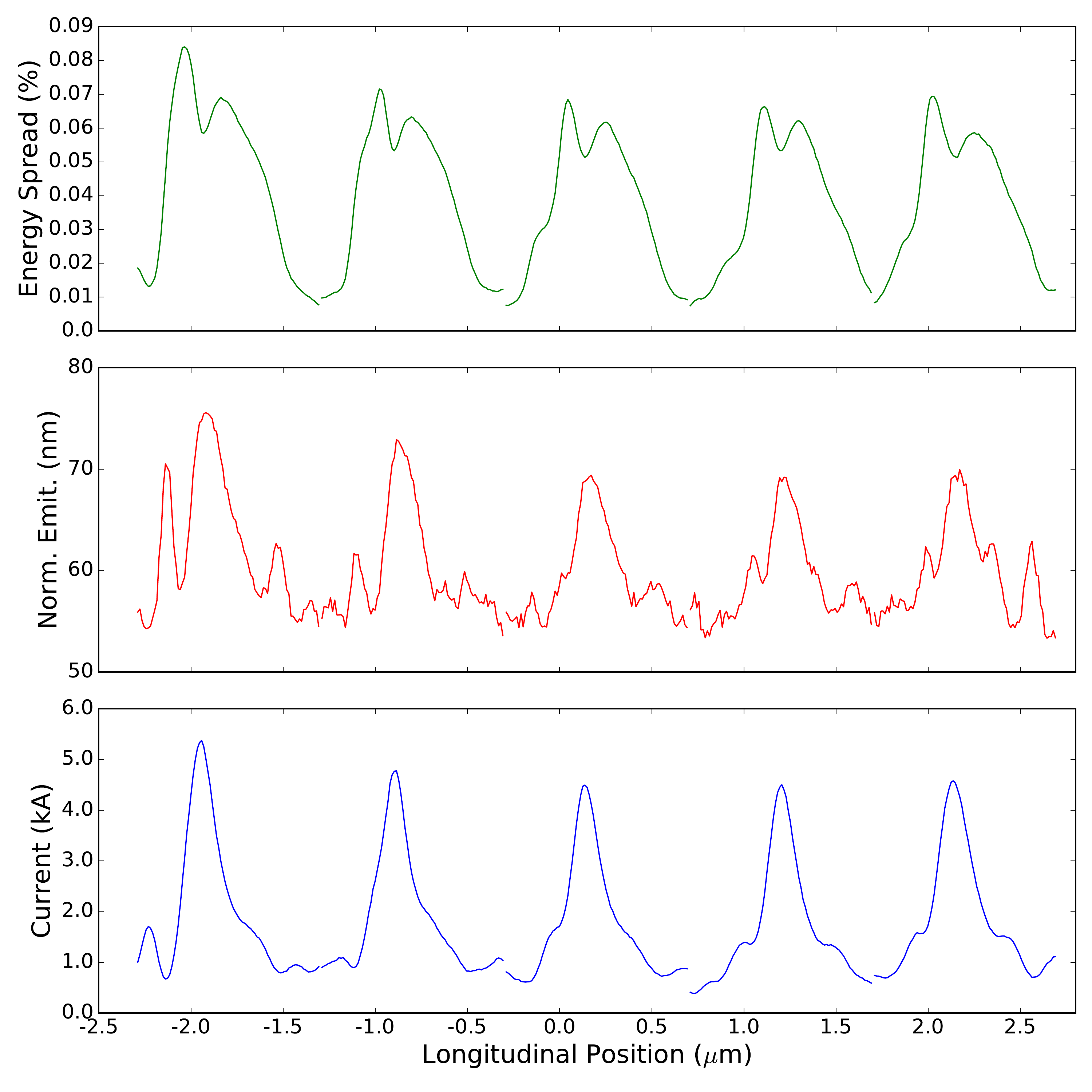}
    \caption{The five final current spikes are shown (bottom) underneath their slice normalized emittance (middle) and slice energy spread (top). It can be seen that the slice energy spreads remain well below $0.1\%$ and the emittance is largely maintained at 70 nm inside of the 4 kA current spikes. }
    \label{fig:Spikes}
\end{figure}

The structure of the current spikes is explored further in Figure \ref{fig:Spikes}. We have taken a 1 $\mu$m window around each spike and plotted them directly next to each other showing the current profile, emittance, and energy spread inside of each spike. The spikes have a FWHM length of approximately 250 nm, or $\sim$250 SXR wavelengths. As discussed below, it can be seen that the slice emittance has been well-preserved at a level which will not compromise FEL performance, remaining at the 70 nm level for most of the spikes. The relative energy spread obtained is well below $10^{-3}$, yielding the desired parameters for the UC-XFEL. 

\subsection{3D CSR Effects During Compression}

\begin{figure}[h!]
    \centering
    \includegraphics[scale=0.7]{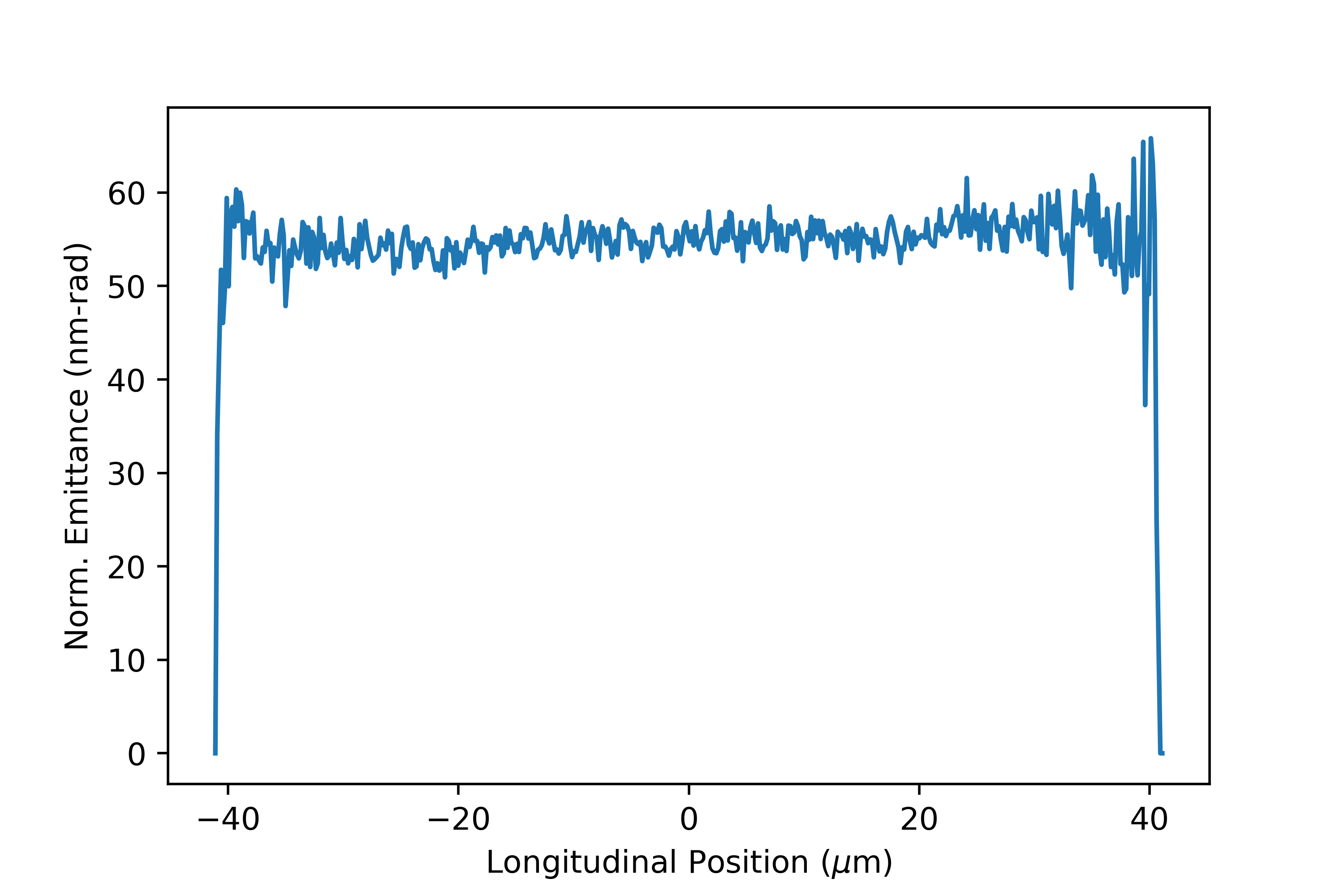}
    \caption{The slice emittance after propagation through the first bunch compressor simulated by GPT using a 3D CSR model. }
    \label{fig:gptbc1}
\end{figure}

To ensure our that the analysis of the compression schemes is valid we have used GPT to study both bunch compressors with 3D CSR effects included. The first compressor is a relatively standard configuration, so no significant differences were expected. Consistent with this assumption, the 3D model of GPT predicts a normalized emittance of 61 nm-rad after BC1, to be compared with the 65 nm-rad figure predicted by \texttt{elegant}. Similarly, the slice emittance is unaffected as it was in the 1D simulations. We show the slice emittance produced by GPT in Figure \ref{fig:gptbc1}. This is consistent with the observations of Ref. \cite{Brynes_2018}, in which it was shown that the 1D model employed by \texttt{elegant} tends to overestimate CSR induced emittance growth relative to both 3D models (GPT, CSRTrack) and experiment. 

\begin{figure}[h!]
    \centering
    \includegraphics[scale=0.6]{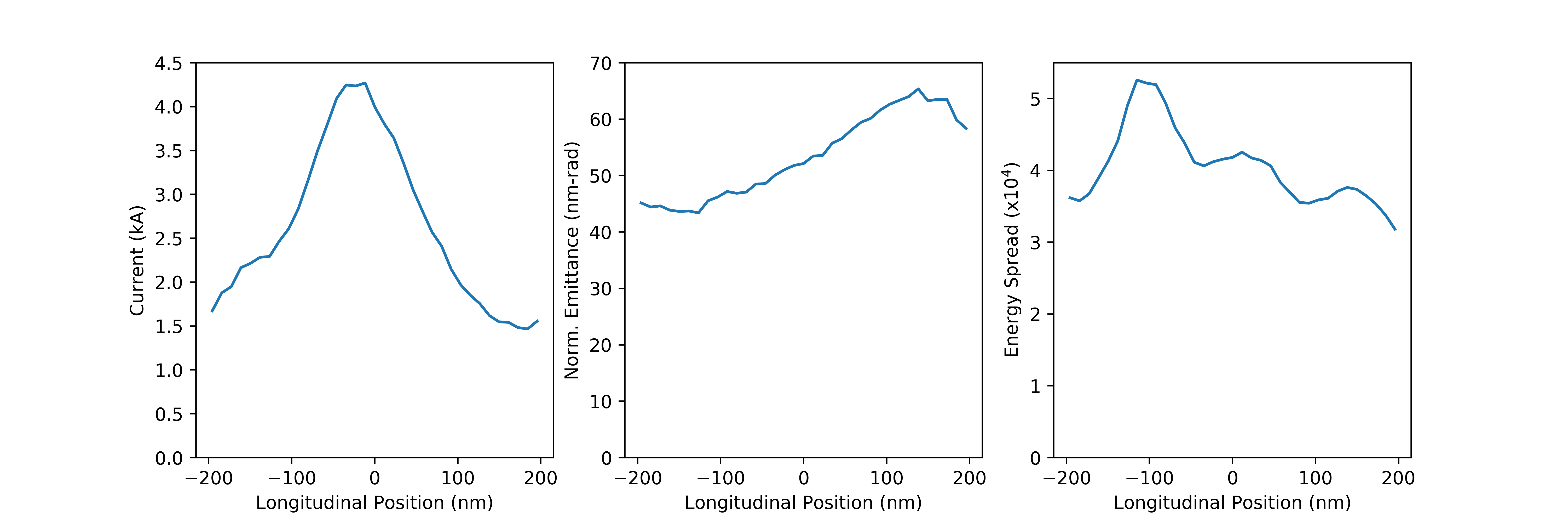}
    \caption{The results of a full 3D simulation of the final bunch compresssor are shown for one current spike. }
    \label{fig:gptbc2}
\end{figure}

Indeed, with the exceedingly short lengths of the micro-bunches produced by BC2, it is reasonable to expect that the accuracy of the 1D model may be compromised. This is commonly evaluated using the Derbenev criterion \cite{Derbenev1995}, which requires that $\sigma_x (R\sigma_z^2)^{-1/3}<<1$; at the end of the final chicane, where the beam is tightly focused, this parameter is maximized at $~0.03$ (note that it is much smaller in BC1). Despite this small value, we have performed simulations of BC2 using the 3D CSR model included in GPT. We simulated the entire bunch through the compressor, and for clarity we show the results for a 400 nm window around one of the current spikes in Figure \ref{fig:gptbc2}. We have observed two interesting results from these studies. First, the 1D model employed by \texttt{elegant} overestimates the CSR effects, as in the case of the first compressor. More surprisingly, the complicated 3D CSR fields rearrange the particles in the slices of the beam such that the final emittance profile in the current spike resembles a linear ramp. In some places the emittance has even dropped from the original slice emittance value, and in the center of the spike with the highest current the emittance is roughly the original $55$ nm-rad. We note that this is not a violation of Liouville's theorem, as it is due to a removal of correlations in the transverse phase space.. All told, there does not seem to be a noticeable dilution of the brightness as observed in \texttt{elegant}, This complex behavior underlying possible improvements to be exploited from a 3D CSR analysis will be the focus of future studies. However, to preserve the well understood start-to-end suite of codes, GPT-\texttt{elegant}-GENESIS, we take the conservative results from \texttt{elegant} in which slight brightness dilutions are observed, as our   predictions for the FEL simulation input.

\subsection{Micro-bunching Instability Considerations}

Conventional XFELs are plagued by the so-called micro-bunching instability (MBI) \cite{borland2001start, BORLAND2002268, heifets2002coherent}, whereby a very low energy spread beam from the photoinjector has induced energy modulations that are amplified by collective effects.  A combination of longitudinal space charge (LSC) forces during acceleration, CSR in the bunch compressors, and the added longitudinal rearrangements due to the momentum compaction of the compressor bends, act together to produce unstable growth of the energy modulations. Specifically, the sequence of effects is that, first,  LSC forces coupled with beam shot noise lead to high-frequency modulation of the beam energy. This is followed by the initial bends in the compressor which convert this energy modulation into density modulation. The bends contribute further energy modulation due to enhanced CSR. This amplified density modulation drives an even stronger LSC-induced energy modulation in the next linac. 

This first-pass design for the UC-XFEL beamline does not display indications that MBI will be a notable problem, as evidenced by the results shown in the previous section. This result can be understood using the basic outline of the development of MBI detailed above. Although the photoinjector beam produces an exceedingly low energy spread, which leads to significant susceptibility to MBI if not increased with a laser heater \cite{Huang2004}, the very high accelerating gradients employed in the linac, combined with the relatively low final beam energy, yield a very short total linac length -- between 8 and 13 m active length. This short interaction distance does not permit LSC energy modulation to develop to the same extent it can in other, full-scale instruments such as the LCLS, which has a total linac length nearly two orders of magnitude larger than the UC-XFEL. The mitigation of MBI is yet further aided by the small momentum compaction associated with the IFEL-based bunch compressor. The near-elimination of MBI (as well as IBS) in a very high brightness beam-based XFEL is a key advantage inherent in the compactness of the UC-XFEL approach. 

\subsection{Uncorrelated Energy Spread Considerations in ESASE Compressors\label{subsec:espread}}

Predicting the uncorrelated energy spread in simulation is a particularly difficult task, as those physical phenomena (\textit{e.g} MBI and IBS) which directly affect the energy spread are some of the most difficult  to accurately model. Indeed, recent start-to-end studies of the LCLS beamline showed simulations consistently under-estimating the uncorrelated energy spread relative to experimental measurements \cite{qiang2017start}. For this reason, it is important that any conceptual design, such as that we present here, is robust to under-estimation of the slice energy spread. In this section we thus place an upper bound on the allowable energy spread in the UC-XFEL and also discuss the unique benefits of using an ESASE compressor in this respect. 

\begin{figure}[h!]
    \centering
    \includegraphics[scale=0.7]{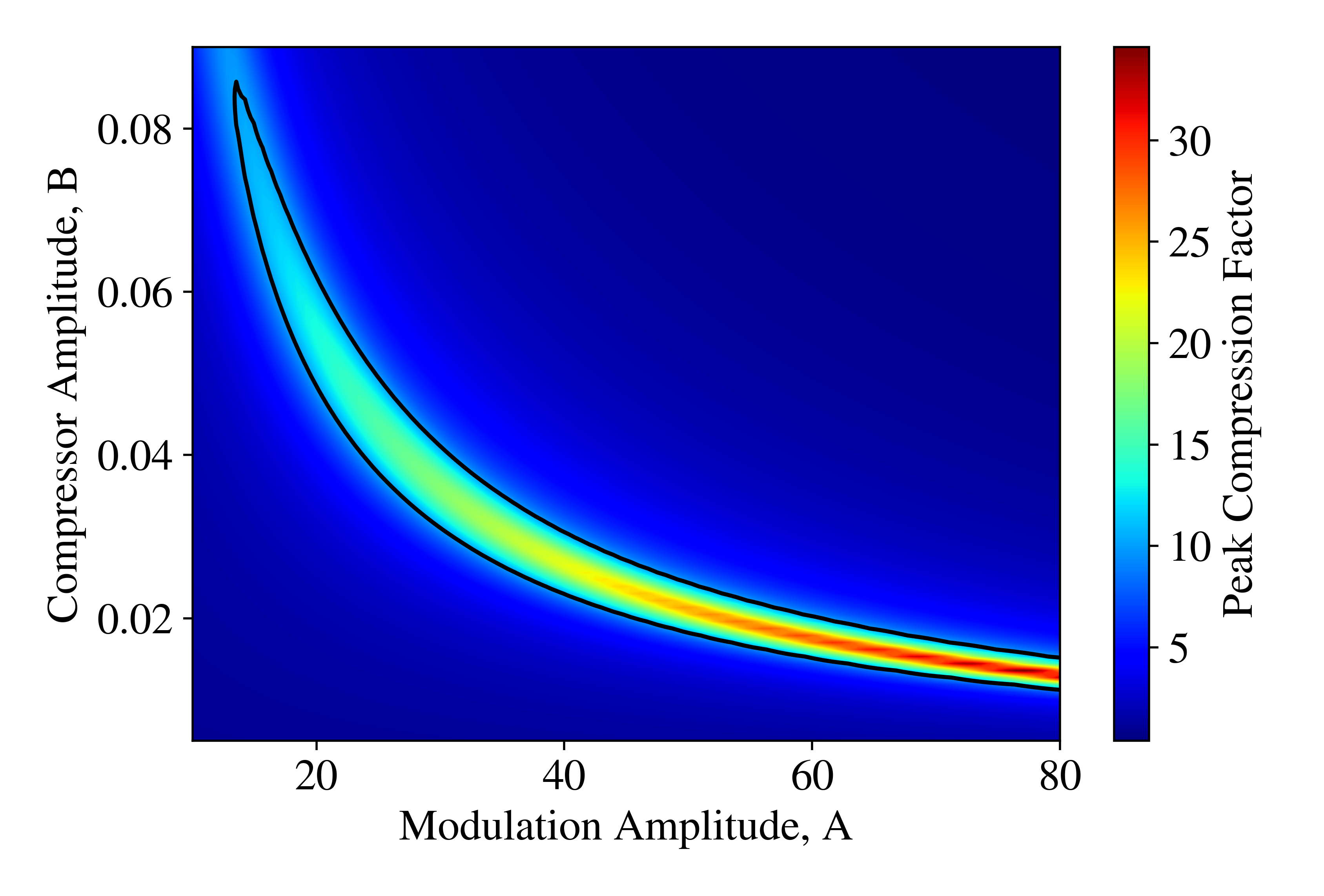}
    \caption{The compression strength of an ESASE compressor is plotted against the normalized modulation and compression amplitudes. The black contour line denotes compression by a factor of 10.}
    \label{fig:ESASECompression}
\end{figure}

As we discussed briefly in Section \ref{sec:operatingParameters}, our ESASE design utilizes a particularly large energy modulation relative to the uncorrelated energy spread of the beam, as quantified by the parameter $A=\Delta\gamma/\sigma_\gamma$. For our current design $A\simeq 34$, a choice is motivated by a desire to minimize the $R_{56}$ of the compressor (through choice of the parameter $B= k_LR_{56}\sigma_\gamma/\gamma$). Because of this large value of $A$ our final energy spread is dominated by the modulation rather than the uncorrelated spread,  and upstream increases in the energy spread do not necessarily imply increases in the final energy spread found in the high-current micro-bunches. For reference we show in Figure \ref{fig:ESASECompression} the peak compression factor as a function of both the modulation amplitude and the compressor amplitude. We have included a black contour line indicating compression by a factor of 10. This line is multi-valued, with the upper portion indicating over-compression and the lower portion indicating under-compression. As previously mentioned, we avoid over-compression due to CSR considerations. 

This figure shows clearly that compression by a factor of 10 is possible for $A$ as low as about 13. If we consider variations of the energy spread while fixing the modulation amplitude, we can thus tolerate nearly three times as much uncorrelated energy spread as we currently obtain from our analysis without compromising our compressor strength or our energy spread. Additionally, in moving from $A\approx 34$ to $A\approx 13$ the value of $B$ increases by close to a factor of three itself, meaning that there would only be a small change in the momentum compaction of the compressor since $B$ scales linearly with the uncorrelated spread. This is important, since the chicane $R_{56}$  is what determines the strength of collective effects. Additionally, as we noted before, moving to a lower value for $A$ has the additional benefit of lengthening the final current spikes, which mitigates potentially harmful slippage effects in the XFEL. A larger uncorrelated energy spread also serves to suppress the already small effect of MBI. 

We also note that one can tolerate even more than three times the currently determined uncorrelated spread by also increasing the physical modulation amplitude $\Delta\gamma$. There is some freedom to do this if necessary, since in the current design  $\sigma_\gamma/\gamma = 6\times 10^{-4}$ and the FEL permits energy spreads as high as $1\times 10^{-3}$, as is discussed in the FEL simulation sections below. This implies that one can tolerate a  five-fold increase in the uncorrelated energy spread without compromising the performance of the FEL. Such a change would require a re-optimization of the chicane. Even then, the expected effects would be beneficial as the larger modulation amplitude would imply a smaller momentum compaction and therefore weaker collective effects. This is a critical conceptual advantage of the ESASE compression scheme: the eventual energy spread in the FEL is nearly independent of the beam's uncorrelated energy spread, as long as the uncorrelated spread does not surpass a threshold specified by the requirements of the machine. 

\section{Short-period Undulators}

We consider here two different beam energies and two different undulator designs for the SXR and HXR examples discussed below. The SXR case operates near 10 {\AA} using a $U_e=$ 1 GeV electron beam and conservative undulator parameters: a $B_0=1$ T peak field with a $\lambda_u=6.5$ mm period. This is comparable to previously demonstrated hybrid Halbach undulators \cite{OShea2010},  which have measured a $B_0=1.1$ T peak field within a 2 mm gap for a 7 mm period undulator at room temperature, expected to increase to 1.33 T when cryogenically cooled to 30 K \cite{OShea2016}. This technology, based on Pr or Dy material, is by now well-diffused and mature \cite{Murokh2014, OShea2016, Tanaka2019}. 

\begin{figure}[h!]
    \centering
    \includegraphics[width=1.0\linewidth]{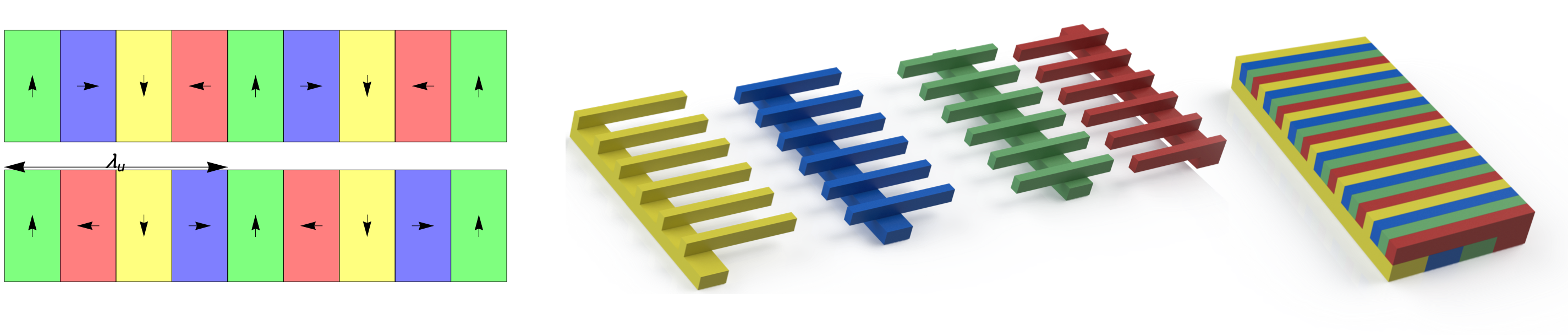}
    \caption{Rendering of a proposed undulator and assembly approach in the style of \cite{Majernik2019Comb}, showing a standard Halbach geometry with manufacturing techniques adapted to mm-period scale.}
    \label{fig:Combover}
\end{figure}

The HXR case examines the generation of x-rays near 1.6 {\AA} using a $U_e = 1.6$ GeV electron beam and a more ambitious undulator design; combining the concepts of cryogenically cooled, hybrid Halbach undulators with comb fabrication \cite{Majernik2019Comb} to produce a 1 T peak field in a 3 mm period undulator. Below this period length, the feasibility of the undulator faces challenges attendant with use of MEMS fabrication techniques \cite{Harrison2012, Harrison2014}.   A rendered schematic indicating a proposed Halbach geometry is shown in Figure \ref{fig:Combover}. While these undulators have not yet been realized, they have been studied in detail, and at the periods under discussion, the challenges of fabrication and tuning should be met. 

\begin{table}[h!]
\centering
\begin{tabular}{|l|l|l|l|}
\hline
Parameter                  & Units    & Soft x-ray        & Hard x-ray        \\ \hline
Undulator period, $\lambda_u$                     & mm       & 6.5                 & 3.0                 \\ 
Peak undulator field, $B_0$                    & T        & 1.0                 & 1.0                 \\ 
Undulator parameter, $K_u$                          &          & 0.60             & 0.28            \\ \hline
\end{tabular}
\caption{Summary of undulator specifications for soft and hard x-ray UC-XFEL simulations.}
\label{tab:undulatorSummary}
\end{table}

\begin{figure}[h!]
    \centering
    \includegraphics[width=1.0\linewidth]{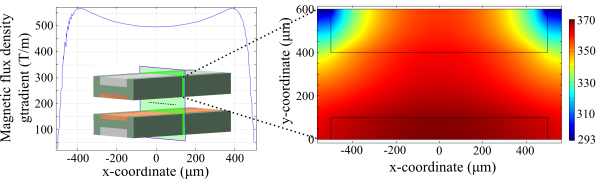}
    \caption{Geometry and simulated performance of modified, scaled-down Panofsky quadrupole for intra-undulator focusing in UC-XFEL. The right side shows a map of the device temperature under continuous operation in units of degrees Kelvin.}
    \label{fig:panofsky}
\end{figure}

Such undulators (Table \ref{tab:undulatorSummary}) have notable systematic issues to address in implementation. The undulator parameter is naturally low, and while the loss is mitigated somewhat in the scaling of the coupling parameter $\theta_{\max}=K_u/\gamma$, the electron-radiation coupling can be impacted, particularly at very low $K_u$. Further, short-period undulators are compact and have narrow gaps, which requires that the beam be well-focused during traversal of the undulator. Small focused spot sizes also will yield high beam density, potentially improving XFEL gain. Several innovative approaches enable the use of very strong (up to 3 kT/m gradient) focusing quadrupoles in  micro electro-mechanical systems (MEMS) structures integrated with the undulator \cite{Harrison2015}. Recently, this work has concentrated on a modified Panofsky quadrupole geometry, which naturally may be placed inside the gap of the cryogenic undulator, and operated cryogenically to obtain up to 500 T/m gradient. A simulation of this device, which is currently under development at UCLA, is shown in Figure \ref{fig:panofsky}. Note the major simplifying modification of this miniaturized device is the lack of current layers on vertically oriented side walls. 

As will be seen through discussion of the FEL simulations, the optimized $\beta$-function needed for this system is in the range of 60-80 cm. For the 1 GeV beam in the SXR case, this can be accomplished by focusing with an intra-undulator 40 cm-period focus-defocus (FD) quadrupole lattice. The peak field gradient in this case is $B'=$375 T/m, which is within the capability of the modified Panofsky quadrupole introduced above. One may also consider a VISA-style focusing schemes \cite{Carr2001}, with a pair of (opposing) vertically polarized permanent magnets placed inside the Halbach array, to arrive at a field gradient of $B'>150$ T/m. Lengthening the lattice period would enable focusing adequate to optimize the XFEL performance.

We note that for sub-cm $\lambda_u$ undulators the narrow, mm-scale gap (including possible focusing elements) produces an effect that can have large impact on the UC-XFEL: resistive wall wakefields. With the low energy scale of the instrument, energy loss due to these wakefields, which is enhanced by the nearness of the material boundaries, can be a serious issue. It is fortunate that the bunch-train format associated with the IFEL ESASE method serves to mitigate these wakefields, as discussed in the following section.

\section{Resistive Wall Wakefields} 

\begin{figure}[h!]
    \centering
    \includegraphics[width=0.9\linewidth]{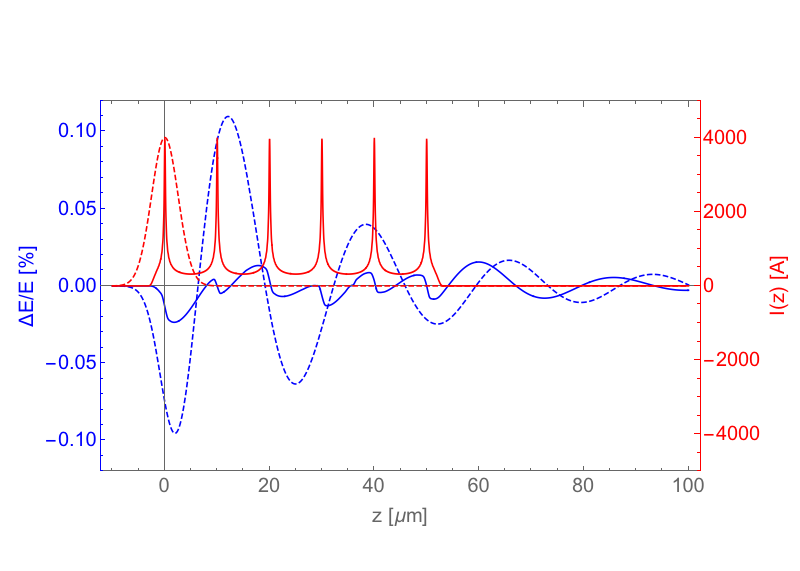}
    \caption{The relative energy loss for micro-bunch train (solid) and a single bunch (dashed) as they transit a cryogenic (RRR = 100) copper flat wall structure with gap size of 2 mm and an overall length of one meter. Both beams have an energy of 1 GeV, total charge of 85 pC and peak current of 4 kA, but the train splits the charge over 6 micro-bunches with 10 $\mu$m inter-bunch spacing. The energy loss as a percent is in blue and the beam current is in red.}
    \label{fig:combinedWakePlot}
\end{figure}

Resistive wall wakefields are an important consideration for these high current beams in a small gap undulator and, as such, have been investigated in some detail. A recent article \cite{Stupakov2015} discusses a regime of resistive wall wakes particularly relevant to the UC-XFEL parameter space: a cryogenic, flat wall geometry with ultra-short beams. The details of this analysis are included in \ref{app:resistiveWakes} but the results for the particular case of the SXR UC-XFEL are shown in Figure \ref{fig:combinedWakePlot}. The scenario under consideration -- that of a cryogenic resistive wall response to a micro-bunched beam -- has some similarities with the case of the proposed MaRIE-style XFEL discussed in \cite{Carlsten2019}. However, in our case the undulator is very different, having short period, and thus substantially smaller gap, and is also operated at cryogenic temperatures.  We find that the impact of these wakes on FEL gain length in UC-XFEL, despite the disadvantages in geometry and beam energy as compared to the MaRIE example, is negligible by virtue of the dramatically shorter saturation length.

To illustrate the behavior of the resistive wall effect in this unique case, we analyze an idealized train of six IFEL micro-bunches, produced from the parameters of section \ref{sec:operatingParameters}, with 10 $\mu$m inter-bunch spacing passes through the cryogenic (RRR = 100) copper flat wall structure with gap size of 2 mm and an overall length of one meter. Each micro-bunch is approximately a gaussian with $\sigma_z$ =  424 nm and a peak current of 4 kA with an inter-bunch current of approximately 320 A. All particles have energy $U_e$ = 1 GeV. We compare this result to that obtained from examining a beam with the same total charge in a single, longer pulse, also with a peak current of 4 kA. It can be seen that the resistive wall wake amplitude is strongly suppressed by the use of a micro-bunch train instead of a single, fully compressed spike. This is because the wake is notably dissipated in the time between micro-bunches. This damping does not strongly assert itself, however, during the passage of the single, full charge bunch. For the micro-bunched case, as we will see in the following section, the energy spread induced by these wakes does not significantly impact the XFEL performance. 

\section{Soft X-ray FEL Performance}
Two representative cases of the UC-XFEL are considered using the 3D, time-dependent FEL simulation code, GENESIS 1.3 \cite{Reiche1999}. The first is the SXR case, where the performance of the system is based on self-consistent simulations of the component parts of the XFEL system, \textit{i.e.} the GPT and \texttt{elegant} simulation suite output. The second is the HXR scenario, which requires some improvements in both the undulator and beam performance over those demonstrated either in simulation or in experiment. We begin with the detailed discussion of the SXR case. 

\begin{figure}[h]
    \centering
    \includegraphics[width=\linewidth] {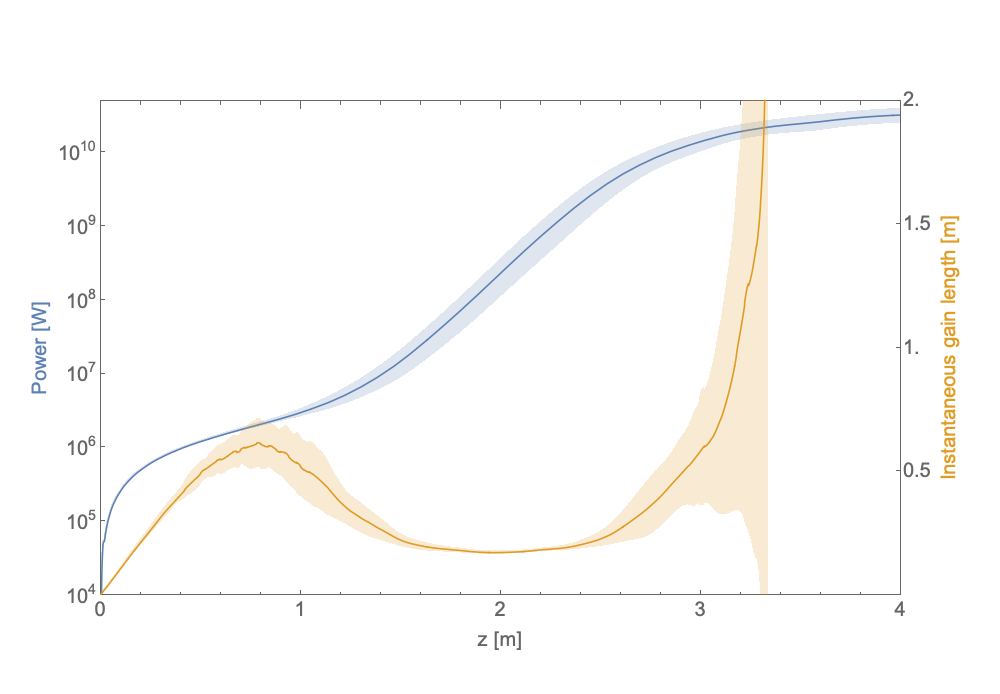}
    \caption{Plots showing the peak power (blue) and instantaneous gain length (orange) for GENESIS simulations of the SXR case detailed in Table \ref{tab:SXRSummary}. The shaded area corresponds to one standard deviation based on 16 SASE averaged runs.}
    \label{fig:SXRplots}
\end{figure}

The beam and undulator parameters used in the XFEL simulations, as well as other performance characteristics, are summarized in Table \ref{tab:SXRSummary} for a single ESASE micro-bunch. In both cases, uncorrelated longitudinally gaussian beams were generated based on the moments expected from the accelerator and beamline simulations discussed above. Due to the large inter-bunch spacing (10 $\mu$m) relative to the slippage length (100's of nm), each micro-bunch can be treated as independent in the simulations. The final results presented in Table \ref{tab:SXRSummary} and in Figure \ref{fig:SXRplots} are each based on 16 time dependent, SASE averaged GENESIS runs. The SXR case presented here is thus a result of a complete simulation of the subcomponents of the system. 

The spatial characteristics of the beam critically include both its longitudinal and transverse properties. We have discussed the approach to transverse focusing above; this produces a beam with small spot size ($<5$ $\mu$m), presenting challenges in measurement that may be overcome using novel optical methods \cite{PotylitsynIBIC2018,Sukhikh2017} as well as coherent imaging \cite{MarinelliCDI} or high resolution wire-scanners \cite{borrelli2018}. The methods employed in coherent imaging used to reconstruct the spatial profile are borrowed from those used in x-ray science. In order to employ them in beam diagnosis, one should imprint coherence onto the emission process (coherent transition radiation in \cite{MarinelliCDI}). This is accomplished naturally in the UC-XFEL, through the tight micro-bunching imposed in the ESASE section. With each micro-bunch having rms length of $0.42$ $\mu$m, transition radiation emission on harmonics of the ESASE laser frequency up to $\lambda_\mathrm{TR}=2\pi\sigma_z\simeq 2.5$ $\mu$m (4th harmonic of $\lambda_L$ for the SXR case) are coherently emitted. By observing the diffractive far-field intensity of the CTR at this wavelength, one may robustly reconstruct the beam spot profile. 

For the longitudinal diagnostics, one may extend the range of existing RF deflector-based methods \cite{Xdeflector} straightforwardly by using the same K$_\mathrm{a}$-band (34.27 GHz) multi-MW source available for the longitudinal phase space linearizer to create a sweeping measurement that can reach femtosecond resolution. In addition, to see the details of the micro-bunches at a finer length scale, a hybrid IFEL/deflector-based approach termed \textit{attosweeper} can be employed \cite{Attosweeper},  Finally, we note that by direct observation of the form of the CTR spectrum over a range of wavelengths from $\sim\lambda_L$ down to partially coherent harmonics, one may reconstruct the form of the micro-bunch train \cite{Tremaine1998,Liu1998}.  The discussion of deployment of diagnostics and their locations is given in \ref{app:diagnostics}. 

\begin{table}[h!]
\centering
\begin{tabular}{|l|l|l|}
\hline
Parameter                  & Units    & Value        \\ \hline
Energy                     & GeV      & 1.0                 \\ 
Energy spread              & \%       & 0.1              \\ 
Micro-bunch charge                     & pC       & 14.2     \\
Micro-bunch rms length, $\sigma_z$                   & nm       & 424               \\ 
Peak current               & kA       & 4.0                  \\ 
Normalized emittance, ($\epsilon_{n,x}$, $\epsilon_{n,y}$) & nm-rad   & (80, 60)          \\ 
Mean spot size, $\sigma_r$             & $\mu$m       & 4.9              \\ \hline 
Undulator period, $\lambda_u$                     & mm       & 6.5                                \\ 
Peak undulator field, $B_0$                    & T        & 1.0                                \\ 
Undulator parameter, $K_u$                          &          & 0.60                      \\  
Undulator length & m & 4 \\ \hline
Radiation fundamental, $\lambda_1$                  & {\AA} & 10.0                \\ 
Photon energy              & keV      & 1.2                \\ 
Gain length, ${L}_{\mathrm{g, 3D}}$                & m        & 0.21             \\ 
Radiation peak power                 & GW       & 25                \\ 
Radiation rms bandwidth &  \% & 0.046 \\
Radiation pulse energy/$\mu$bunch               & $\mu$J       & 19.2              \\
$\mu$bunch count &  & 6 \\
Radiation pulse energy/train               & $\mu$J       & 115.2              \\ 
Number of photons/train     &     & $6\times 10^{11}$                \\ 
$\rho$                   &    $10^{-3}$      & 3.1            \\ 
$\rho_{\mathrm{3D}}$                   &    $10^{-3}$      & 1.4 \\ 
$L_\mathrm{g, 3D}/L_\mathrm{g, 1D}$                  &          & 2.2  \\ \hline
\end{tabular}
\caption{Summary of parameters and results for soft x-ray UC-XFEL simulations.}
\label{tab:SXRSummary}
\end{table}

The value of $\rho$ calculated from Equation \ref{eq:rhodefn} is derived from one-dimensional, idealized theory. We have optimized through simulation the various values of emittance, energy spread and focusing $\beta$-function. It is instructive to revisit the definitions introduced concerning this optimization procedure in light of the final scenario identified through simulation. We have, for the case discussed: $\eta_\epsilon=0.45$, $\eta_\gamma=0.31$ $\eta_\beta=0.35$. These tuning parameters indicate the complex trade-offs that one must make when optimizing an FEL design. The value of the gain parameter which is instead deduced empirically from simulation or experiment through determination of the observed gain length $L_\mathrm{g, 3D}=(\mathrm{d}(\ln P)/\mathrm{d}z)^{-1}$, quantitatively takes into account all effects contributing to its value not included in the 1D theoretical value. This parameter is often termed $\rho_\mathrm{3D}$, and it is compared to $\rho$ in Table \ref{tab:SXRSummary}. These effects include energy spread, finite emittance, radiation and  diffraction, and space-charge, as discussed in Refs. \cite{Marcus2011} and \cite{Xie2000}; for completeness we note that in our simulation analysis the effects of slippage are also included. The value of $\rho_\mathrm{3D}$, obtained through examination of simulation data shown in Figure \ref{fig:SXRplots} (via use of the relation $L_\mathrm{g, 3D}=\lambda_u/4\pi\sqrt{3}\rho_\mathrm{3D}$) is compared to the calculated $\rho$ in Table \ref{tab:SXRSummary}. It is decreased by a factor of 2.2. Of the increase in gain length, approximately fifty percent is due to diffraction.  


\begin{figure}[]
    \centering
    \includegraphics[width=1.0\linewidth] {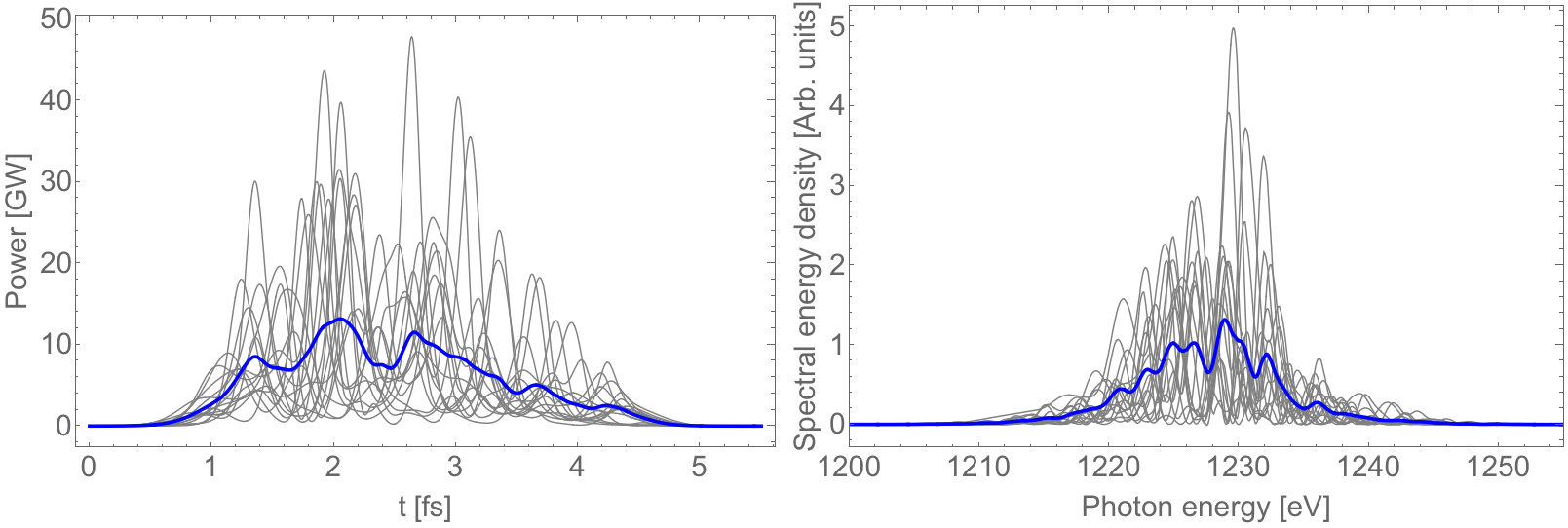}
    \caption{The temporal power profile (left) and energy density spectra (right) at the undulator exit for each of the 16 runs (gray) and the average result (blue). The ensemble-average rms bandwidth is 0.046\%.}
    \label{fig:SXR-spectra}
\end{figure}

With all of these effects included, the overall performance of the SXR case for the UC-XFEL is quite robust, producing the desired few percent of energy obtainable from present km-scale instruments. In this regard, one should also appreciate the multi-pulse format of the X-rays as well as the number of photons in deciding how to evaluate the utility of the UC-XFEL in application. The SXR FEL proceeds to saturation in a remarkable 4 m of undulator length with an rms photon energy spread, 0.046\%, consistent with typical approximations \cite{huang2007} (see Figure \ref{fig:SXR-spectra}, which also shows the temporal power profile at undulator exit), with three-dimensional gain length $L_\mathrm{g, 3D}$ also predicted well by the scaling laws given in Ref. \cite{Marcus2011}. It should be noted in this regard that the observed cooperation length determined by $L_{c}=\lambda_r/4\pi\sqrt{3}\rho_\mathrm{3D}$ is, at 33 nm, smaller than the rms electron micro-bunch length.  It is known that if the micro-bunch length approaches $2\pi L_c$ (and up to $4\pi L_c$) in single-spike mode \cite{Rosenzweig2008}, that the output fluctuations of the FEL are stabilized above multi-mode long-pulse SASE \cite{HalavanauPC}, diminishing to $\sim$20 percent. The slippage inherent in the cooperation length also has a slight suppressing effect on the final SXR photon output and associated value of $\rho_\mathrm{3D}$.  This micro-bunch-derived effect is less important in a HXR FEL case, as discussed in the following section.  

As we operate with a relatively low electron energy, $U_e$, and high value of $\rho_\mathrm{3D}$ in this case, the efficiency of converting beam energy to photon energy is quite high, 
\begin{equation}  
\eta_\mathrm{FEL} =  \frac{N_\gamma \hbar\omega_r}{N_e U_e} = 1.35\times 10^{-3},
\label{eq:efficiency}
\end{equation}
which is nearly identical to $\rho_\mathrm{3D}$, as one might expect. The electron beam energy needed to create one photon of 1.2 keV is, with this efficiency, 890 keV. By comparison, one needs over 7 MeV of energy in an LCLS-based scenario to create an equal-energy photon. In order to compare wall-plug power needs for the UC-XFEL with those of existing sources, one must specify the beam loading level utilized; however, in the linear collider study of Ref. \cite{bane2018advanced}, it is found that a highly efficient system based on cryogenic RF acceleration is possible. Thus one may see that the energetic cost of x-ray photons in the UC-XFEL may be notably smaller than in existing full-scale sources. 

\begin{figure}[h]
    \centering
    \includegraphics[scale=0.475]{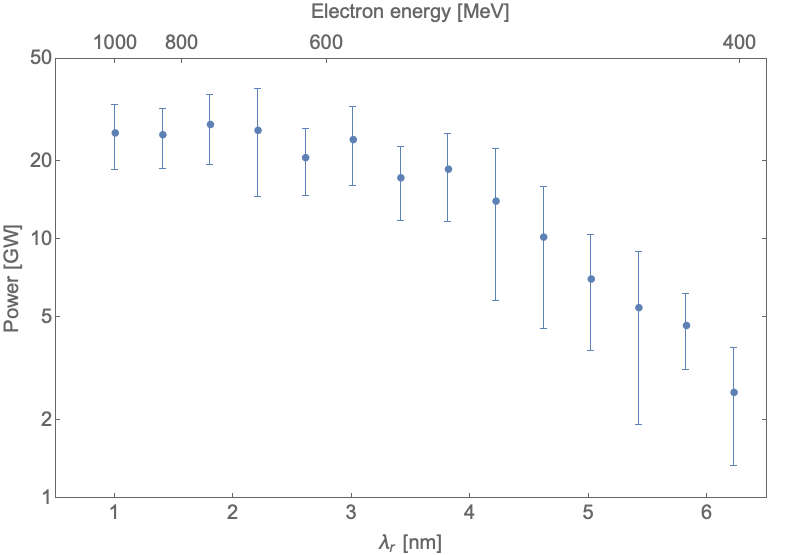}
    \caption{Demonstration of the tunable range of the SXR UC-XFEL described in Table \ref{tab:SXRSummary}. The electron energy range scales from the minimum constrained by the bunching process to the full, 1 GeV energy. With the exception of energy and matched spot size, parameters are otherwise held constant. Each instance shown gives the mean peak power and one standard deviation error bars from 16 SASE averaged GENESIS simulations.}
    \label{fig:SXRTunability}
\end{figure}

The model used for evaluating the SXR case is also useful in exploring the wavelength tunability of this UC-XFEL system, As noted above, the energy of the beam at the undulator can be shifted by changing the acceleration downstream of the IFEL modulator, as well as the settings of the final chicane. The performance of the SXR over the relevant tuning range is shown in Figure \ref{fig:SXRTunability}. It can be seen that one may access wavelengths up to $\sim$4 nm before there is a significant loss of power. This power loss is driven in the example SXR scenario by diffraction and slippage effects. 

\section{Extending UC-XFEL to Hard X-ray Lasing}
A detailed analysis of  the SXR case was made possible by the fact that the sub-components needed have been identified through design and simulation work. Simulations of the injector have already produced the emittance and beam brightness needed for the case summarized in Table \ref{tab:HXRSummary}, at $\epsilon_n=50$ nm. However, for hard x-ray lasing, some progress in downstream systems must be made. Critically, the emittance of the beam after compression must be improved; this implies a renewed emphasis on the performance of the compression system. Further, the undulator performance with $\lambda_u=3$ mm must be improved to reach the level needed for HXR lasing. This awaits further development of concepts such as the comb Halbach array. 
 
 Returning to the issue of optimized compression, with a radiation wavelength shorter by a factor of 6.5, the HXR case can utilize a beam with notably smaller micro-bunch length. As such, we examine a case where the ESASE laser system operates at $\lambda_L=3.2$ $\mu$m, thus lowering the micro-bunch charge (the peak current remaining at $I_p=4$ kA) and mitigating collective effects in compression as well as those due to the resistive wall wakes. We note that for the HXR case that the beam energy is also increased to 1.6 GeV; this permits lasing approximately at the original LCLS design wavelength $\lambda_r=1.6$ {\AA}. 

\begin{table}[h!]
\centering
\begin{tabular}{|l|l|l|l|}
\hline
Parameter                  & Units     & Value        \\ \hline
Energy                     & GeV                  & 1.6                 \\ 
Energy spread              & \%                   & 0.03              \\ 
Micro-bunch charge                     & pC                 & 4.7               \\ 
Micro-bunch rms length, $\sigma_z$                   & nm               & 140               \\ 
Peak current               & kA                     & 4.0                 \\ 
Normalized emittance, ($\epsilon_{n,x}$, $\epsilon_{n,y}$) & nm-rad      & (50, 50)          \\ 
Mean spot size, $\sigma_r$             & $\mu$m                      & 4.1               \\ \hline 
Undulator period, $\lambda_u$                     & mm                        & 3.0                 \\
Peak undulator field, $B_0$                    & T                         & 1.0                 \\ 
Undulator parameter, $K_u$                          &                      & 0.28            \\  
Undulator length & m & 6 \\ \hline
Radiation fundamental, $\lambda_r$                  & {\AA}           & 1.6               \\ 
Photon energy              & keV               & 7.8               \\ 
Gain length, $\mathrm{L}_{\mathrm{g, 3D}}$                & m               & 0.33              \\ 
Radiation peak power                 & GW                   & 8.3                \\ 
Radiation pulse energy/$\mu$bunch               & $\mu$J                 & 2.4               \\ 
$\mu$bunch count &  & 6 \\
Radiation pulse energy/train               & $\mu$J                 & 14.4               \\ 
Number of photons/train     &     & $1.13\times 10^{10}$                \\
$\rho$                   &    $10^{-3}$                  & 0.78           \\ 
$\rho_{\mathrm{3D}}$                   &    $10^{-3}$       & 0.42  \\ 
${L}_{\mathrm{g, 3D}}/{L}_{\mathrm{g, 1D}}$                  &           & 1.9  \\ \hline
\end{tabular}
\caption{Summary of parameters and results for hard x-ray UC-XFEL simulations.}
\label{tab:HXRSummary}
\end{table}

\begin{figure}[h!]
    \centering
    \includegraphics[width=\linewidth] {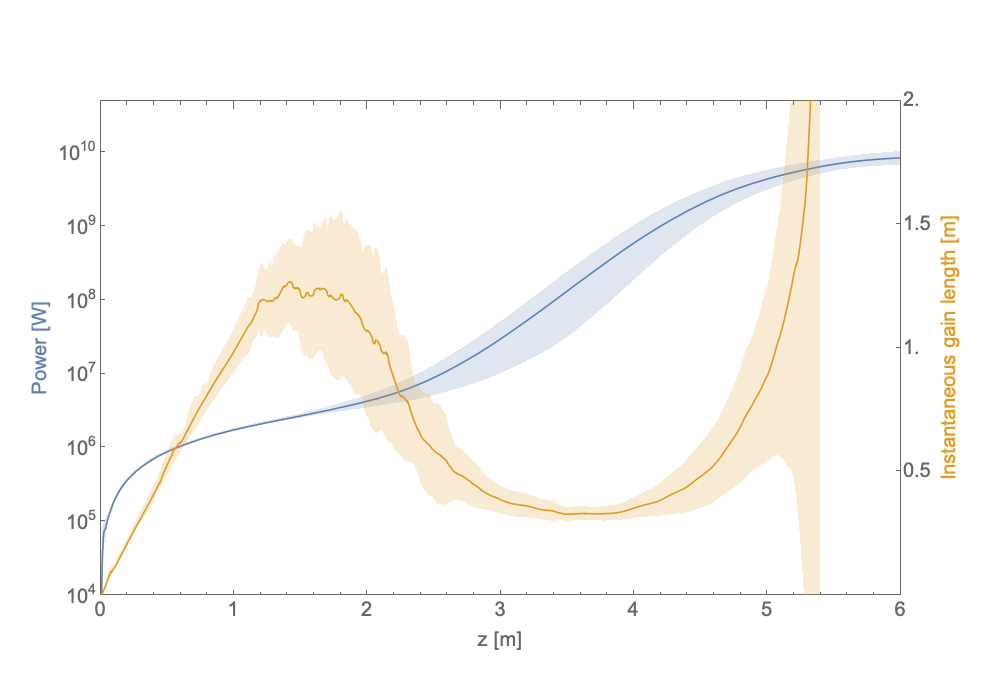}
    \caption{Plots showing the peak power (blue) and instantaneous gain length (orange) for GENESIS simulations of the HXR case detailed in Table \ref{tab:HXRSummary}. The shaded area corresponds to one standard deviation based on 16 SASE averaged runs.}
    \label{fig:HXRplots}
\end{figure}

The beam and undulator parameters used in this study are shown in Table \ref{tab:HXRSummary}, along with a summary of the results obtained from GENESIS. The simulation output is given graphically in Figure \ref{fig:HXRplots}, which shows the predicted performance for the case under study.  The gain length is near 30 cm, with attendant saturation, at 8.3 GW (15 $\mu$J/pulse train), obtained in less than 6 m. It can be seen that the slippage is well controlled despite the reduction of the micro-bunch length. As can be ascertained by the length of the accelerator and the undulator, the overall length of the system is foreseen to be expanded by approximately 50 percent over the SXR case. 

This study, as introduced above, does not yet correspond to a self-consistent parameter set obtained from simulation of all component parameters. Small improvements in the injector 6D brightness performance are needed, and these have already been obtained in simulation \cite{Rosenzweig2019}. A path to the undulator design, utilizing the comb Halbach approach, is also outlined above, but has yet to be realized. The  advances needed to reach the performance outlined here are found in the compression system. As such, promising design studies to mitigate emittance growth in the compressors, exploiting a deeper understanding of the computationally challenging effects of 3D CSR, are ongoing. 

\section{X-ray Optics}

While the downstream x-ray optics of a XFEL might experience nonlinear effects due to the enormous peak power density, the dominant factor for damage is the peak energy density \cite{LCLSCDR}. This concern drives many of the design considerations for the UC-XFEL x-ray optics. It is instructive to review some of the historical record as well as the current state-of-the-art in this area. The x-ray optic transport system for an XFEL faces some extra challenges with respect to those encountered in the field of synchrotron radiation. As the beam is almost fully coherent, as well as possessing a high energy density, two major problems have to be well controlled: wavefront preservation and surface damage. If one needs to further monochromatize the radiation, especially in the SXR spectral region, the beam stretching effect caused by a grating monochromator should be minimized to preserve a transform limited pulse. Non-ideal effects can be mitigated by designing a monochromator in which the dominant effect limiting the resolving power is the number of illuminated lines on the grating.  

The main problem to confront in avoiding deterioration of the x-ray beam quality is the degradation of the wavefront due to mirror imperfections. To quantify this effect, the Strehl ratio (SR) \cite{mahajan1983}, is often used. The SR represents the ratio between the obtained or simulated peak intensity and what is available from a perfect optical system. It therefore may take on values between 0 and 1. The SR can be expressed as:
\begin{equation}
    \mathrm{SR} \simeq \mathrm{e}^{-(2\pi\phi)^2} \simeq 1-(2\pi\phi)^2. 
\end{equation}
Here $\phi$ is the  $\lambda_r$-dependent phase error introduced by the non-ideal optics on the wavefront. In the case of a grazing incidence mirror, with rms shape deviation from the desired profile $\delta h$ and grazing angle-of-incidence $\theta$, this phase error can be expressed as
\begin{equation}
    \phi =  \frac{2\delta h \sin\theta}{\lambda_r} . 
\end{equation}
From the equations above, one can calculate the maximum tolerable shape error to achieve a given SR. For a general system with $N$ separate optics this shape error is:
\begin{equation}
   \delta h = \frac{\lambda_r\sqrt{1-\mathrm{SR}}}{\sqrt{N}4\pi\sin\theta} . 
\end{equation}

One must next examine restrictions on the optics' shape errors. The Maréchal criterion \cite{marechal1970}, is often used to assess the quality of an optical system. It states that an adequate optical system must possess  SR $\ge$ 0.8, {\sl i.e.} 80 percent of the available photons are actually focused within the ideal spot.  This is a useful criterion if the optical system is employed to deliver light at a focus. Away from the focus, one needs a more rigorous criterion. It has been demonstrated, during the studies for the upgrade of LCLS, that a SR $\ge$ 0.97 is required to maintain the beam uniformity out of focus. 

These concepts permit us to consider a possible scenario for the UC-XFEL. The limited available space for the optics, while posing a potential problem for optical damage, as described below, may be advantageous in limiting the effect of the shape errors on the wavefront. In fact, the low divergence of the beam and the vicinity of the mirrors to the source, allows the use of very shallow grazing incidence angle on relatively short mirrors. From the above discussion, it is evident that this permits relaxed requirements on the shape error needed to achieve a SR in excess of 0.97. As an example, a mirror located 5 m from the end of the undulators, mirrors having angles as shown in Table \ref{tab:Xrayoptics}, require the rms shape errors shown for Strehl ratios of 0.97 and 0.8. The angles of incidence in this example are chosen to maintain the requirements for shape errors to an achievable value with the current state of the art mirror polishing capabilities, while keeping the x-ray flux below the damage threshold and avoiding excessive optical system lengths, as described below.

\begin{table}[h!]
\centering
\begin{tabular}{|l|l|l|l|l|l|}
\hline
Parameter & Units                 & SXR  & SXR & HXR      & HXR \\ \hline
Photon energy                     & eV                 & 300 & 1200 & 3000 & 7800                 \\ 
Angle of incidence        & mrad                & 7 & 7 & 2  & 2                \\ 
Footprint on the mirror (FWHM)        & mm                & 65 &	16	& 60	& 25                \\ 
rms shape errors, SR $>$ 0.97 & nm	& 8.1 &	2.0	& 2.8	& 1.1 \\ 
rms shape errors, SR $>$ 0.8 & nm	  & 21 &	5&	7.3	& 2.8 \\ \hline
\end{tabular}
\caption{Overview of x-ray optics design requirements.}
\label{tab:Xrayoptics}
\end{table}

\begin{figure}[h!]
    \centering
    \includegraphics[width=\linewidth] {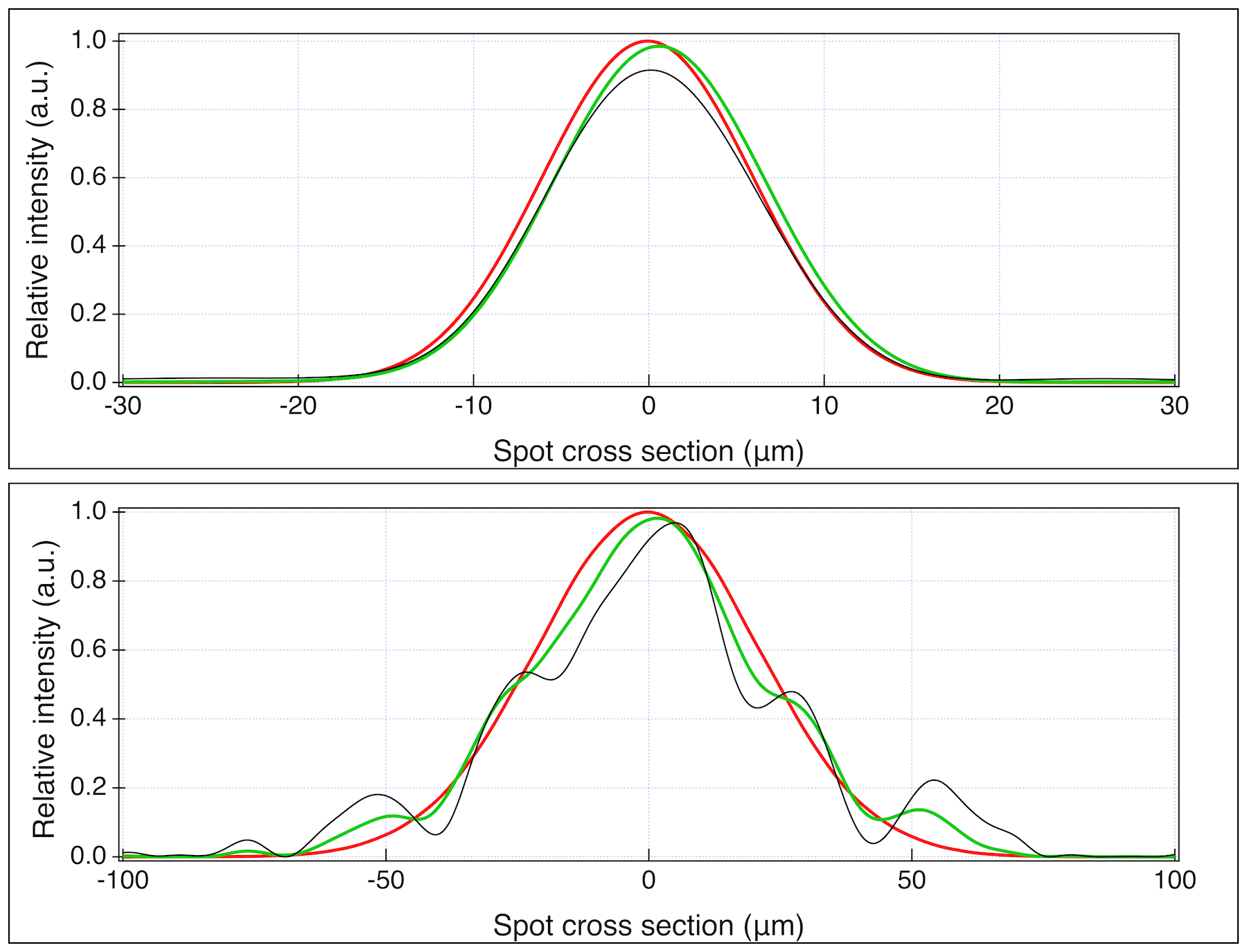}
    \caption{(top) In-focus spot profile at 1.2 keV with a 1:1 focusing optics in the cases of: red, perfect mirror; green; 2 nm rms shape error mirror, corresponding to SR=0.97; black: 5 nm rms shape error, corresponding to SR=0.8. (bottom) Out of focus spot obtained by bending the mirrors. Color code is the same as for upper figure. }
    \label{fig:Xrayspots}
\end{figure}

 The mirrors envisioned for this optical system are pre-shaped plane elliptical mirrors mounted in the so-called Kirkpatrick-Baez configuration \cite{Kirkpatrick48}. Each mirror focuses along its tangential direction and the two mirrors are installed one after the other at respective 90 degree angles. The mirrors are installed on a two-actuator bender to preserve the elliptical profile over various focal distances. For such short mirrors and these types of elliptical profiles, shape errors of 1 nm rms are achievable with state-of-the-art polishing. Further, bending mechanisms able to maintain this mirror quality have been designed at both the LCLS or ESRF. With 5 m focal length mirrors as described here, the expected beam profiles, both in and out of focus, are shown in Figure \ref{fig:Xrayspots}.
 
With the dual purposes of characterizing the wave front of the radiation produced by the undulators and of optimizing the beamline performance in mind, this system is an ideal candidate for employing a wave front sensor (WFS).
In the recent years, wave front sensors, in both the soft and hard x-ray regime \cite{grizolli2017, idir2010, mercere2003} have been developed and optimized, reaching accuracies of the order of $\lambda$/50 rms. This is sufficient to enable measurement of a Strehl ratio in excess of 0.95. With a highly simplified operating principle, a WFS is based on a 2D periodic array of holes located on either a converging or diverging beam (\textit{i.e.} not in focus). A x-ray direct imager, or a scintillator coupled with a visible camera, is located downstream of the pin-hole array and records the distribution pattern due to the radiation passing through the holes. By properly selecting the spacing and diameter of the holes, one may obtain, or not, an interference pattern.  In the first case, the WFS will work on a shearing (or Talbot) principle. In the second case, it functions according to the Hartmann (or Shack-Hartmann) principle. In both cases, one measures the difference between the ideal distribution of the image of the holes array on the camera and the actual one observed. This provides a direct measurement of the distortion of the wavefront, and, indirectly, the measurement of the characteristics of the source. 
For the hard X-ray measurement, in particular, a diamond based beam splitter can be used \cite{marathe2014}. This can provide non-invasive WFS able to either optimize the undulator chain performance or the mirror focus. In the SXR case, a WFS has been successfully used on FEL beamlines \cite{floter2010, liu2018}. Nonetheless, in this region, the WFS can only be invasive, or, if the sample is semi-transparent, it can be located downstream of the focal spot.

\begin{figure}[h!]
    \centering
    \includegraphics[width=0.9\linewidth]{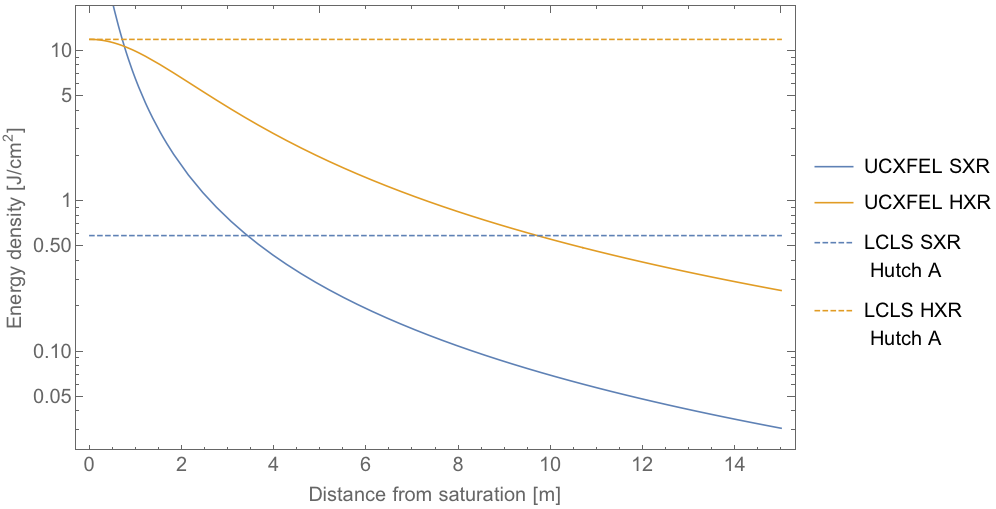}
    \caption{Plot of the peak energy density of the fundamental radiation produced by the SXR and HXR UC-XFEL configurations as a function of distance downstream from the undulator. Also shown are the peak energy densities experienced by optics at the LCLS hutch A for their SXR and HXR configurations.}
    \label{fig:radPeakPower}
\end{figure}

The other major potential problem related to optical transport of ultra-intense and ultra-short-pulse x-ray sources is damage to the optical surface. All the energy absorbed by the mirror in the short time of a pulse length cannot be thermally dissipated. Therefore, the material on the surface should be able to dissipate the energy in the crystalline atomic structure, without breaking of bonds or reaching the instantaneous melting temperature.  To address these concerns, the LCLS utilizes two optics hutches, indicated as A and B, at distances of 50 and 400 m downstream of the undulator \cite{LCLSCDR}.  The x-ray pulse carried 3 mJ of FEL fundamental radiation at 0.8 keV in the LCLS SXR configuration and 2.5 mJ of 8.3 keV photons in the corresponding HXR configuration, resulting in peak energy densities of 0.59 and 11.9 J/cm$^{2}$ in hutch A and 0.01 and 0.57 J/cm$^{2}$ in hutch B for SXR and HXR respectively. Comparing these values with the results obtained from GENESIS simulations, and summarized in Tables \ref{tab:SXRSummary} and \ref{tab:HXRSummary}, the peak energy as a function of distance has been plotted for the UC-XFEL SXR and HXR cases, and compared with the values calculated for the LCLS case (Figure \ref{fig:radPeakPower}). In each case, the energy of the full pulse train has been used since the inter-pulse timing is compared to typical electronic thermalization times \cite{ThermalMetal}. The plot illustrates that both UC-XFEL cases drop to energy density levels below those in LCLS hutch A in less than 5 meters, facilitating the use of focusing optics and harmonic rejection on an unattenuated beam in a very compact footprint. In this regard, it is estimated that the instrumentation in the X-ray measurement instrument hutch may be maintained below 3.5 meters in total length.

The energy density delivered to an optical surface is responsible for the potential damage of the optical element. Several completed and ongoing studies on damage have been performed. These results are not conclusive, in the sense not all the mechanisms involved in the deterioration of the optical properties are fully understood. However, for low repetition rate machines such as the UC-XFEL, a conservative approach is to consider 1 eV/atom as the damage threshold for most materials, with a safe limit for operation being 0.1 eV/atom. Details on how to calculate the adsorbed dose can be found in the LCLS conceptual design report \cite{LCLSCDR} and in \cite{Cocco_2015}. For the present discussion, we have calculated the absorbed dose for some optics that may be envisioned to be used with the UC-XFEL. For simplicity, and to consider a conservative case, all the optics are located at  5 m downstream of the source. The optics considered are: the K-B focusing mirrors for the SXR range with two potential coatings: a harmonic rejection system composed of 4 mirrors to reduce, by 6 orders of magnitude, the contamination of the third harmonic at 400 eV; a diffraction grating with blaze profile and blaze angle of 0.40; a K-B mirror system for hard x-ray; a Si(111) crystal monochromator; and a zone plate for HXR nano-focusing.  Table \ref{tab:XrayopticsEnergyAbsorption}, gives the values calculated using the maximum power per bunch of 0.114 mJ for the SXR at 1200 eV and 0.039 mJ at 7800 eV. One can immediately see that in almost all cases the optical components operate at safe levels. The only exception is the Au zone plate. This is because high-$Z$ materials tend to absorb the radiation within a few nm of the surface, depositing all the energy within a few atomic layers. Tests carried out at the LCLS \cite{XrayDavid} have confirmed that Au-based zone plates do not survive the FEL radiation, while a diamond-based zone plate does. Fortunately, the technology for developing diamond zone plates has advanced notably and it is now possible to obtain small spots (near to 100 nm) without compromising flux or wavefront requirements.  Diamond-based solutions operating in cryogenically cooled conditions are also favored for x-ray optics thanks to an exceptionally  high thermal conductivity and a very small coefficient of thermal expansion for temperatures T $\le$ 100 deg K \cite{Stoupin2010, Stoupin2012}.  Current work in these areas at UC-Merced is based on in-house developed heat transfer, dynamic diffraction and thermo-mechanical analysis codes that are now being integrated to form a complete tool for systematic photo-thermal-mechanical characterization and analysis of thermal loading effects on crystal optics.  

\begin{table}[h!]
\centering
\begin{tabular}{|l|l|l|l|l|}
\hline
Optics type                & $\theta_i$ $[^{\circ}]$  & 	Coating & Photon energy [eV]     & 	Absorbed dose  \\ 
~                & ~  & 	~     & ~ &	[meV/atom] \\ \hline
KB mirror                & 0.40  & 	B$_4$C  & 1200      & 	0.88  \\ \hline
KB mirror                & 0.40  & 	Pt  & 1200      & 	12.1  \\ \hline
Harmonic suppressor     & 2.0  & 	Si  & 400      & 	12.7  \\ \hline
Grating (0.40 blaze angle) & 1.0  & 	B$_4$C  & 1200      & 	9.7   \\ \hline
KB mirror & 0.11  & 	W & 7800      & 	12.1  \\ \hline
Si(111) monochromator	& 75.31 &	Si &	7800  & 4.39  \\ \hline
Zone plate &	90.0	 & Diamond &	7800  &	0.2  \\ \hline
Zone plate &	90.0	 & Au &	7800 &	103 \\ \hline
\end{tabular}
\caption{Survey of x-ray optics designs, showing optic type, angle of incidence $\theta_i$ and associated energy absorption.  The nominal pulse energies of 0.114 and 0.039 mJ for the SXR and HXR cases, respectively, are used}
\label{tab:XrayopticsEnergyAbsorption}
\end{table}	

\section{Brief Survey of UC-XFEL Applications}

We have introduced above, at the level of detail needed to establish feasibility, a new design paradigm for an affordable x-ray free-electron laser that will make available coherent, ultra-fast sources to the broad scientific community in a ubiquitous fashion, similar to the way that optical lasers are currently employed routinely in laboratories. The dominance of present XFELs in single-shot HXR imaging is not in doubt; this is an important but relatively small subset of possible applications. However, as dictated by operational limitations, current XFELs tend to produce one-of-a-kind results in a short period of running – on the scale of a few days. Relaxing these restrictions to make coherent x-ray sources widely distributed will give a new, open path to significant  results in x-ray-based science and applications. This is particularly true in the SXR regime, where the demand for photon flux is lessened. The model we envision is that the UC-XFEL becomes a ubiquitous technology, with many such instruments available to the user community. We now proceed to give examples of the promise brought by the flexibility of use of the proposed instrument; these applications represent a small sample of those enabled by a UC-XFEL.  

\subsection{Physics Applications}
The UC-XFEL has a unique temporal format that naturally permits attosecond-level synchronization of the x-ray pulses with the phase of an infrared laser.  In this regard, we note that the overall length of the pulse train and its periodicity (that of the IFEL laser wavelength) are chosen in this paper on the basis of optimizing the FEL gain. This produces, for example, in the SXR FEL presented, a train of six pulses separated by 10 microns each. There is some flexibility in these choices, and the optimization may be re-examined to arrive at a different number of pulses.  The synchronized X-ray pulse train opens new possibilities in experiments involving coherent x-rays and IR light as both pump and probe \cite{Weisshaupt2017} well beyond the capabilities of high harmonic generation sources \cite{Popmintchev2012}. With such sub-cycle synchronization, one may be able to obtain dynamical information with unprecedented resolution, as recently demonstrated in the XLEAP experiment at SLAC \cite{exleap}, which was also a first test of the ESASE scheme, albeit in modified form, for XFEL use. Further, the compactness of the XFEL and the high gain predicted from the presented UC-XFEL approach may enable a new frontier in recirculating FEL systems such as regenerative amplifiers. Such an x-ray FEL could produce very large fluxes of x-rays with highly monochromatic radiation in the $\sim$4 keV range, a spectral region ideal for performing pump-probe condensed matter investigations, including THz excitation \cite{Kubacka1333} derived from the electron beam itself \cite{Cook2009}. Systems and behaviors under study could include ferro- and anti-ferromagnetism, multi-ferroics, charge and spin density waves, high $T_c$  superconductors, and topological insulators. As the properties listed are sensitive to a wide variety of parameters (temperature, pressure, doping, external fields), comprehensive and time-intensive studies would represent a critical advantage in their success. 

As a relevant example from condensed matter physics, soft x-rays can resonantly excite core electrons to valence levels in both oxygen and transition metals, which makes them an ideal probe to investigate, among other materials, transition metal oxides.  Transition metal oxide materials continue to play an outsized role in condensed matter research as they exhibit high temperature superconductivity, multi-ferroicity, and colossal magneto-resistance among other high-impact phenomena. Because it is largely the valence electrons in these materials that are responsible for their behavior, resonant soft x-ray scattering has become a necessary tool to investigate ordering phenomena that are too subtle to be detected by other means. For example, soft x-rays have played a major role in identifying stripe and charge order in the copper-oxide superconductors and orbital order in magneto-resistive manganites  \cite{Tokura_2006}. 

The UC-XFEL may offer a next generation of experiments, where these orders can not only be detected, but manipulated with light. One of the major goals in this line of research is to disentangle the roles of different degrees of freedom in a strongly correlated system that give rise to several coexisting orders. In many of these compounds, it is not known whether orders coexist due to cooperation between the two (or more) orders or whether they are in competition. For instance, in copper-oxide superconductors, one can use an incident pump pulse to melt the sample and monitor the charge order response with soft x-rays to observe enhanced or diminished behavior, or changes in resonance frequency or wavelength.  This would yield crucial information about proposed competition/cooperation scenarios. Furthermore, one can coherently drive a particular vibration or spin mode by tuning the incident energy of the pump laser pulse and use this excitation to coherently control the charge or orbital order. These kinds of experiments tend to be extremely challenging at present due to the large amount of experimental preparation needed during allocated beam time, a constraint mitigated by the UC-XFEL.

It should also be noted that through tuning the energy chirp of the electron beam through linac phase and tunable wakefield effects \cite{Fu2015,Deng2014} before the IFEL modulator, that one may obtain a single attosecond pulse of x-rays from the UC-XFEL. A variant of this approach also utilizing beam self-fields for micro-bunching was shown in the XLEAP experiments at the LCLS recently \cite{exleap}.  This capability will be exploited in a wide variety of scientific applications. Indeed, as the micro-bunch current and charge in the UC-XFEL are foreseen to be similar to that shown in XLEAP, the obtainable pulse energy should be similar ($\sim$10 $\mu$J \cite{exleap}). One may also produce two pulses in this manner, with x-rays obtained that enable nonlinear scenarios such as attosecond pump–attosecond probe experiments \cite{Inoue1492}, with time scales shorter than the Auger recombination permitting the study of  ultrafast phase transition dynamics.  Finally, we note that flexibility in the ESASE modulating laser wavelength is desirable, as one may adjust the timing between x-ray pulses, and thus stroboscopically probe different periodic processes. 

The UC-XFEL has a number of particular properties that can be exploited to push the frontiers of XFEL physics \textit{per se}. In the SXR spectral region, each micro-bunch in the electron beam train may reach single spike lasing, and will thus display a high level of first order coherence. Further, the train could be used to increase the coherence even at higher orders \cite{GlauberCoherence} (\textit{e.g. fluctuation suppression} in second order) by coupling the emission of the different micro-bunches. This may be realised by splitting the amplifier into sections and inserting at the end of each section, a long-period undulator segment being resonant at a wavelength comparable to the micro-bunch separation. The light from previous micro-bunches seeds the following, a process which may be repeated until the  final spike is seeded in all the amplifier sections. The second-order coherence should be  greatly improved by this coupling, as shown recently in a seeded XFEL \cite{Gorobtsov}.  Finally, we note that that a class of mode-locked XFEL pulse trains may be produced if one introduces short chicanes between undulator sections to shift the radiation forward by a micro-bunch, as proposed in \cite{ModeLockXFEL}. This scheme may give temporal and spectral characteristics displaying advantages of mode-locked oscillators.

A new frontier in the HXR regime has been recently observed in the stimulated emission of x-rays in Mn-based systems pumped by XFEL pulses \cite{StimXray}. This mechanism has produced very narrow line widths starting from an initial population inversion on a 6.6 keV transition. Recently, it has been proposed to utilize this effect in creating an x-ray laser oscillator \cite{HalavanauXLO}. Such an experimental program, given its multiple layers of complexity, would be an excellent candidate for development at an HXR UC-XFEL, where iterative experimentation is permitted. 

Finally, we note that as a complex large instrument, the XFEL has required significant human resources to operate and optimize in use. Recent trends in machine learning have had a notable, positive impact on the needs for expert operators \cite{Scheinker2019,Durissimo}. The UC-XFEL may serve as a platform form exploring advanced methods in machine learning for free-electron laser \cite{FELML} and associated acceleration and beam transport systems \cite{EmmaML2018}. Conversely, the refinement of such methods should  permit much smaller teams of expert personnel to operate the UC-XFEL. 

\subsection{Chemistry}
The introduction of such an operational mode in the soft x-ray spectral region may open the way to exploratory investigations of the basic underpinnings of chemistry. With liberal access to a soft x-ray FEL, allowing exploitation of absorption edges in oxygen and nitrogen, experiments may be performed that drive theoretical and simulation efforts for comprehensive understanding of the electron dynamics involved in chemical reaction \cite{FemtosecImaging}, relying on the ultra-short x-ray pulse duration to create ``molecular movies". Resolving the dynamics of charge transfer processes illuminates the making and breaking of chemical bonds and, by tuning the x-ray energy to be resonant with the target atom, molecular dynamics can be revealed \cite{Erk2013,McFarland2014}. This element-targeted approach can even be used to probe the spin dynamics of excited states \cite{Zhang2014}. Such techniques are useful in the study of systems relevant to solar cells, batteries, and photocatalysts \cite{Siefermann2014}. The unique pump-probe synchronization characteristics of a UC-XFEL may be used to directly observe photoexcited electronic states, as recently demonstrated with XFEL-derived soft x-rays for both x-ray absorption spectroscopy and resonant inelastic scattering methods \cite{Ismail2020}.

\subsection{Biology}
The possibilities associated with an on-campus coherent hard x-rays with high availability likewise promise high impact in biological and biomedical research \cite{ProteinStructure}, as evidenced by the work of the UCLA DOE Imaging Institute \cite{DOEimaging1}, which was an early adopter of the XFEL in biology. In addition, the availability of highly increased access would permit the promising start to coherent imaging of cell structures in bacterial \cite{ProteinCrystal} and cancer studies begun at the LCLS to proceed to a mature level of success that has not been possible with the current generation of XFELs. The ``diffraction-before-destruction" method leverages the enormous peak power of the XFEL pulses to extract a useful diffraction pattern before the sample is destroyed by the deposited energy. The speed with which this occurs allows samples to be studied in more natural conditions, \emph{e.g.} not requiring cryo-cooling. A further application of the UC-XFEL is to crystallography of ``nanocrystals'', comprised of only a few unit cells. This technique was demonstrated by \cite{Chapman2011} using nanocrystals as small as 200 nm, even facilitating \emph{in vivo} crystallography of naturally occurring nanocrystals within their originating cells \cite{Sawaya2014}. This is also a particular advantage for many medically relevant structures which are difficult to crystallize; obviating the need for large crystals opens the door to structure-driven drug design \cite{Fenalti2015,Zhang2015}. Also, as with the chemistry application, ``molecular movies'' of biological systems can also elucidate their operation \cite{Barends2015}.

Perhaps the most compelling current application of x-ray FELs is in imaging viruses -- in particular the SARS-CoV-2 virus responsible for the 2020 pandemic. The use of XFELs in various aspects of virus imaging, including crystallography and single particle imaging (SPI), has drawn much attention \cite{virology2019}, predating the current crisis. While SPI in general has envisioned use of photon fluxes exceeding what are predicted by this analysis by a factor of 3 to 15, this drawback can be mitigated by better control of background, mounting strategies, and target hit-rate. A dedicated UC-XFEL would be extremely helpful in this regard. Further, XFELs remain an attractive option for virus imaging as one may use samples in a natural state (in contrast to high resolution cryo-EM), and their time structure permits exploration of non-equilibrium conditions, such as application of chemical or electromagnetic stimuli \cite{ProteinMech}. The exploration of non-equilibrium response of virus structures may also take advantage of the laser-locked pulse-train structure of the UC-XFEL. Finally, we note that the analysis of \cite{virology2019} indicates that the optimum photon energy for viral imaging is near 4 keV; we have recently begun an analysis of a UC-XFEL operated at this energy.

\subsection{Ultrafast Imaging}
As a further example, from the hard x-ray regime, one may envision introducing entirely new techniques in coherent diffraction imaging (CDI) \cite{Miao530}, but these must be developed through dedicated, time-intensive experiments. Realization of these initiatives are not possible without lengthy delays with the limited beamtime now available at XFELs. As a current example, we can point to x-ray ptychography, which enables diffractive imaging of extended samples by raster-scanning across the illuminating x-ray beam, permitting more general application of  CDI techniques. As this is a time-intensive technique, a migration towards table-top sources such as HHG has recently been noted \cite{Tadesse2019,Sandberg2007,Ravasio2009,Gardner2017}. The availability of a UC-XFEL source would give essential advantages in the time needed for employing this technique in scientific applications. 

Another application of UC-XFEL is based on its emerging  ability to produce attosecond x-ray pulses, as recently experimentally demonstrated \cite{AttoFermi}. Attosecond science has been transforming our understanding of electron dynamics in atoms, molecules and solids \cite{KrauszRMP,CorkumAtto,Agostini_2004}. However, to date, almost all of the attoscience experiments have been based on spectroscopic measurements because attosecond pulses have intrinsically broad spectra due to the uncertainty principle, and are incompatible with conventional imaging systems. Recently, a novel CDI  method has been proposed to overcome this major obstacle \cite{rana2019ptychographic}. Using simulated attosecond pulses, it has demonstrated that the spectrum, probes and spectral images of extended objects can be simultaneously reconstructed from a set of ptychographic diffraction patterns. This method was recently experimentally validated by successful reconstruction of the spectrum, 17 probes and 17 spectral images using a broad spectrum optical source \cite{rana2019ptychographic}. The combination of this powerful method with the UC-XFEL may significantly reduce the data acquisition time for spectro-imaging experiments by harnessing the considerable x-ray flux available. Potentially enabled applications  range from visualizing attosecond electron dynamics to imaging materials and biological samples at the nanometer scale.

\subsection{Industrial Applications}
One may also consider advanced industrial uses for the UC-XFEL. In applications of x-rays to modern nano-electronic devices, the objects being imaged have reached the point where it is no longer possible to image entire devices with their intricated interconnections non-destructively. This is due to the small feature sizes and the complex three-dimensional structures present on a chip. This unmet need in metrology represents a serious problem for device production and quality control. It has been recently demonstrated, however, that with a coherent source one may employ x-ray ptychography to create three-dimensional images of integrated circuits with a lateral resolution in all directions as low as 14.6 nm \cite{Holler2017}. This method relies on x-ray fluxes that are notably smaller than the maximum available at existing XFELS, and thus the availability and lower intensity of the UC-XFEL would represent a significant step forward in widespread adoption of this technique. This application would advance the state-of-the-art in chip inspection and also permit reverse engineering and inspection, opening the door to use in device security. 

\section{Conclusions and Future Directions}
The principles behind an ultra-compact XFEL have been presented here, as well as the details of how these principles are manifested in the beam and radiation physics involved, and the advanced technologies and methods -- such as machine learning -- employed. With the vision of realizing this potentially paradigm shifting instrument stated, it is useful to discuss the near-future research directions that are needed to permit progress towards the goal of the UC-XFEL. It is important to note that this research impacts not only the outlook for the UC-XFEL, but also the future plans for more traditional free-electron lasers. We review these research areas by subject, as with the preceding discussions. 

As noted, the production of very high brightness electron beams is the linchpin of the UC-XFEL recipe presented. The development of the critical component, that of the cryogenic RF photoinjector, is presently underway at UCLA, in an effort undertaken by a UCLA-SLAC-INFN-LANL collaboration with wide-ranging  interests in high gradient acceleration \cite{bane2018advanced}. This experimental effort is aimed primarily at demonstrating the operation of the C-band cryogenic photoinjector discussed above, operated at 27 deg K, and testing the production and diagnosis (particularly the measurement of very low emittances \cite{Renkai2012}) of high brightness beams.  This program has many relevant technical challenges such as the integration of a cryogenic solenoid. Further, issues such as dark current emission and control using coatings and active sweeping are being examined experimentally. Cryogenic cathode operation, using a dedicated 0.5 cell source with cathode load-lock, is being investigated in the context of this program. As noted previously, this new approach to photoinjectors can enhance the outlook for linear collider sources, and may also be utilized for larger scale XFELs, for wakefield acceleration experiments, and for next generation relativistic electron microscopy and diffraction instruments \cite{Rosenzweig2019, Li2014}. The development of this generation of sources is of strong interest to the NSF Science and Technology Center known as the Center for Bright Beams. 

The experimental development of linear accelerator structures is proceeding at SLAC, with notable experimental activity performed in parallel at LANL and UCLA. This work is presently concentrated on cryogenic operation at 77 K. The linac structures constructed for current experiments employ novel brazeless joining techniques.   Efforts are underway in support of the development of the independent manifold coupling and, importantly, to implement transverse higher order mode damping in the linac cells. This initiative is aimed at three applications that utilize multiple bunches per RF pulse: the linear collider; a MaRIE-class XFEL \cite{Carlsten2018}; and a high averarge flux inverse Compton scattering (ICS) source. Transverse  damping in these cases is needed to avoid multi-bunch BBU. Once developed, a BBU-stabilized linac may also be applied the UC-XFEL, potentially enabling kHz average repetition rate, thus giving over $10^{13}$ HXR photons/second, enabling many high average flux applications.  The advantages conferred on these applications include a much diminished spatial footprint, and associated mitigation of instabilities such as MBI and also BBU. The development of the K$_\mathrm{a}$-band RF sources for the compact linearizer and potential RF sweeper application is now proceeding in the context of the XLS collaboration. 

Along with beam production and acceleration, advanced work on pulse compression is now being performed. As noted above, the immediate challenge in this work is to further optimize the performance of the first UC-XFEL compressor. These investigations entail a more detailed understanding, using advanced computational models, of mitigating 3D CSR effects.  This is accompanied by a renewed emphasis on the ESASE analysis of the second compressor, where by taking into account finite-laser spot size effects, one may increase the brightness of the electron beam micro-bunches, at a cost of some efficiency in the bunching process.  Advanced work on the ESASE technique is being applied to exploratory research on MaRIE-class XFELs. Along with the advances in beam brightness and compact acceleration techniques, the study of the efficacy of micro-bunching for final compression can permit greatly enhanced performance -- in particular in increasing the power output -- of such very high-energy photon XFELs to be obtained.  This would be an important development in this high-profile application area \cite{Carlsten2019}.  The current work here also has potential impact on XLEAP-like scenarios for existing XFELs \cite{exleap}.  Finally, we note that optically bunched high-peak-current beams may be effectively used to excite TV/m-amplitude wakefield acceleration waves in plasma \cite{resonantPWFA}.

The development of cryogenic, sub-cm-to-cm period undulators has continued in recent years, with notable progress reported at a number of laboratories \cite{OShea2016, HuangJC2017, CryoUndCouprie}. This technology is by now mature enough to be applied in a UC-XFEL. Other standard approaches to undulator development may include superconducting undulators, where designs with periods nearing those needed have been studied \cite{Mishra17}. At few-mm period length, development work on novel variations on short period Halbach arrays is being pursued \cite{Majernik2019Triangular,Majernik2019Comb}. At the mm-to-sub-mm scale, there has been progress reported on mesoscale machined undulators \cite{Peterson2014, Harrison2014}; designs so far do not reach the peak fields needed, and the innovations introduced  at the few mm-period may be necessary in this scaled down device. MEMS fabricated systems applied to the UC-XFEL now concentrate mainly on micro-quadrupole development, with micro-undulators less emphasized, due to their impracticality in scaling to mesoscale -- the methods of MEMS fabrication become unwieldy at this dimensional scale.  Successful development of shorter period undulators may extend the wavelength reach of XFELs.  Finally, we note that for many applications, a helically-polarized, very short-period undulator would be advantageous. Research efforts into this option are in their early stages \cite{Tan_19}. 

The performance of the XFEL remains the subject of further refinement and optimization. The parameter sets employed here were driven by the current availability of beam sources, compression schemes, instability theory, and advanced undulators. As these evolve, the UC-XFEL will also be re-evaluated. This evolution is ongoing in the present development of the HXR case, where emphasis is being placed on studies of the compressor systems. It is also desirable to examine in further detail both space-charge \cite{Gadjev2017} and quantum mechanical \cite{Debus_2019} effects on XFEL performance, as well as two-color schemes \cite{twocolorFEL}. In addition to single operating point optimizations, one must explore further the reproducibility and wavelength tunability of the UC-XFEL.  The applications of x-ray pulse trains and the FEL physics associated with this format, concentrating on the production of low-fluctuation, single-spike-per-micro-bunch operation, are of interest. Further, the suitability of the x-ray pulse trains in pump-probe experiments \cite{Inoue1492} must be investigated. These issues are central to UC-XFEL development, as well as systems with shared emphasis on this format (MaRIE, XLEAP). 

The problems of designing appropriately compact x-ray optical systems at relevant SXR and HXR wavelengths are under study by the current collaboration, with work ongoing at UC-Merced and UC-Berkeley as well as connected institutions. This activity is performed in tandem with LCLS-II-directed research and development, and thus is synergistic with current efforts on large-scale XFELs. Based on the model of the UC-XFEL photon output in both soft and hard x-ray cases, this effort seeks to confront the design challenge of creating an x-ray optics system  to deliver the photons to experimental users that is commensurately as compact as the UC-XFEL. With the photon flux per pulse, and repetition rate (up to 120 Hz) of the UC-XFEL may have problems in thermal loading similar to those of existing XFELs, due to absorption of x-rays on crystal optics \cite{Halavanau2019}. There are issues that are specific to the UC-XFEL as well. These include the effects of the pulse-train format on x-ray optics, which also are of concern for two-pulse x-ray pump-probe experiments.  

Investigations of applications of the UC-XFEL in advanced scientific, medical and industrial applications are still in the early phase of development. The survey of such applications given above illustrates the central theme of these investigations well -- to identify uses of the UC-XFEL that would benefit from a much more liberal availability of the instrument, or by the unique aspects of the UC-XFEL, or both. These motivations are also present in other projects, such as EuPRAXIA, or by the emerging proposed x-ray sources, based on super-radiant emission from nano-bunching of low charge, low energy beams, at Arizona State University \cite{Graves:FEL2017-TUB03}. The ASU initiative involves a series of workshops \cite{ASUWeb} on the use of the source. Despite notable differences in approach, these discussions aid the maturation of the science program outlined here. 

The initiative described in this article is now beginning its path to realization. The beam, FEL physics, accelerator and x-ray experimental technology development efforts are presently centered at UCLA, with research efforts taking place at the SAMURAI Laboratory. The suite of initiatives reviewed in the above discussion, encompassing high brightness electron beam production and measurements, high gradient acceleration, beam compression (particularly emphasizing ESASE and coherent radiation-based beam diagnosis), and initial FEL and ICS radiation production are to take place in a large (18 m length) bunker that is now nearing completion. SAMURAI has the  radio-frequency, laser, vacuum, and ultra-fast experimental measurement infrastructure for proceeding to the next steps in making the vision of the UC-XFEL a reality. The first UC-XFEL, upon realization, is envisioned to serve as a template for the diffusion of such instruments to many universities, industrial laboratories, and medical research facilities.

\section{Acknowledgments}
This work was performed with support of: the National Science Foundation, a NSF Science and Technology, through the Center for Bright Beams, grant no. PHY-1549132; through a NSF Science and Technology Center, STROBE, grant no. DMR 1548924; and through NSF Award 1350034. Support was also obtained from the US Dept. of Energy, Division of High Energy Physics, under contracts no. DE-SC0009914 and DE-SC0020409. The authors would also like to acknowledge the Moore Foundation for its support of the initial workshop held in 2019 at UCLA on the UC-XFEL concept \cite{UCXFELworkshop}.

\begin{appendix}
\section{Resistive wall wakes}\label{app:resistiveWakes}
The resistive wall wakes arising from the cryogenic beam pipe walls in the UC-XFEL undulator are calculated using the approach of \cite{Stupakov2015}; this more advanced wake analysis is required since the interaction will necessitate inclusion of the material response in the anomalous skin effect (ASE) regime. In instances where the mean free path of electrons in the resistive material, $l$, is not much larger than the classical skin depth, $\delta_\mathrm{NSE}$, ASE must be considered to accurately calculate the wakes. The classical skin depth is defined as:

\begin{equation}  
\delta_\mathrm{NSE} =  \sqrt{\frac{2 c}{Z_0 \sigma_c \omega}},
\label{eq:classicalSkinDepth}
\end{equation}

\noindent where $Z_0$ is the free space impedance, and $\sigma_c$ is the conductivity of the material, and $\omega$ is the frequency. At cryogenic temperatures $\sigma_c$ increases, and use of short beams extends the relevant range of $\omega$: both of these serve to reduce $\delta_\mathrm{NSE}$, causing the classical model to break down \cite{Reuter1948, Dingle1953}. Using the Drude model

\begin{equation}  
\sigma_c = \frac{\omega_p^2 l}{Z_0 c v_f},
\label{eq:drude}
\end{equation}

\noindent where $\omega_p$ is the plasma frequency and $v_f$ is the Fermi velocity, relating $l$ and $\sigma_c$. Finally, the effect on conductivity at cryogenic temperatures is encapsulated by the residual resistivity ratio, RRR, defined as:

\begin{equation}  
\mathrm{RRR} \equiv \sigma_c (4 \: \mathrm{K}) / \sigma_c (293 \: \mathrm{K}).
\label{eq:RRRdef}
\end{equation}

\noindent Copper is considered as the resistive material which has relevant constants \cite{Kittel2005}: $n = 8.5 \times 10^{28} \: \mathrm{m}^{-3} $, $v_f = 1.6 \times 10^6 \: \mathrm{m/s}$, $\omega_p = 1.7 \times 10^{16} \: \mathrm{rad/s}$, and $\sigma_c (293 \: \mathrm{K}) = 5.7 \times 10^7 \: (\mathrm{\Omega \cdot m})^{-1}$. The complex conjugate of the ASE surface impedance \cite{Reuter1948, Dingle1953} is:

\begin{equation}
\hspace*{-0.5cm} 
\zeta^*(\omega) \approx -\frac{i \omega l}{c} \left ( \frac{-u}{\pi} \int_0^\infty \ln \left (  1 + \frac{ \left (\frac{3 i l^2  u^{-3} }{2 \delta_\mathrm{NSE}^2} \right ) (2 t^{-3}((1+t^2)\tan^{-1}(t)-t))}{t^2} \right ) \mathrm{d} t \right )^{-1}
\label{eq:ASEImpedance}
\end{equation}

\noindent with $u$ defined as $1 + i \omega l/v_f$.

The beam impedance of a resistive flat wall is given as \cite{Bane2005}:

\begin{equation}  
Z(k) = \frac{Z_0}{2 \pi a} \int_0^\infty \frac{\mathrm{d} q}{\cosh(q)} \left ( \frac{\cosh(q)}{\zeta(k)} - \frac{i k a \sinh{q}}{q} \right )^{-1},
\label{eq:flatWallImpedance}
\end{equation}

\begin{figure}[h]
    \centering
    \includegraphics[width=0.8\linewidth]{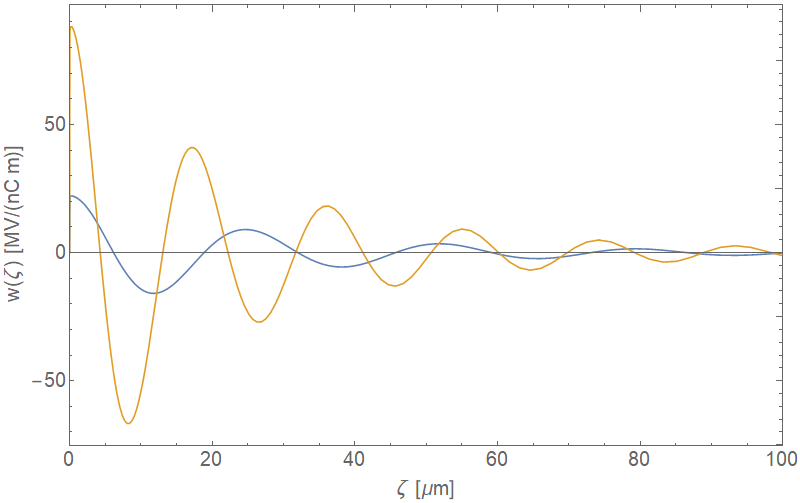}
    \caption{Resistive wall wakefields in cryogenic (RRR = 100) copper flat wall structures for gap sizes of 2 mm (blue) and 1 mm (orange).}
    \label{fig:wakePlot}
\end{figure}

\noindent where the wave-number, $k = \omega/c$, and $a$ is the half-gap between the two flat surfaces. The wake, $w(\zeta)$, at a position $\zeta$ behind the driving electron, is given by

\begin{equation}  
w(\zeta) = \frac{c}{2 \pi} \int_{-\infty}^\infty Z(k) \mathrm{e}^{-i k \zeta} \mathrm{d} k .
\label{eq:wakePlot}
\end{equation}

\noindent The spatial wakes converted from the beam impedances are given in Figure \ref{fig:wakePlot} for specific cases of 1 and 2 mm gaps with RRR = 100 copper walls. By convolving the wake function with the beam profile, the wake induced relative energy variation $\eta = \Delta E / E$ is calculated at a point along the bunch $\zeta$:

\begin{equation}  
\eta(\zeta) = - \frac{e L}{c U_e} \int_{-\infty}^\infty I(s) w(\zeta - s) \mathrm{d} s
\label{eq:energyDelta}
\end{equation}

\noindent where $e$ is the electron charge, $L$ is the length of the parallel plates, $U_e$ is the initial beam energy, and $I(\zeta)$ is the bunch's current profile. 
The specific results for the SXR UC-XFEL case are shown in Figure \ref{fig:combinedWakePlot}.

\section{Beamline diagnostics} \label{app:diagnostics}
The realization of the UC-XFEL, inherently rooted in advances in high quality beam generation, requires manifold considerations of diagnostics capable of precision beam characterization. In this section we enumerate the diagnostic components needed for the UC-XFEL in support of the discussions in Sections 4-8, effectively tracing the relevant beam phase space characteristics of each stage. The diagnostics encompass a mix of established techniques and advanced methods scaled to the challenging nature of the generated beams' characteristics. While the exact deployment of these systems is subject to optimization during design and implementation, this discussion provides the context needed to appreciate the scope of the diagnostic systems needed. It further permits an understanding of their layout and impact on the footprint of the UC-XFEL and forms a basis of the cost estimates given in \ref{app:cost}. 

The optimization of the beam after the injector requires a full characterization of the beam at the RF gun exit. Chief among these are the emittance, the bunch length, the beam energy spectrum, and beam charge. The emittance at this point is measured using a slit- or pepper-pot-based system \cite{andersonemit,marxemit} that obviates space-charge effects and gives the needed nm-rad resolution. This system requires a drift which will necessitate removing the power in the first linac section before a beam profile monitor that is encountered just downstream of the linac. It should be noted that twenty beam profile monitors based on both Ce:YAG \cite{murokhyag} and transition radiation will be employed in the beam diagnostic system during beam acceleration, compression, and transport. These are complemented by use of compact position monitors for locating the beam centroid in the accelerator and undulator \cite{KULKE1988241}. These systems are non-destructive, have negligible insertion length, and also serve to measure the charge. 

In the lower energy region after the gun, we plan to use a coherent transition radiation-based measurement system, in which the shape of the beam is inferred from autocorrelation of the emitted THz radiation from a TR foil \cite{laisievers1}. Beams of this approximate energy, charge, and length have been measured using this technique previously \cite{MUROKH1998, Andonian:2007}. The energy spectrum after the gun will be obtained using a very compact spectrometer magnet (insertion length 25 cm), as used in previous experiments at UCLA \cite{FUKASAWA20142}. The charge obtained from the gun is determined non-destructively with high precision using an integrating current transformer \cite{WuICT}, which again has a negligible insertion length.

During acceleration in the linac up to the first compressor, as well as beyond, to the second compressor, the beam must be examined for proper transverse dynamics (matching to the focusing, emittance preservation, monitoring of halo) using the methods described above. The longitudinal phase space input to the first chicane is measured \cite{Moody2009} using a combination of a short (23 cm) C-band deflector and the initial bend of the chicane \cite{Anderson2003}. This is needed to optimize the longitudinal phase space with the K$_\mathrm{a}$-band linearizer cavity. After characterization, the beam bunch length after compression is examined through beam-based radiation analysis, specifically coherent diffraction radiation interferometry, to allow for non-invasive monitoring using a real-time interferometer \cite{RTI2012} well-suited to its characteristics at 400 MeV, and with a sub-picosecond pulse. 

The phase space after the IFEL modulator (second chicane) must be examined using a more powerful RF deflector in K$_\mathrm{a}$-band, along with a compact bend magnet spectrometer. 
The 15~cm long K$_\mathrm{a}$-band deflector operates at 35.7~GHz, and is driven by a 4~MW source that has the capability to provide transverse deflecting voltage on the order of 2~MV. The design is scaled from an existing X-band deflecting cavity operational at the BNL-ATF. At this point in the transport the beam parameters include an energy of 400~MeV, $\sigma_r\sim35\mu m$. At full power and screen located at the appropriate phase advance from the cavity, the deflector will provide a temporal resolution of $\sim$1~fs. 
This measurement provides a method for determining IFEL modulation amplitudes as well as characterization of the effects of MBI and CSR on the longitudinal phase space.  This diagnostic permits the tuning of the second compressor input, and provides a powerful tool for machine-learning algorithms to optimize system performance.  

After the final chicane, measurement of the microbunching is accomplished with a compact coherent transition radiation measurement system \cite{Tremaine1998}. This measurement involves spectral resolution of infrared to far-infrared signals on the fundamental and, given the high degree of bunching expected, also the harmonics of the bunching laser wavelength \cite{ROSENZWEIG1995}. This measurement may be augmented in two possible ways currently under study for their feasibility: the addition of a K$_\mathrm{a}$-band deflector, or employing a version of the ``attoscope" method \cite{Attosweeper, WEIKUM2018369}. Each of these diagnostic systems would add approximately one meter of insertion length. 

Ensuring the electron beam alignment and matching conditions are met at the undulator is critical for optimizing FEL performance. Past experience, particularly obtained from the Visible to Infrared SASE Amplifier (VISA) FEL at Brookhaven National Laboratory \cite{Murokh:2003} provides a foundation and expertise to address this challenge. Similar to the UC-XFEL, the VISA undulator was a 4 m undulator that incorporated a strong focusing quadrupole lattice. The VISA FEL demonstrated the shortest gain length for IR FELs which was made possible by an array of intra-undulator diagnostics for both e-beam and radiation characterization \cite{Andonian:2005}. In the UC-XFEL, a similar approach is considered.

The UC-XFEL relies on transverse profile monitors for two purposes. They provide the two-dimensional beam distribution imaged from a scintillating screen or from optical transition radiation (OTR) from a metal foil. The information is critical to get the appropriate beam size matching condition at the undulator entrance, as well as matching into various section of the transport, \textit{e.g.} compression stages. The extreme nature of the UC-XFEL mandates a matched beam size as small as $\sim$5 $\mu$m at the undulator entrance. Standard scintillator monitors resolutions are not suitable as the resolution depends on beam intensity, which leads to performance degradation and limited resolution at higher intensity \cite{murokhyag}. OTR imaging can provide high resolution, however care must be exercised due to micro-bunching effects which introduce coherence. In order to measure few $\mu$m scale beams, advanced optical methods are required. In this case, a high resolution imaging system includes a thin radiator foil ($\sim$10 $\mu$m) at the focal plane of an objective lens. The collected light is directed to an imaging lens providing 2 $\mu$m optical resolution \cite{barber:2020}. The objective and foil are co-located on a longitudinal translation carrier, allowing transverse imaging along the beam path. The objective is infinite conjugate so only the objective and foil are moved while none of the other optical elements need adjustment to maintain focus. The entire OTR imaging layout has been demonstrated to show sub 3-$\mu$m beam resolutions \cite{Barber:thesis}.

The transverse profile monitors provide a coarse measurement of the beam centroid trajectory. However, for $\sim$ $\mu$m level precision needed in the undulator, high-resolution cavity beam position monitors (BPMs) are required \cite{WALSTON20071}. While wire scanners can also provide the required resolution, they are destructive, relatively large-footprint diagnostic systems, and require multi-shot operation. The cavity BPMs provide a fast method of orbit corrections and used at premiere FEL facilities worldwide. The cavity BPMs will be mounted at the entrance and exit of each undulator section, providing the necessary feedback for orbit correction using steering magnets.

In addition to the transverse profile monitors, the diagnostic insert in the undulator sections incorporates a second stop-position, again similar to VISA FEL. The second stop position mounts a crystal to reflect the FEL radiation out of the vacuum port to an x-ray diagnostic station to measure FEL properties such as pulse energy, spectrum, and angular distribution. The crystal is also located on the longitudinal carrier, allowing for precise measurements of the FEL gain curve during commissioning and optimization. 

It should be noted that operational experience beginning with proof-of-principle experiments such as VISA FEL, to facilities such as the LCLS, shows that the radiation diagnostic provides a critical, highly sensitive measurement needed for FEL optimization. System tuning beginning with an emphasis on electron beam alignment and matching at the beginning of the undulator increasingly relies on FEL light output to make adjustments later in the undulator. 

The development of a complete diagnostic system as outlined above is a necessary condition for using machine learning to make the UC-XFEL system tuning tractable \cite{FELML}. The digitization of relevant signals and subsequent analysis packages are necessary for incorporation into machine-learning algorithms for beam optimization and performance enhancing feedback. This feedback is to be paired with a powerful start-to-end simulation model run on a parallel computing platform, that provides a dynamic training set which can evolve to provide an increasing capability for fast system optimization as the UC-XFEL is run in its myriad possible variations.

Thus a computer that both controls the UC-XFEL by neural networks and has high capabilities for running PIC codes for operational real-time feedback is required. Such a computer would require a robust base with a high-end CPU. For neural networks of the caliber under consideration, a high-end desktop GPU would be the most cost-effective solution. Since PIC codes require large proportions of RAM, a workstation GPU would be the best choice. Additional workstation-grade GPUs may be appended to further expedite simulations, possibly necessitating a server-grade motherboard and CPU. Once the process of hardware implementation is complete, a multi-person programming and integration effort is foreseen.

\section{Representative cost estimates} \label{app:cost}
This section lists cost estimates for the construction of the $n^\mathrm{th}$ SXR UC-XFEL, that is after initial research development costs for components are no longer relevant. It is important to distinguish these estimates from those associated with first generation versions which will have different needs as a test-bed for the UC-XFEL concept. These costs are summarized in Table \ref{tab:costEstimates-summary} with detailed estimates broken out by subsystem in subsequent tables. The building-level infrastructure estimate has been extrapolated from existing XFELs. These costs, which scale with the instrument dimensions, are thus a small fraction of those associated with the full scale XFEL.

\begin{table}[]
\centering
\caption{Soft x-ray UC-XFEL cost estimate summary}
\label{tab:costEstimates-summary}
\begin{tabular}{|l|l|}
\hline
\textbf{Subsystem} & \textbf{Cost [k\$]} \\ \hline
RF & 15980 \\ \hline
Cryogenics & 840 \\ \hline
Magnets & 1650 \\ \hline
Diagnostics & 2505 \\ \hline
Lasers & 1615 \\ \hline
X-ray optics & 1285 \\ \hline
X-ray instruments & 1360 \\ \hline
Building/infrastructure & 7080 \\ \hline

\textbf{Total} & 32315 \\ \hline
\end{tabular}
\end{table}

\begin{table}[]
\caption{Cost estimates for the RF systems of a soft x-ray UC-XFEL}
\label{tab:costEstimates-RF}
\begin{tabular}{|l|l|l|l|}
\hline
\textbf{Item \phantom{thistextsetsthecolumnsizexx}}                                                                        & \textbf{Count} & \textbf{Unit cost [k\$]} & \textbf{Total cost [k\$]} \\ \hline
Klystron (50 MW)                                                     & 5     & 350                      & 1750                      \\ \hline
Modulator with chiller                                                      & 5     & 500                      & 2500                      \\ \hline
Waveguide network per klystron & 5     & 100                      & 500                       \\
\phantom{    }(phase shifters, isolators, & & & \\
\phantom{    }RF windows, etc.) & & & \\ \hline
SLED                                                                        & 5     & 400                      & 2000                      \\ \hline
Linac section (per meter)                                                   & 8     & 700                      & 5600                      \\ \hline
Gun                                                                         & 1     & 700                      & 700                       \\ \hline
K$_\mathrm{a}$-band linearizer                                                                  & 1     & 500                      & 500                       \\ \hline
K$_\mathrm{a}$-band klystron, modulator, etc.                                      & 1     & 1000                     & 1000                      \\ \hline
Passive dechirper                                                           & 1     & 750                      & 750                       \\ \hline
Low level c-band                                                            &      &                       & 120                       \\ \hline
Mid level c-band amplification                        &      &                       & 200                       \\ \hline
RF detectors and digitizers                                  &      &                       & 100                       \\ \hline
43.5 GHz network analyzer                                                   & 1     & 160                      & 160                       \\ \hline
Vector signal analyzer                                                      & 1     & 100                      & 100                       \\ \hline
\textbf{RF subtotal}                                                                          &       &                          & \textbf{15980}                     \\ \hline
\end{tabular}
\end{table}

\begin{table}[]
\caption{Cost estimates for the cryogenic systems of a soft x-ray UC-XFEL}
\label{tab:costEstimates-cryo}
\begin{tabular}{|l|l|l|l|}
\hline
\textbf{Item \phantom{thistextsetsthecolumnsizexx}}                                                                        & \textbf{Count} & \textbf{Unit cost [k\$]} & \textbf{Total cost [k\$]} \\ \hline
Cryocooler                                                                  & 2     & 20                       & 40                        \\ \hline
Cryostats                                                   &      &                       & 300                       \\ \hline
LN$_2$ reliquification plant                                                   &      &                       & 500                       \\ \hline
\textbf{Cryogenics subtotal}                                                                          &       &                          & \textbf{840}                     \\ \hline
\end{tabular}
\end{table}

\begin{table}[]
\caption{Cost estimates for the magnets of a soft x-ray UC-XFEL}
\label{tab:costEstimates-magnets}
\begin{tabular}{|l|l|l|l|}
\hline
\textbf{Item \phantom{thistextsetsthecolumnsizexx}}                                                                        & \textbf{Count} & \textbf{Unit cost [k\$]} & \textbf{Total cost [k\$]} \\ \hline
Cryo-solenoid                                                               & 1     & 30                       & 30                        \\ \hline
Quads                                                                       & 20    & 8                        & 160                       \\ \hline
Steering magnets                                                            & 20    & 4                        & 80                        \\ \hline
Spectrometer                                                                & 1     & 40                       & 40                        \\ \hline
Large power supply                                                          & 10    & 4                        & 40                        \\ \hline
Small power supply                                                          & 40    & 2                        & 80                        \\ \hline
Undulator (per meter)                                                       & 5     & 200                      & 1000                      \\ \hline
Upstream chicanes                                                           & 2     & 40                       & 80                        \\ \hline
Downstream chicane                                                          & 1     & 40                       & 40                        \\ \hline
IFEL modulator                                                              & 1     & 100                      & 100                       \\ \hline
\textbf{Magnets subtotal}                                                                          &       &                          & \textbf{1650}                     \\ \hline
\end{tabular}
\end{table}

\begin{table}[]
\caption{Cost estimates for the diagnostics of a soft x-ray UC-XFEL}
\label{tab:costEstimates-diagnostics}
\begin{tabular}{|l|l|l|l|}
\hline
\textbf{Item \phantom{thistextsetsthecolumnsizexx}}                                                                        & \textbf{Count} & \textbf{Unit cost [k\$]} & \textbf{Total cost [k\$]} \\ \hline
Cameras, YAG, pop-ins, etc.                                                 & 20    & 7                        & 140                       \\ \hline
ICT                                                                         & 1     & 5                        & 5                         \\ \hline
CTR diagnostic                                                              & 2     & 50                       & 100                       \\ \hline
X-ray CCD                                                                   & 5     & 100                      & 500                       \\ \hline
X-ray streak camera                                                         & 1     & 200                      & 200                       \\ \hline
X-ray detector                                                              & 3     & 20                       & 60                        \\ \hline
X-band deflector                          & 1     & 1500                     & 1500                      \\
\phantom{    }(klystron, modulator, and cavity) & & & \\ \hline
\textbf{Diagnostics subtotal}                                                                          &       &                          & \textbf{2505}                     \\ \hline
\end{tabular}
\end{table}

\begin{table}[]
\caption{Cost estimates for the lasers of a soft x-ray UC-XFEL}
\label{tab:costEstimates-lasers}
\begin{tabular}{|l|l|l|l|}
\hline
\textbf{Item \phantom{thistextsetsthecolumnsizexx}}                                                                        & \textbf{Count} & \textbf{Unit cost [k\$]} & \textbf{Total cost [k\$]} \\ \hline
Photoinjector laser                                                         & 1     & 250                      & 250                       \\ \hline
IFEL laser                                                                  & 1     & 1000                     & 1000                      \\ \hline
UV conversion and beam shaping                                              & 1     & 15                       & 15                        \\ \hline
Optics                                                                      &     &                         & 30                        \\ \hline
Opto-mechanics                                                              &     &                         & 30                        \\ \hline
Laser vacuum isolation                                                      &     &                         & 15                        \\ \hline
TW laser optics                                                             &     &                         & 30                        \\ \hline
TW laser optics vacuum transport                                            &     &                         & 45                        \\ \hline
Laser compressor and vacuum box                                      & 1     & 100                      & 100                       \\ \hline
Optical tables                                                              & 5     & 20                       & 100                       \\ \hline

\textbf{Lasers subtotal}                                                                          &       &                          & \textbf{1615}                     \\ \hline
\end{tabular}
\end{table}

\begin{table}[]
\caption{Cost estimates for the x-ray optics of a soft x-ray UC-XFEL}
\label{tab:costEstimates-xrayOptics}
\begin{tabular}{|l|l|l|l|}
\hline
\textbf{Item \phantom{thistextsetsthecolumnsizexx}}                                                                        & \textbf{Count} & \textbf{Unit cost [k\$]} & \textbf{Total cost [k\$]} \\ \hline
K-B mirrors (Substrates, chamber,                   &      &                       & 675                       \\ 
\phantom{    }actuators, controllers) & & & \\ \hline
Photon transport (Shutters,                         &      &                       & 210                       \\ 
\phantom{    }collimators, aperture) & & &  \\ \hline
Diagnostics (2x wavefront sensor,           &      &                       & 400                       \\ 
\phantom{    }motorized stage, cameras, YAGs) & & & \\ \hline 

\textbf{X-ray optics subtotal}                                                                          &       &                          & \textbf{1285}                     \\ \hline
\end{tabular}
\end{table}

\begin{table}[]
\caption{Cost estimates for two user endstation instruments (coherent diffraction and AMO-directed)}
\label{tab:costEstimates-instruments}
\begin{tabular}{|l|l|l|l|}
\hline
\textbf{Item \phantom{thistextsetsthecolumnsizexx}}                                                                        & \textbf{Count} & \textbf{Unit cost [k\$]} & \textbf{Total cost [k\$]} \\ \hline
Vacuum chambers                   &   2   &   65                    & 130                       \\  \hline
Supports                        &     &                       & 40                      \\  \hline
Vacuum systems          &      &                       & 70                      \\ \hline 
Sample injector (CDI)          &  1    &   90                    & 90                     \\ \hline 
Sample injector and catcher (CDI)          &  1    &   100                    & 100                 \\ \hline 
Sample manipulator (AMO)          &  1    &   40                    & 40                 \\ \hline 
Electron analyzer (AMO)          &  1    &   250                    & 250                 \\ \hline 
Laser injector systems          &  1    &   80                    & 80                 \\
\phantom{    }for pump/probe (AMO)          &     &                       &                  \\ \hline
Detector (CDI)          &  1    &   220                    & 220                 \\ \hline 
Detectors for          &  2    &   120                    & 240                 \\ 
\phantom{    } correlation experiments (AMO)         &     &                       &                  \\ \hline
Miscellaneous equipment        &      &                       & 100                 \\ \hline

\textbf{Endstation instruments subtotal}                                                                          &       &                          & \textbf{1360}                     \\ \hline
\end{tabular}
\end{table}

\begin{table}[]
\caption{Cost estimates for the infrastructure of a soft x-ray UC-XFEL}
\label{tab:costEstimates-infrastructure}
\begin{tabular}{|l|l|l|l|}
\hline
\textbf{Item \phantom{thistextsetsthecolumnsizexx}}                                                                        & \textbf{Count} & \textbf{Unit cost [k\$]} & \textbf{Total cost [k\$]} \\ \hline
Building/civil construction                                                      &    &                      & 4950                       \\ \hline
End-user experimental hutches        &      &                       & 300                       \\ 
\phantom{    }(shutters, shielding, interlock, etc.) & & & \\ \hline
Motion control systems                                                      & 5     & 100                      & 500                       \\ \hline
Vacuum systems                                                              & 20    & 10                       & 200                       \\ \hline
Cables                                                                      &    &                         & 100                       \\ \hline
Computerized control                            &      &                      & 1000                      \\
\phantom{    }and machine learning & & & \\ \hline
Beamline support structures                                                 &     &                         & 30                        \\ \hline

\textbf{Infrastructure subtotal}                                                                          &       &                          & \textbf{7080}                     \\ \hline
\end{tabular}
\end{table}

\section{RF System Tolerance Studies}

The compressor systems we have described above must not only work at their design point but to be robust also maintain their functionality when exposed to errors in the RF systems. The most potentially problematic of these errors are deviations from the design phase and gradient. Here we will treat each individual linac section as having an independent, gaussian distributed error in each of these variables of standard deviations 0.1\% in the linac gradient and 0.1 degree in the linac phase; these values are representative of those achieved in FEL injector systems to date. Instead of performing one tolerance scan of the whole SXR beamline, we separate these studies into independent studies of the first bunch compressor and second bunch compressor. This is for two purposes: first, in moving between the SXR and HXR configurations, as well as performing fine-tuning of the output wavelength, the beamline through the first compressor will be largely unchanged, so knowledge of the individual tolerance of the first compressor is valuable. Further, we will find that the individual ESASE compressor is largely insensitive to RF fluctuations in the linacs following the first compressor.

\subsection{First Bunch Compressor}

\begin{figure}[h!]
    \centering
    \includegraphics[scale=0.435]{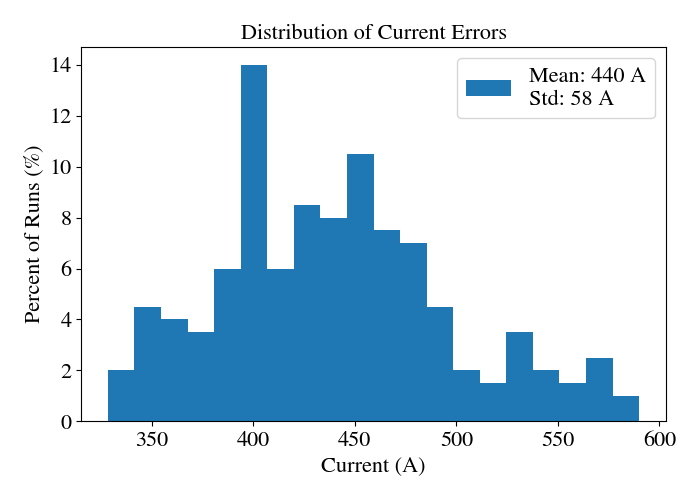}
    \includegraphics[scale=0.435]{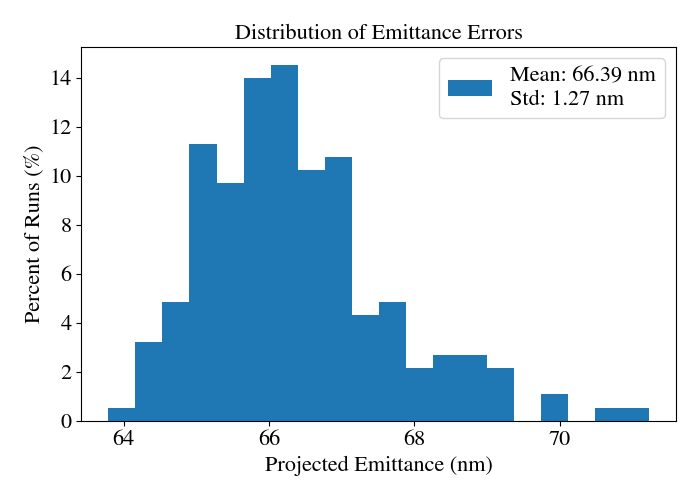}
    \caption{Variations in the current and projected emittance after the first bunch compressor are plotted for 200 simulations varying the linac gradient and phase randomly about the design values. The current value is the average of the current over the central 40 $\mu$m region of the bunch. }
    \label{fig:BC1rftol}
\end{figure}

To study the tolerance of the first bunch compressor we ran 200 simulations and recorded the two primary variable parameters: the average current in the central 40 $\mu$m region of the bunch and the projected emittance over the full bunch. In the nominal design these values are roughly 400 A and 65 nm. We show the results of these studies in Figure \ref{fig:BC1rftol}. The projected emittance variations are quite small, with a standard deviation just above 1 nm. The current variations seem potentially more difficult with a standard deviation of 58 A. Considering the relative insensitivity of the output ESASE micro-bunch currents to the existing variations in the nominally flat-top 400 A profile, fluctuations of the average current at this level should not pose any meaningful problem. 

\subsection{Second Bunch Compressor}

\begin{figure}[h!]
    \centering
    \includegraphics[scale=0.435]{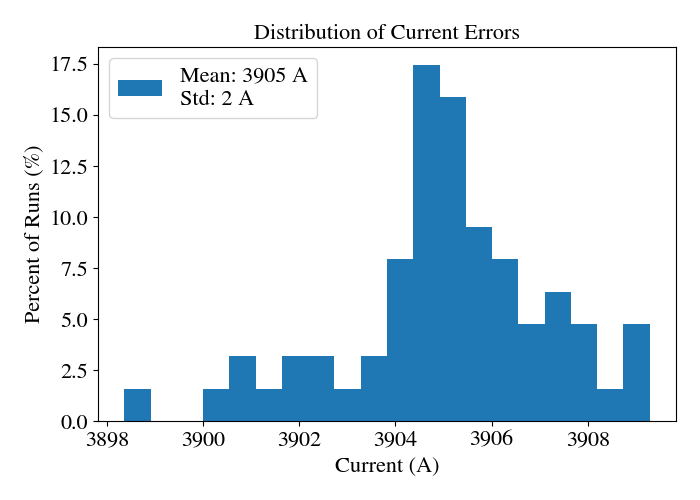}
    \includegraphics[scale=0.435]{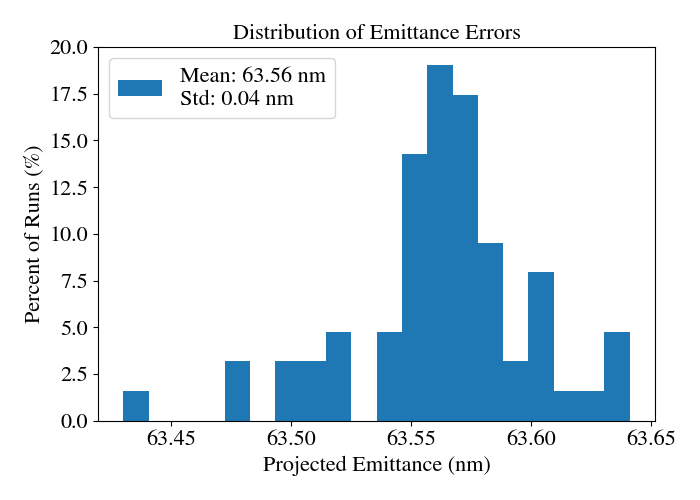}
    \caption{Variations in the mean peak microbunch current and slice emittance after the first bunch compressor are plotted for 80 simulations varying the linac gradient and phase randomly about the design values after the first compressor. }
    \label{fig:BC2rftol}
\end{figure}

As an intermediate step we ran a similar tolerance study on the output of the second bunch compressor assuming ideal performance in the first compresssor. This involved 80 simulations allowing the five meter-long structures after the modulator to have independent gaussian distributed errors as described above. For each run we obtained the average of the peak currents and emittances at the locations of the peak currents for each current spike and averaged them. We show the results of this study in Figure \ref{fig:BC2rftol}. This data shows very clearly that the ESASE compressor is extremely tolerant of errors in the preceding linac structures: the emittance varies by not even a tenth of a nanometer and the current varies by single amperes. The slightly lower mean current of 3.9 kA is a result of using fewer particles for these simulations to speed up the tolerance study; this should not affect the qualitative result of the simulations which is that the ESASE compressor is insensitive to gradient and phase fluctuations at this level. 

\end{appendix}

\section*{References}
\bibliographystyle{unsrt}
\bibliography{references} 

\begin{thebibliography}{100}

\bibitem{PellegriniRMP}
C.~Pellegrini, A.~Marinelli, and S.~Reiche.
\newblock The physics of x-ray free-electron lasers.
\newblock {\em Rev. Mod. Phys.}, 88:015006, Mar 2016.

\bibitem{Bostedt2016}
Christoph Bostedt, S{\'{e}}bastien Boutet, David~M Fritz, Zhirong Huang, Hae~Ja
  Lee, Henrik~T Lemke, Aymeric Robert, William~F Schlotter, Joshua~J Turner,
  and Garth~J Williams.
\newblock {Linac Coherent Light Source: The first five years}.
\newblock {\em Rev. Mod. Phys.}, 88(1):15007, mar 2016.

\bibitem{Emma2010}
P.~Emma, R.~Akre, J.~Arthur, R.~Bionta, C.~Bostedt, J.~Bozek, A.~Brachmann,
  P.~Bucksbaum, R.~Coffee, F.~J. Decker, Y.~Ding, D.~Dowell, S.~Edstrom,
  A.~Fisher, J.~Frisch, S.~Gilevich, J.~Hastings, G.~Hays, Ph. Hering,
  Z.~Huang, R.~Iverson, H.~Loos, M.~Messerschmidt, A.~Miahnahri, S.~Moeller,
  H.~D. Nuhn, G.~Pile, D.~Ratner, J.~Rzepiela, D.~Schultz, T.~Smith, P.~Stefan,
  H.~Tompkins, J.~Turner, J.~Welch, W.~White, J.~Wu, G.~Yocky, and J.~Galayda.
\newblock First lasing and operation of an {\aa}ngstrom-wavelength
  free-electron laser.
\newblock {\em Nature Photonics}, 4(9):641--647, 2010.

\bibitem{Geloni2017}
Gianluca Geloni, Zhirong Huang, and Claudio Pellegrini.
\newblock Chapter 1: The physics and status of x-ray free-electron lasers.
\newblock In {\em X-Ray Free Electron Lasers: Applications in Materials,
  Chemistry and Biology}, pages 1--44. The Royal Society of Chemistry, 2017.

\bibitem{LCLS-II-CDR}
LCLS-II Design~Study Group et~al.
\newblock Lcls-ii conceptual design report.
\newblock Technical report, Report LCLSII-1.1-DR-0001-R0, SLAC, 2014.

\bibitem{Sheffield2017}
Richard~L Sheffield, Cris~W Barnes, and John~P Tapia.
\newblock {Matter-Radiation Interactions in Extremes (MaRIE): Project
  Overview}.
\newblock Technical report, Los Alamos National Lab.(LANL), Los Alamos, NM
  (United States), 2017.

\bibitem{Russell2015}
Steven~John Russell, Bruce~Eric Carlsten, Leanne~Delma Duffy, Frank~L Krawczyk,
  I~V Lewellen, W~John, and Richard~L Sheffield.
\newblock {MaRIE XFEL pre-conceptual reference design injector}.
\newblock Technical report, Los Alamos National Lab.(LANL), Los Alamos, NM
  (United States), 2015.

\bibitem{Lewellen2015}
John Lewellen, Kip Bishofberger, Nikolai Yampolsky, Frank Krawczyk, Leanne
  Duffy, Steven Russell, Bruce Carlsten, Quinn Marksteiner, Dinh Nguyen, and
  Richard Sheffield.
\newblock {Status of the MaRIE X-FEL accelerator design}.
\newblock {\em 6th International Particle Accelerator Conference}, 2015.

\bibitem{Pile2011}
David Pile.
\newblock {First light from SACLA}.
\newblock {\em Nature Photonics}, 5(8):456--457, 2011.

\bibitem{Kang2017}
Heung-Sik Kang, Chang-Ki Min, Hoon Heo, Changbum Kim, Haeryong Yang, Gyujin
  Kim, Inhyuk Nam, Soung~Youl Baek, Hyo-Jin Choi, Geonyeong Mun, Byoung~Ryul
  Park, Young~Jin Suh, Dong~Cheol Shin, Jinyul Hu, Juho Hong, Seonghoon Jung,
  Sang-Hee Kim, KwangHoon Kim, Donghyun Na, Soung~Soo Park, Yong~Jung Park,
  Jang-Hui Han, Young~Gyu Jung, Seong~Hun Jeong, Hong~Gi Lee, Sangbong Lee,
  Sojeong Lee, Woul-Woo Lee, Bonggi Oh, Hyung~Suck Suh, Yong~Woon Parc, Sung-Ju
  Park, Min~Ho Kim, Nam-Suk Jung, Young-Chan Kim, Mong-Soo Lee, Bong-Ho Lee,
  Chi-Won Sung, Ik-Su Mok, Jung-Moo Yang, Chae-Soon Lee, Hocheol Shin, Ji~Hwa
  Kim, Yongsam Kim, Jae~Hyuk Lee, Sang-Youn Park, Jangwoo Kim, Jaeku Park,
  Intae Eom, Seungyu Rah, Sunam Kim, Ki~Hyun Nam, Jaehyun Park, Jaehun Park,
  Sangsoo Kim, Soonam Kwon, Sang~Han Park, Kyung~Sook Kim, Hyojung Hyun,
  Seung~Nam Kim, Seonghan Kim, Sun-min Hwang, Myong~Jin Kim, Chae-yong Lim,
  Chung-Jong Yu, Bong-Soo Kim, Tai-Hee Kang, Kwang-Woo Kim, Seung-Hwan Kim,
  Hee-Seock Lee, Heung-Soo Lee, Ki-Hyeon Park, Tae-Yeong Koo, Dong-Eon Kim, and
  In~Soo Ko.
\newblock {Hard X-ray free-electron laser with femtosecond-scale timing
  jitter}.
\newblock {\em Nature Photonics}, 11(11):708--713, nov 2017.

\bibitem{European-XFEL-TDR}
Massimo Altarelli, Reinhard Brinkmann, Majed Chergui, Winfried Decking, Barry
  Dobson, Stefan D{\"u}sterer, Gerhard Gr{\"u}bel, Walter Graeff, Heinz
  Graafsma, Janos Hajdu, et~al.
\newblock The european x-ray free-electron laser technical design report.
\newblock {\em DESY}, 97(2006):4, 2006.

\bibitem{SwissFEL-CDR}
Romain Ganter.
\newblock Swissfel - conceptual design report.
\newblock Technical report, Paul Scherrer Institute (PSI), 2010.

\bibitem{FERMI-CDR}
FERMI@Elettra~Design team.
\newblock Fermi@elettra - conceptual design report.
\newblock Technical report, Sincrotrone Trieste, 2007.

\bibitem{Schreiber2015}
Siegfried Schreiber and Bart Faatz.
\newblock The free-electron laser flash.
\newblock {\em High Power Laser Science and Engineering}, 3, 2015.

\bibitem{Pellegrini2016a}
C.~Pellegrini.
\newblock {X-ray free-electron lasers: From dreams to reality}.
\newblock {\em Physica Scripta}, 2016(T169), 2016.

\bibitem{RMPLPA}
E.~Esarey, C.~B. Schroeder, and W.~P. Leemans.
\newblock Physics of laser-driven plasma-based electron accelerators.
\newblock {\em Rev. Mod. Phys.}, 81:1229--1285, Aug 2009.

\bibitem{DLARMP}
R.~Joel England, Robert~J. Noble, Karl Bane, David~H. Dowell, Cho-Kuen Ng,
  James~E. Spencer, Sami Tantawi, Ziran Wu, Robert~L. Byer, Edgar Peralta, Ken
  Soong, Chia-Ming Chang, Behnam Montazeri, Stephen~J. Wolf, Benjamin Cowan,
  Jay Dawson, Wei Gai, Peter Hommelhoff, Yen-Chieh Huang, Chunguang Jing,
  Christopher McGuinness, Robert~B. Palmer, Brian Naranjo, James Rosenzweig,
  Gil Travish, Amit Mizrahi, Levi Schachter, Christopher Sears, Gregory~R.
  Werner, and Rodney~B. Yoder.
\newblock Dielectric laser accelerators.
\newblock {\em Rev. Mod. Phys.}, 86:1337--1389, Dec 2014.

\bibitem{SRhistory}
Arthur~L. Robinson.
\newblock History of synchrotron radiation.
\newblock {\em Synchrotron Radiation News}, 28(4):4--9, 2015.

\bibitem{Harrison2012}
Jere Harrison, Abhijeet Joshi, Jonathan Lake, Rob Candler, and Pietro Musumeci.
\newblock {Surface-micromachined magnetic undulator with period length between
  10 um and 1 mm for advanced light sources}.
\newblock {\em Phys. Rev. ST Accel. Beams}, 15(7):70703, jul 2012.

\bibitem{Bonifacio1984}
R~Bonifacio, C~Pellegrini, and L.~M. Narducci.
\newblock {Collective instabilities and high-gain regime in a free electron
  laser}.
\newblock {\em Optics Communications}, 50(6):373--378, jul 1984.

\bibitem{KIM198671}
Kwang-Je Kim.
\newblock Brightness, coherence and propagation characteristics of synchrotron
  radiation.
\newblock {\em Nuclear Instruments and Methods in Physics Research Section A:
  Accelerators, Spectrometers, Detectors and Associated Equipment}, 246(1):71
  -- 76, 1986.

\bibitem{Fraser1985}
J.~S. {Fraser}, R.~L. {Sheffield}, E.~R. {Gray}, and G.~W. {Rodenz}.
\newblock High-brightness photoemitier injector for electron accelerators.
\newblock {\em IEEE Transactions on Nuclear Science}, 32(5):1791--1793, 1985.

\bibitem{Akre2008}
R~Akre, D~Dowell, P~Emma, J~Frisch, S~Gilevich, G~Hays, Ph. Hering, R~Iverson,
  C~Limborg-Deprey, H~Loos, A~Miahnahri, J~Schmerge, J~Turner, J~Welch,
  W~White, and J~Wu.
\newblock {Commissioning the Linac Coherent Light Source injector}.
\newblock {\em Phys. Rev. ST Accel. Beams}, 11(3):30703, mar 2008.

\bibitem{ChambersASE}
R.~G. Chambers and Otto~Robert Frisch.
\newblock The anomalous skin effect.
\newblock {\em Proceedings of the Royal Society of London. Series A.
  Mathematical and Physical Sciences}, 215(1123):481--497, 1952.

\bibitem{Rosenzweig2018}
J~B Rosenzweig, A~Cahill, B~Carlsten, G~Castorina, M~Croia, C~Emma, A~Fukusawa,
  B~Spataro, D~Alesini, V~Dolgashev, M~Ferrario, G~Lawler, R~Li, C~Limborg,
  J~Maxson, P~Musumeci, R~Pompili, S~Tantawi, and O~Williams.
\newblock {Ultra-high brightness electron beams from very-high field cryogenic
  radiofrequency photocathode sources}.
\newblock {\em Nuclear Instruments and Methods in Physics Research Section A:
  Accelerators, Spectrometers, Detectors and Associated Equipment},
  909:224--228, 2018.

\bibitem{Rosenzweig2019}
J~B Rosenzweig, A~Cahill, V~Dolgashev, C~Emma, A~Fukasawa, R~Li, C~Limborg,
  J~Maxson, P~Musumeci, A~Nause, R~Pakter, R~Pompili, R~Roussel, B~Spataro, and
  S~Tantawi.
\newblock {Next generation high brightness electron beams from ultrahigh field
  cryogenic rf photocathode sources}.
\newblock {\em Phys. Rev. Accel. Beams}, 22(2):23403, feb 2019.

\bibitem{LCLSCDR}
J.~Arthur, P.~Anfinrud, P~Audebert, K.~Bane, I.~Ben-Zvi, V.~Bharadwaj,
  R.~Bionta, and P.~Bolton.
\newblock {Linac Coherent Light Source (LCLS) Conceptual Design Report}.
\newblock {\em SLAC-R-593}, 2002.

\bibitem{Prat2019}
E.~Prat, P.~Dijkstal, M.~Aiba, S.~Bettoni, P.~Craievich, E.~Ferrari,
  R.~Ischebeck, F.~L\"ohl, A.~Malyzhenkov, G.~L. Orlandi, S.~Reiche, and
  T.~Schietinger.
\newblock Generation and characterization of intense ultralow-emittance
  electron beams for compact x-ray free-electron lasers.
\newblock {\em Phys. Rev. Lett.}, 123:234801, Dec 2019.

\bibitem{tantawi2018distributed}
Sami Tantawi, Mamdouh Nasr, Zenghai Li, Cecile Limborg, and Philipp Borchard.
\newblock Distributed coupling accelerator structures: A new paradigm for high
  gradient linacs, 2018.

\bibitem{Robles2019}
River Robles and James Rosenzweig.
\newblock Compression of ultra-high brightness beams for a compact x-ray
  free-electron laser.
\newblock {\em Instruments}, 3(4), 2019.

\bibitem{Zholents2005}
Alexander~A. Zholents.
\newblock Method of an enhanced self-amplified spontaneous emission for x-ray
  free electron lasers.
\newblock {\em Phys. Rev. ST Accel. Beams}, 8:040701, Apr 2005.

\bibitem{SashaTrieste}
A.A. Zholents, W.M. Fawley, P.~Emma, Z.~Huang, and G.~Stupakov.
\newblock Current-enhanced sase using an optical laser and its application to
  the lcls.
\newblock In {\em FEL Conf. Proc. 2004: 582-585, 2004}. JaCOW, 2004.

\bibitem{Carlsten2019}
Bruce~E. Carlsten, Petr~M. Anisimov, Cris~W. Barnes, Quinn~R. Marksteiner,
  River~R. Robles, and Nikolai Yampolsky.
\newblock High-brightness beam technology development for a future dynamic
  mesoscale materials science capability.
\newblock {\em Instruments}, 3(4), 2019.

\bibitem{Popmintchev2012}
Tenio Popmintchev, Ming-Chang Chen, Dimitar Popmintchev, Paul Arpin, Susannah
  Brown, Skirmantas Ali{\v{s}}auskas, Giedrius Andriukaitis, Tadas
  Bal{\v{c}}iunas, Oliver~D M{\"{u}}cke, Audrius Pugzlys, Andrius
  Baltu{\v{s}}ka, Bonggu Shim, Samuel~E Schrauth, Alexander Gaeta, Carlos
  Hern{\'{a}}ndez-Garc$\backslash$'$\backslash$ia, Luis Plaja, Andreas Becker,
  Agnieszka Jaron-Becker, Margaret~M Murnane, and Henry~C Kapteyn.
\newblock {Bright Coherent Ultrahigh Harmonics in the keV X-ray Regime from
  Mid-Infrared Femtosecond Lasers}.
\newblock {\em Science}, 336(6086):1287--1291, 2012.

\bibitem{Gadjev2019}
I~Gadjev, N~Sudar, M~Babzien, J~Duris, P~Hoang, M~Fedurin, K~Kusche, R~Malone,
  P~Musumeci, M~Palmer, I~Pogorelsky, M~Polyanskiy, Y~Sakai, C~Swinson,
  O~Williams, and J~B Rosenzweig.
\newblock {An inverse free electron laser acceleration-driven Compton
  scattering X-ray source}.
\newblock {\em Scientific Reports}, 9(1):532, 2019.

\bibitem{Zholents2004}
A.~A. Zholents.
\newblock Current-enhanced sase using an optical laser and its application to
  the lcls.
\newblock {\em SLAC-PUB-10713}, 9 2004.

\bibitem{LBNLbrightness}
M.~Gullans, G.~Penn, J.~S. Wurtele, and M.~Zolotorev.
\newblock Three-dimensional analysis of free-electron laser performance using
  brightness scaled variables.
\newblock {\em Phys. Rev. ST Accel. Beams}, 11:060701, Jun 2008.

\bibitem{Xie1995}
Ming Xie.
\newblock {Design optimization for an X-ray free electron laser driven by SLAC
  linac}.
\newblock In {\em Proceedings Particle Accelerator Conference}, volume~1, pages
  183--185. IEEE, 1995.

\bibitem{Xie2000}
Ming Xie.
\newblock {Exact and variational solutions of 3D eigenmodes in high gain FELs}.
\newblock {\em Nuclear Instruments and Methods in Physics Research Section A:
  Accelerators, Spectrometers, Detectors and Associated Equipment},
  445(1):59--66, 2000.

\bibitem{Marcus2011}
G~Marcus, E~Hemsing, and J~Rosenzweig.
\newblock {Gain length fitting formula for free-electron lasers with strong
  space-charge effects}.
\newblock {\em Phys. Rev. ST Accel. Beams}, 14(8):80702, aug 2011.

\bibitem{Cahill2018b}
A.~D. Cahill, J.~B. Rosenzweig, V.~A. Dolgashev, Z.~Li, S.~G. Tantawi, and
  S.~Weathersby.
\newblock rf losses in a high gradient cryogenic copper cavity.
\newblock {\em Phys. Rev. Accel. Beams}, 21:061301, Jun 2018.

\bibitem{Cahill2018}
A~D Cahill, J~B Rosenzweig, V~A Dolgashev, Z~Li, S~G Tantawi, and S~Weathersby.
\newblock rf losses in a high gradient cryogenic copper cavity.
\newblock {\em Phys. Rev. Accel. Beams}, 21(6):61301, jun 2018.

\bibitem{jrecscaling}
J.~Rosenzweig and E.~Colby.
\newblock Charge and wavelength scaling of rf photoinjector designs.
\newblock {\em AIP Conference Proceedings}, 335(1):724--737, 1995.

\bibitem{KIM1989201}
Kwang-Je Kim.
\newblock Rf and space-charge effects in laser-driven rf electron guns.
\newblock {\em Nuclear Instruments and Methods in Physics Research Section A:
  Accelerators, Spectrometers, Detectors and Associated Equipment}, 275(2):201
  -- 218, 1989.

\bibitem{ZHuang2005}
Z.~Huang, D.~Dowell, P.~Emma, C.~Limborg-Deprey, G.~Stupakov, and J.~Wu.
\newblock Uncorrelated energy spread and longitudinal emittance of a
  photoinjector beam.
\newblock In {\em Proc. 2005 Particle Accelerator Conference, Knoxville,
  Tennessee 2005}, Particle Accelerator Conference, pages 3570--3572, Geneva,
  Switzerland, Jun. 2005. JACoW Publishing.

\bibitem{Rosenzweig1994}
J.B. Rosenzweig and L.~Serafini.
\newblock Transverse particle motion in radio-frequency linear accelerators.
\newblock {\em Physical review. E, Statistical physics, plasmas, fluids, and
  related interdisciplinary topics}, 49:1599--1602, 03 1994.

\bibitem{GPTSite}
S.~B. {van der Geer} and M.~J. {de Loos}.
\newblock Applications of the general particle tracer code.
\newblock In {\em Proceedings of the 1997 Particle Accelerator Conference (Cat.
  No.97CH36167)}, volume~2, pages 2577--2579 vol.2, May 1997.

\bibitem{Vecchione2013}
T~Vecchione, D~Dowell, W~Wan, J~Feng, and H~A Padmore.
\newblock {Quantum efficiency and transverse momentum from metals}.
\newblock {\em Proceedings of FEL2013,(Geneva, Switzerland: JACoW, 2013)}, page
  424, 2013.

\bibitem{dowellqemete}
David~H. Dowell and John~F. Schmerge.
\newblock Quantum efficiency and thermal emittance of metal photocathodes.
\newblock {\em Phys. Rev. ST Accel. Beams}, 12:074201, Jul 2009.

\bibitem{cryoemission}
Jun Feng, J.~Nasiatka, Weishi Wan, Siddharth Karkare, John Smedley, and
  Howard~A. Padmore.
\newblock Thermal limit to the intrinsic emittance from metal photocathodes.
\newblock {\em Applied Physics Letters}, 107(13):134101, 2015.

\bibitem{pierce2020role}
Christopher~M. Pierce, Matthew~B. Andorf, Edmond Lu, Matthew Gordon, Young-Kee
  Kim, Colwyn Gulliford, Ivan~V. Bazarov, Jared~M. Maxson, Nora~P. Norvell,
  Bruce~M. Dunham, and Tor~O. Raubenheimer.
\newblock The role of low intrinsic emittance in modern photoinjector
  brightness, 2020.

\bibitem{Bae2018}
Jai~Kwan Bae, Ivan Bazarov, Pietro Musumeci, Siddharth Karkare, Howard Padmore,
  and Jared Maxson.
\newblock Brightness of femtosecond nonequilibrium photoemission in metallic
  photocathodes at wavelengths near the photoemission threshold.
\newblock {\em Journal of Applied Physics}, 124(24):244903, 2018.

\bibitem{bane2018advanced}
Karl~L. Bane, Timothy~L. Barklow, Martin Breidenbach, Craig~P. Burkhart,
  Eric~A. Fauve, Alysson~R. Gold, Vincent Heloin, Zenghai Li, Emilio~A. Nanni,
  Mamdouh Nasr, Marco Oriunno, James~McEwan Paterson, Michael~E. Peskin, Tor~O.
  Raubenheimer, and Sami~G. Tantawi.
\newblock An advanced ncrf linac concept for a high energy e$^+$e$^-$ linear
  collider, 2018.

\bibitem{Brinkmann2001}
R~Brinkmann, Y~Derbenev, and K~Fl{\"{o}}ttmann.
\newblock {A low emittance, flat-beam electron source for linear colliders}.
\newblock {\em Phys. Rev. ST Accel. Beams}, 4(5):53501, may 2001.

\bibitem{Shintake_1992}
Tsumoru Shintake.
\newblock The choke mode cavity.
\newblock {\em Japanese Journal of Applied Physics}, 31(Part 2, No.
  11A):L1567--L1570, nov 1992.

\bibitem{MOSNIER198781}
A.~Mosnier.
\newblock Analyse de la stabilite de faisceau dans un accelerateur lineaire
  avec passages multiples en prenant en compte les effets de toutes les
  cavites.
\newblock {\em Nuclear Instruments and Methods in Physics Research Section A:
  Accelerators, Spectrometers, Detectors and Associated Equipment}, 257(2):81
  -- 90, 1987.

\bibitem{perez_huang_voter_2018}
Danny Perez, Rao Huang, and Arthur~F. Voter.
\newblock Long-time molecular dynamics simulations on massively parallel
  platforms: A comparison of parallel replica dynamics and parallel trajectory
  splicing.
\newblock {\em Journal of Materials Research}, 33(7):813–822, 2018.

\bibitem{Theodore2006}
N.~D. {Theodore}, B.~C. {Holloway}, D.~M. {Manos}, R.~{Moore}, C.~{Hernandez},
  T.~{Wang}, and H.~F. {Dylla}.
\newblock Nitrogen-implanted silicon oxynitride: A coating for suppressing
  field emission from stainless steel used in high-voltage applications.
\newblock {\em IEEE Transactions on Plasma Science}, 34(4):1074--1079, Aug
  2006.

\bibitem{bambade2019international}
Philip Bambade, Tim Barklow, Ties Behnke, Mikael Berggren, James Brau, Philip
  Burrows, Dmitri Denisov, Angeles Faus-Golfe, Brian Foster, Keisuke Fujii,
  Juan Fuster, Frank Gaede, Paul Grannis, Christophe Grojean, Andrew Hutton,
  Benno List, Jenny List, Shinichiro Michizono, Akiya Miyamoto, Olivier Napoly,
  Michael Peskin, Roman Poeschl, Frank Simon, Jan Strube, Junping Tian, Maksym
  Titov, Marcel Vos, Andrew White, Graham Wilson, Akira Yamamoto, Hitoshi
  Yamamoto, and Kaoru Yokoya.
\newblock The international linear collider: A global project, 2019.

\bibitem{Leemans2006}
W.~P. Leemans, B.~Nagler, A.~J Gonsalves, Cs. T{\'o}th, K.~Nakamura, C.~G.~R.
  Geddes, E.~Esarey, C.~B. Schroeder, and S.~M. Hooker.
\newblock Gev electron beams from a centimetre-scale accelerator.
\newblock {\em Nature physics}, 2(10):696--699, 2006.

\bibitem{Fuchs2009}
Matthias Fuchs, Raphael Weingartner, Antonia Popp, Zsuzsanna Major, Stefan
  Becker, Jens Osterhoff, Isabella Cortrie, Benno Zeitler, Rainer
  H{\"{o}}rlein, George~D Tsakiris, Ulrich Schramm, Tom~P Rowlands-Rees,
  Simon~M Hooker, Dietrich Habs, Ferenc Krausz, Stefan Karsch, and Florian
  Gr{\"{u}}ner.
\newblock {Laser-driven soft-X-ray undulator source}.
\newblock {\em Nature Physics}, 5(11):826--829, 2009.

\bibitem{Labat2018}
M.~Labat, A.~Loulergue, T.~Andre, I.~A. Andriyash, A.~Ghaith, M.~Khojoyan,
  F.~Marteau, M.~Vall\'eau, F.~Briquez, C.~Benabderrahmane, O.~Marcouill\'e,
  C.~Evain, and M.~E. Couprie.
\newblock Robustness of a plasma acceleration based free electron laser.
\newblock {\em Phys. Rev. Accel. Beams}, 21:114802, Nov 2018.

\bibitem{VanTilborg2017}
J.~van Tilborg, S.~K. Barber, F.~Isono, C.~B. Schroeder, E.~Esarey, and W.~P.
  Leemans.
\newblock {Free-electron lasers driven by laser plasma accelerators}.
\newblock {\em AIP Conference Proceedings}, 1812, 2017.

\bibitem{LPAICS}
K.~Ta~Phuoc, S.~Corde, C.~Thaury, V.~Malka, A.~Tafzi, J.~P. Goddet, R.~C. Shah,
  S.~Sebban, and A.~Rousse.
\newblock All-optical compton gamma-ray source.
\newblock {\em Nature Photonics}, 6(5):308--311, 2012.

\bibitem{Corde2013}
S~Corde, K~Ta Phuoc, G~Lambert, R~Fitour, V~Malka, and A~Rousse.
\newblock {Femtosecond x rays from laser-plasma accelerators}.
\newblock {\em REVIEWS OF MODERN PHYSICS}, 85(January-March), 2013.

\bibitem{Wang2016}
W.~T. Wang, W.~T. Li, J.~S. Liu, Z.~J. Zhang, R.~Qi, C.~H. Yu, J.~Q. Liu,
  M.~Fang, Z.~Y. Qin, C.~Wang, Y.~Xu, F.~X. Wu, Y.~X. Leng, R.~X. Li, and Z.~Z.
  Xu.
\newblock {High-Brightness High-Energy Electron Beams from a Laser Wakefield
  Accelerator via Energy Chirp Control}.
\newblock {\em Physical Review Letters}, 117(12), 2016.

\bibitem{HuangTGU}
Zhirong Huang, Yuantao Ding, and Carl~B. Schroeder.
\newblock Compact x-ray free-electron laser from a laser-plasma accelerator
  using a transverse-gradient undulator.
\newblock {\em Phys. Rev. Lett.}, 109:204801, Nov 2012.

\bibitem{Majernik19}
N.~Majernik, S.~K. Barber, J.~van Tilborg, J.~B. Rosenzweig, and W.~P. Leemans.
\newblock Optimization of low aspect ratio, iron dominated dipole magnets.
\newblock {\em Phys. Rev. Accel. Beams}, 22:032401, Mar 2019.

\bibitem{Maier2012}
AR~Maier, Atoosa Meseck, Sven Reiche, CB~Schroeder, Thorben Seggebrock, and
  Florian Gruener.
\newblock Demonstration scheme for a laser-plasma-driven free-electron laser.
\newblock {\em Physical Review X}, 2(3):031019, 2012.

\bibitem{Barber2018}
S.~K. Barber, J.~{Van Tilborg}, C.~B. Schroeder, R.~Lehe, H.~E. Tsai, K.~K.
  Swanson, S.~Steinke, K.~Nakamura, C.~G.R. Geddes, C.~Benedetti, E.~Esarey,
  and W.~P. Leemans.
\newblock {Parametric emittance measurements of electron beams produced by a
  laser plasma accelerator}.
\newblock {\em Plasma Physics and Controlled Fusion}, 60(5), 2018.

\bibitem{Hidding2012}
B~Hidding, G~Pretzler, J~B Rosenzweig, T~K{\"{o}}nigstein, D~Schiller, and D~L
  Bruhwiler.
\newblock {Ultracold Electron Bunch Generation via Plasma Photocathode Emission
  and Acceleration in a Beam-Driven Plasma Blowout}.
\newblock {\em Phys. Rev. Lett.}, 108(3):35001, jan 2012.

\bibitem{Yunfeng2013}
Y.~Xi, B.~Hidding, D.~Bruhwiler, G.~Pretzler, and J.~B. Rosenzweig.
\newblock Hybrid modeling of relativistic underdense plasma photocathode
  injectors.
\newblock {\em Phys. Rev. ST Accel. Beams}, 16:031303, Mar 2013.

\bibitem{Deng2019}
A~Deng, O~S Karger, T~Heinemann, A~Knetsch, P~Scherkl, G~G Manahan, A~Beaton,
  D~Ullmann, G~Wittig, A~F Habib, Y~Xi, M~D Litos, B~D O'Shea, S~Gessner, C~I
  Clarke, S~Z Green, C~A Lindstr{\o}m, E~Adli, R~Zgadzaj, M~C Downer,
  G~Andonian, A~Murokh, D~L Bruhwiler, J~R Cary, M~J Hogan, V~Yakimenko, J~B
  Rosenzweig, and B~Hidding.
\newblock {Generation and acceleration of electron bunches from a plasma
  photocathode}.
\newblock {\em Nature Physics}, 15(11):1156--1160, 2019.

\bibitem{FERRARIO2018}
M.~Ferrario, D.~Alesini, M.P. Anania, M.~Artioli, A.~Bacci, S.~Bartocci,
  R.~Bedogni, M.~Bellaveglia, A.~Biagioni, F.~Bisesto, F.~Brandi,
  E.~Brentegani, F.~Broggi, B.~Buonomo, P.L. Campana, G.~Campogiani,
  C.~Cannaos, S.~Cantarella, F.~Cardelli, M.~Carpanese, M.~Castellano,
  G.~Castorina, N.~Catalan Lasheras, E.~Chiadroni, A.~Cianchi, R.~Cimino,
  F.~Ciocci, D.~Cirrincione, G.A.P. Cirrone, R.~Clementi, M.~Coreno,
  R.~Corsini, M.~Croia, A.~Curcio, G.~Costa, C.~Curatolo, G.~Cuttone,
  S.~Dabagov, G.~Dattoli, G.~D’Auria, I.~Debrot, M.~Diomede, A.~Drago,
  D.~Di Giovenale, S.~Di Mitri, G.~Di Pirro, A.~Esposito, M.~Faiferri,
  L.~Ficcadenti, F.~Filippi, O.~Frasciello, A.~Gallo, A.~Ghigo, L.~Giannessi,
  A.~Giribono, L.~Gizzi, A.~Grudiev, S.~Guiducci, P.~Koester, S.~Incremona,
  F.~Iungo, L.~Labate, A.~Latina, S.~Licciardi, V.~Lollo, S.~Lupi, R.~Manca,
  A.~Marcelli, M.~Marini, A.~Marocchino, M.~Marongiu, V.~Martinelli,
  C.~Masciovecchio, C.~Mastino, A.~Michelotti, C.~Milardi, V.~Minicozzi,
  F.~Mira, S.~Morante, A.~Mostacci, F.~Nguyen, S.~Pagnutti, L.~Pellegrino,
  A.~Petralia, V.~Petrillo, L.~Piersanti, S.~Pioli, D.~Polese, R.~Pompili,
  F.~Pusceddu, A.~Ricci, R.~Ricci, R.~Rochow, S.~Romeo, J.B. Rosenzweig,
  M.~Rossetti Conti, A.R. Rossi, U.~Rotundo, L.~Sabbatini, E.~Sabia, O.~Sans
  Plannell, D.~Schulte, J.~Scifo, V.~Scuderi, L.~Serafini, B.~Spataro,
  A.~Stecchi, A.~Stella, V.~Shpakov, F.~Stellato, E.~Turco, C.~Vaccarezza,
  A.~Vacchi, A.~Vannozzi, A.~Variola, S.~Vescovi, F.~Villa, W.~Wuensch,
  A.~Zigler, and M.~Zobov.
\newblock Eupraxia at sparclab design study towards a compact fel facility at
  lnf.
\newblock {\em Nuclear Instruments and Methods in Physics Research Section A},
  909:134, 2018.
\newblock 3rd European Advanced Accelerator Concepts workshop (EAAC2017).

\bibitem{smith1986intense}
Todd~I Smith.
\newblock Intense low emittance linac beams for free electron lasers, 1986.

\bibitem{dowell1996boeing}
DH~Dowell, JL~Adamski, TD~Hayward, CG~Parazzoli, and AM~Vetter.
\newblock The boeing photocathode accelerator magnetic pulse compression and
  energy recovery experiment.
\newblock {\em Nuclear Instruments and Methods in Physics Research Section A:
  Accelerators, Spectrometers, Detectors and Associated Equipment},
  375(1-3):108--111, 1996.

\bibitem{behtouei2020sw}
Mostafa Behtouei, Luigi Faillace, Bruno Spataro, Alessandro Variola, and Mauro
  Migliorati.
\newblock A sw ka-band linearizer structure with minimum surface electric field
  for the compact light xls project, 2020.

\bibitem{Borland2000}
M~Borland.
\newblock Elegant: A flexible sdds-compliant code for accelerator simulation.
\newblock {\em 6th International Computational Accelerator Physics Conference},
  LS-287, 8 2000.

\bibitem{huang2005intrabeam}
Z~Huang.
\newblock Intrabeam scattering in an x-ray fel driver.
\newblock Technical report, Stanford Linear Accelerator Center (SLAC), Menlo
  Park, CA, 2005.

\bibitem{Brynes_2018}
A~D Brynes, P~Smorenburg, I~Akkermans, E~Allaria, L~Badano, S~Brussaard,
  M~Danailov, A~Demidovich, G~De~Ninno, D~Gauthier, and et~al.
\newblock Beyond the limits of 1d coherent synchrotron radiation.
\newblock {\em New Journal of Physics}, 20(7):073035, Jul 2018.

\bibitem{Faillace2019}
L.~Faillace, M.~Behtouei, V.A. Dolgashev, B.~Spataro, G.~Torrisi, and
  A.~Variola.
\newblock Compact ultra high-gradient ka-band accelerating structure for
  research, medical and industrial applications.
\newblock In {\em Proc. 10th International Particle Accelerator Conference
  (IPAC'19), Melbourne, Australia, 19-24 May 2019}, number~10 in International
  Partile Accelerator Conference, pages 2842--2845, Geneva, Switzerland, Jun.
  2019. JACoW Publishing.
\newblock https://doi.org/10.18429/JACoW-IPAC2019-WEPRB020.

\bibitem{DiMitri2013}
S.~Di~Mitri, M.~Cornacchia, and S.~Spampinati.
\newblock Cancellation of coherent synchrotron radiation kicks with optics
  balance.
\newblock {\em Phys. Rev. Lett.}, 110:014801, Jan 2013.

\bibitem{Jing2013}
Yichao Jing, Yue Hao, and Vladimir~N. Litvinenko.
\newblock Compensating effect of the coherent synchrotron radiation in bunch
  compressors.
\newblock {\em Phys. Rev. ST Accel. Beams}, 16:060704, Jun 2013.

\bibitem{BANE2016156}
Karl Bane and Gennady Stupakov.
\newblock Dechirper wakefields for short bunches.
\newblock {\em Nuclear Instruments and Methods in Physics Research Section A:
  Accelerators, Spectrometers, Detectors and Associated Equipment}, 820:156 --
  163, 2016.

\bibitem{Hemsing2014}
Erik Hemsing, Gennady Stupakov, Dao Xiang, and Alexander Zholents.
\newblock Beam by design: Laser manipulation of electrons in modern
  accelerators.
\newblock {\em Rev. Mod. Phys.}, 86:897--941, Jul 2014.

\bibitem{Derbenev1995}
Ya.S. Derbenev, J.~Rossbach, E.L. Saldin, and V.D. Shiltsev.
\newblock {Microbunch radiative tail - head interaction}, 9 1995.

\bibitem{borland2001start}
M~Borland, Y-C Chae, S~Milton, R~Soliday, V~Bharadwaj, P~Emma, P~Krejcik,
  C~Limborg, H-D Nuhn, and M~Woodley.
\newblock Start-to-end jitter simulations of the linac coherent light source.
\newblock In {\em PACS2001. Proceedings of the 2001 Particle Accelerator
  Conference (Cat. No. 01CH37268)}, volume~4, pages 2707--2709. IEEE, 2001.

\bibitem{BORLAND2002268}
M.~Borland, Y.C. Chae, P.~Emma, J.W. Lewellen, V.~Bharadwaj, W.M. Fawley,
  P.~Krejcik, C.~Limborg, S.V. Milton, H.-D. Nuhn, R.~Soliday, and M.~Woodley.
\newblock Start-to-end simulation of self-amplified spontaneous emission free
  electron lasers from the gun through the undulator.
\newblock {\em Nuclear Instruments and Methods in Physics Research Section A:
  Accelerators, Spectrometers, Detectors and Associated Equipment}, 483(1):268
  -- 272, 2002.
\newblock Proceedings of the 23rd International Free Electron Laser Confere nce
  and 8th FEL Users Workshop.

\bibitem{heifets2002coherent}
Stupakov Heifets, G~Stupakov, and S~Krinsky.
\newblock Coherent synchrotron radiation instability in a bunch compressor.
\newblock {\em Physical Review Special Topics-Accelerators and Beams},
  5(6):064401, 2002.

\bibitem{Huang2004}
Z~Huang, M~Borland, P~Emma, J~Wu, C~Limborg, G~Stupakov, and J~Welch.
\newblock {Suppression of microbunching instability in the linac coherent light
  source}.
\newblock {\em Phys. Rev. ST Accel. Beams}, 7(7):74401, jul 2004.

\bibitem{qiang2017start}
Ji~Qiang, Y~Ding, P~Emma, Z~Huang, D~Ratner, TO~Raubenheimer, M~Venturini, and
  F~Zhou.
\newblock Start-to-end simulation of the shot-noise driven microbunching
  instability experiment at the linac coherent light source.
\newblock {\em Physical Review Accelerators and Beams}, 20(5):054402, 2017.

\bibitem{OShea2010}
F.~H. O'Shea, G.~Marcus, J.~B. Rosenzweig, M.~Scheer, J.~Bahrdt,
  R.~Weingartner, A.~Gaupp, and F.~Gr{\"{u}}ner.
\newblock {Short period, high field cryogenic undulator for extreme performance
  x-ray free electron lasers}.
\newblock {\em Physical Review Special Topics - Accelerators and Beams},
  13(7):1--12, 2010.

\bibitem{OShea2016}
F.~H. O'Shea, R.~Agustsson, Y.-C. Chen, A.~J. Palmowski, and E.~Spranza.
\newblock {Development of a short period cryogenic undulator at RadiaBeam}.
\newblock {\em Proceedings of NAPAC2016}, 2016.

\bibitem{Murokh2014}
Alex Murokh, Vyacheslav Solovyov, Ron Agustsson, Finn~H. O'Shea, Oleg Chubar,
  Yung Chen, and Thomas Grandsaert.
\newblock {Textured dysprosium and gadolinium poles for high-field,
  short-period hybrid undulators}.
\newblock {\em Nuclear Instruments and Methods in Physics Research, Section A:
  Accelerators, Spectrometers, Detectors and Associated Equipment},
  735:521--527, 2014.

\bibitem{Tanaka2019}
Takashi Tanaka and Akihiro Kagamihata.
\newblock {Demonstration of high-performance pole pieces made of
  monocrystalline dysprosium for short-period undulators}.
\newblock {\em Journal of Synchrotron Radiation}, 26:1220--1225, 2019.

\bibitem{Majernik2019Comb}
N.~Majernik and J.~B. Rosenzweig.
\newblock {Design of Comb Fabricated Halbach Undulators}.
\newblock {\em Instruments}, 3(58), 2019.

\bibitem{Harrison2014}
J.~Harrison, A.~Joshi, Y.~Hwang, O.~Paydar, J.~Lake, P.~Musumeci, and R.~N.
  Candler.
\newblock {Surface-micromachined electromagnets for 100 $\mu$m-scale undulators
  and focusing optics}.
\newblock {\em Physics Procedia}, 52:19--26, 2014.

\bibitem{Harrison2015}
Jere Harrison, Yongha Hwang, Omeed Paydar, and Jimmy Wu.
\newblock {High-gradient microelectromechanical system quadrupole
  electromagnets for particle beam focusing and steering}.
\newblock {\em Physical Review Accelerators and Beams}, 023501:1--10, 2015.

\bibitem{Carr2001}
Roger Carr, Max Cornacchia, Paul Emma, Heinz-Dieter Nuhn, Ben Poling, Robert
  Ruland, Erik Johnson, George Rakowsky, John Skaritka, Steve Lidia, Pat Duffy,
  Marcus Libkind, Pedro Frigola, Alex Murokh, Claudio Pellegrini, James
  Rosenzweig, and Aaron Tremaine.
\newblock Visible-infrared self-amplified spontaneous emission amplifier free
  electron laser undulator.
\newblock {\em Phys. Rev. ST Accel. Beams}, 4:122402, Dec 2001.

\bibitem{Stupakov2015}
G.~Stupakov, K.~L.F. Bane, P.~Emma, and B.~Podobedov.
\newblock {Resistive wall wakefields of short bunches at cryogenic
  temperatures}.
\newblock {\em Physical Review Special Topics - Accelerators and Beams},
  18(3):1--6, 2015.

\bibitem{Reiche1999}
S.~Reiche.
\newblock {GENESIS 1.3: A fully 3D time-dependent FEL simulation code}.
\newblock {\em Nuclear Instruments and Methods in Physics Research, Section A:
  Accelerators, Spectrometers, Detectors and Associated Equipment},
  429(1):243--248, 1999.

\bibitem{PotylitsynIBIC2018}
A.~Potylitsyn, G.~Kube, A.I. Novokshonov, and L.G. Sukhikh.
\newblock {S}patial {R}esolution {I}mprovement of {OTR} {M}onitors by
  {O}ff{-}axis {L}ight {C}ollection.
\newblock In {\em Proc. 7th International Beam Instrumentation Conference
  (IBIC'18), Shanghai, China, 09-13 September 2018}, number~7 in International
  Beam Instrumentation Conference, pages 451--454, Geneva, Switzerland, Jan.
  2019. JACoW Publishing.
\newblock https://doi.org/10.18429/JACoW-IBIC2018-WEPB11.

\bibitem{Sukhikh2017}
L.~G. Sukhikh, G.~Kube, and A.~P. Potylitsyn.
\newblock Simulation of transition radiation based beam imaging from tilted
  targets.
\newblock {\em Phys. Rev. Accel. Beams}, 20:032802, Mar 2017.

\bibitem{MarinelliCDI}
A.~Marinelli, M.~Dunning, S.~Weathersby, E.~Hemsing, D.~Xiang, G.~Andonian,
  F.~O'Shea, Jianwei Miao, C.~Hast, and J.~B. Rosenzweig.
\newblock Single-shot coherent diffraction imaging of microbunched relativistic
  electron beams for free-electron laser applications.
\newblock {\em Phys. Rev. Lett.}, 110:094802, Mar 2013.

\bibitem{borrelli2018}
Simona Borrelli, Gian~Luca Orlandi, Martin Bednarzik, Christian David, Eugenio
  Ferrari, Vitaliy~A Guzenko, Cigdem Ozkan-Loch, Eduard Prat, and Rasmus
  Ischebeck.
\newblock Generation and measurement of sub-micrometer relativistic electron
  beams.
\newblock {\em Communications Physics}, 1(1):1--8, 2018.

\bibitem{Xdeflector}
C.~Behrens, F.~J. Decker, Y.~Ding, V.~A. Dolgashev, J.~Frisch, Z.~Huang,
  P.~Krejcik, H.~Loos, A.~Lutman, T.~J. Maxwell, J.~Turner, J.~Wang, M.~H.
  Wang, J.~Welch, and J.~Wu.
\newblock Few-femtosecond time-resolved measurements of x-ray free-electron
  lasers.
\newblock {\em Nature Communications}, 5(1):3762, 2014.

\bibitem{Attosweeper}
G.~Andonian, E.~Hemsing, D.~Xiang, P.~Musumeci, A.~Murokh, S.~Tochitsky, and
  J.~B. Rosenzweig.
\newblock Longitudinal profile diagnostic scheme with subfemtosecond resolution
  for high-brightness electron beams.
\newblock {\em Phys. Rev. ST Accel. Beams}, 14:072802, Jul 2011.

\bibitem{Tremaine1998}
A.~Tremaine, J.~B. Rosenzweig, S.~Anderson, P.~Frigola, M.~Hogan, A.~Murokh,
  C.~Pellegrini, D.~C. Nguyen, and R.~L. Sheffield.
\newblock Observation of self-amplified spontaneous-emission-induced
  electron-beam microbunching using coherent transition radiation.
\newblock {\em Phys. Rev. Lett.}, 81:5816--5819, Dec 1998.

\bibitem{Liu1998}
Y.~Liu, X.~J. Wang, D.~B. Cline, M.~Babzien, J.~M. Fang, J.~Gallardo,
  K.~Kusche, I.~Pogorelsky, J.~Skaritka, and A.~van Steenbergen.
\newblock Experimental observation of femtosecond electron beam microbunching
  by inverse free-electron-laser acceleration.
\newblock {\em Phys. Rev. Lett.}, 80:4418--4421, May 1998.

\bibitem{huang2007}
Zhirong Huang and Kwang-Je Kim.
\newblock Review of x-ray free-electron laser theory.
\newblock {\em Physical Review Special Topics-Accelerators and Beams},
  10(3):034801, 2007.

\bibitem{Rosenzweig2008}
J~B Rosenzweig, D~Alesini, G~Andonian, M~Boscolo, M~Dunning, L~Faillace,
  M~Ferrario, A~Fukusawa, L~Giannessi, E~Hemsing, G~Marcus, A~Marinelli,
  P~Musumeci, B~O'Shea, L~Palumbo, C~Pellegrini, V~Petrillo, S~Reiche,
  C~Ronsivalle, B~Spataro, and C~Vaccarezza.
\newblock {Generation of ultra-short, high brightness electron beams for
  single-spike SASE FEL operation}.
\newblock {\em Nuclear Instruments and Methods in Physics Research Section A:
  Accelerators, Spectrometers, Detectors and Associated Equipment},
  593(1):39--44, 2008.

\bibitem{HalavanauPC}
{Private communication with A. Halavanau}.

\bibitem{mahajan1983}
Virendra~N Mahajan.
\newblock Strehl ratio for primary aberrations in terms of their aberration
  variance.
\newblock {\em JOSA}, 73(6):860--861, 1983.

\bibitem{marechal1970}
A.~{Mar{\'e}chal} and M.~{Fran{\c{c}}on}.
\newblock {\em {Diffraction: Image structure. Influence of the coherence of
  light.}}
\newblock Masson, Paris, 1970.

\bibitem{Kirkpatrick48}
Paul Kirkpatrick and A.~V. Baez.
\newblock Formation of optical images by x-rays.
\newblock {\em J. Opt. Soc. Am.}, 38(9):766--774, Sep 1948.

\bibitem{grizolli2017}
Walan Grizolli, Xianbo Shi, Tomasz Kolodziej, Yuri Shvyd'ko, and Lahsen
  Assoufid.
\newblock Single-grating talbot imaging for wavefront sensing and x-ray
  metrology.
\newblock In {\em Advances in Metrology for X-Ray and EUV Optics VII}, volume
  10385, page 1038502. International Society for Optics and Photonics, 2017.

\bibitem{idir2010}
Mourad Idir, Pascal Mercere, Mohammed~H Modi, Guillaume Dovillaire, Xavier
  Levecq, Samuel Bucourt, Lionel Escolano, and Paul Sauvageot.
\newblock X-ray active mirror coupled with a hartmann wavefront sensor.
\newblock {\em Nuclear Instruments and Methods in Physics Research Section A:
  Accelerators, Spectrometers, Detectors and Associated Equipment},
  616(2-3):162--171, 2010.

\bibitem{mercere2003}
Pascal Merc{\`e}re, Philippe Zeitoun, Mourad Idir, S{\'e}bastien Le~Pape, Denis
  Douillet, Xavier Levecq, Guillaume Dovillaire, Samuel Bucourt, Kenneth~A
  Goldberg, Patrick~P Naulleau, et~al.
\newblock Hartmann wave-front measurement at 13.4 nm with $\lambda$ euv/120
  accuracy.
\newblock {\em Optics letters}, 28(17):1534--1536, 2003.

\bibitem{marathe2014}
Shashidhara Marathe, Xianbo Shi, Ali~M Khounsary, Michael~J Wojcik, Naresh~G
  Kujala, Albert~T Macrander, and Lahsen Assoufid.
\newblock Development of single grating x-ray talbot interferometer as a
  feedback loop sensor element of an adaptive x-ray mirror system.
\newblock In {\em Adaptive X-Ray Optics III}, volume 9208, page 92080D.
  International Society for Optics and Photonics, 2014.

\bibitem{floter2010}
Bernhard Fl{\"o}ter, Pavle Jurani{\'c}, Svea Kapitzki, Barbara Keitel, Klaus
  Mann, Elke Pl{\"o}njes, Bernd Sch{\"a}fer, and Kai Tiedtke.
\newblock Euv hartmann sensor for wavefront measurements at the free-electron
  laser in hamburg.
\newblock {\em New Journal of Physics}, 12(8):083015, 2010.

\bibitem{liu2018}
Yanwei Liu, Matthew Seaberg, Diling Zhu, Jacek Krzywinski, Frank Seiboth, Corey
  Hardin, Daniele Cocco, Andrew Aquila, Bob Nagler, Hae~Ja Lee, et~al.
\newblock High-accuracy wavefront sensing for x-ray free electron lasers.
\newblock {\em Optica}, 5(8):967--975, 2018.

\bibitem{ThermalMetal}
B.~Y. Mueller and B.~Rethfeld.
\newblock Relaxation dynamics in laser-excited metals under nonequilibrium
  conditions.
\newblock {\em Phys. Rev. B}, 87:035139, Jan 2013.

\bibitem{Cocco_2015}
Daniele Cocco.
\newblock Recent developments in uv optics for ultra-short, ultra-intense
  coherent light sources.
\newblock {\em Photonics}, 2(1):40–49, Jan 2015.

\bibitem{XrayDavid}
C.~David, S.~Gorelick, S.~Rutishauser, J.~Krzywinski, J.~Vila-Comamala, V.~A.
  Guzenko, O.~Bunk, E.~F{\"a}rm, M.~Ritala, M.~Cammarata, D.~M. Fritz,
  R.~Barrett, L.~Samoylova, J.~Gr{\"u}nert, and H.~Sinn.
\newblock Nanofocusing of hard x-ray free electron laser pulses using diamond
  based fresnel zone plates.
\newblock {\em Scientific Reports}, 1(1):57, 2011.

\bibitem{Stoupin2010}
Stanislav Stoupin and Yuri~V. Shvyd'ko.
\newblock Thermal expansion of diamond at low temperatures.
\newblock {\em Phys. Rev. Lett.}, 104:085901, Feb 2010.

\bibitem{Stoupin2012}
S.~Stoupin, A.~M. March, H.~Wen, D.~A. Walko, Y.~Li, E.~M. Dufresne, S.~A.
  Stepanov, K.-J. Kim, Yu.~V. Shvyd'ko, V.~D. Blank, and S.~A. Terentyev.
\newblock Direct observation of dynamics of thermal expansion using pump-probe
  high-energy-resolution x-ray diffraction.
\newblock {\em Phys. Rev. B}, 86:054301, Aug 2012.

\bibitem{Weisshaupt2017}
J~Weisshaupt, A~Rouz{\'{e}}e, M~Woerner, M~J~J Vrakking, T~Elsaesser, E~L
  Shirley, and A~Borgschulte.
\newblock {Ultrafast modulation of electronic structure by coherent phonon
  excitations}.
\newblock {\em Phys. Rev. B}, 95(8):81101, feb 2017.

\bibitem{exleap}
Joseph Duris, Siqi Li, Taran Driver, Elio~G. Champenois, James~P. MacArthur,
  Alberto~A. Lutman, Zhen Zhang, Philipp Rosenberger, Jeff~W. Aldrich, Ryan
  Coffee, Giacomo Coslovich, Franz-Josef Decker, James~M. Glownia, Gregor
  Hartmann, Wolfram Helml, Andrei Kamalov, Jonas Knurr, Jacek Krzywinski,
  Ming-Fu Lin, Jon~P. Marangos, Megan Nantel, Adi Natan, Jordan~T. O’Neal,
  Niranjan Shivaram, Peter Walter, Anna~Li Wang, James~J. Welch, Thomas J.~A.
  Wolf, Joseph~Z. Xu, Matthias~F. Kling, Philip~H. Bucksbaum, Alexander
  Zholents, Zhirong Huang, James~P. Cryan, and Agostino Marinelli.
\newblock Tunable isolated attosecond x-ray pulses with gigawatt peak power
  from a free-electron laser.
\newblock {\em Nature Photonics}, 24:30, January 2020.

\bibitem{Kubacka1333}
T.~Kubacka, J.~A. Johnson, M.~C. Hoffmann, C.~Vicario, S.~de~Jong, P.~Beaud,
  S.~Gr{\"u}bel, S.-W. Huang, L.~Huber, L.~Patthey, Y.-D. Chuang, J.~J. Turner,
  G.~L. Dakovski, W.-S. Lee, M.~P. Minitti, W.~Schlotter, R.~G. Moore, C.~P.
  Hauri, S.~M. Koohpayeh, V.~Scagnoli, G.~Ingold, S.~L. Johnson, and U.~Staub.
\newblock Large-amplitude spin dynamics driven by a thz pulse in resonance with
  an electromagnon.
\newblock {\em Science}, 343(6177):1333--1336, 2014.

\bibitem{Cook2009}
A.~M. Cook, R.~Tikhoplav, S.~Y. Tochitsky, G.~Travish, O.~B. Williams, and
  J.~B. Rosenzweig.
\newblock {Observation of narrow-band terahertz coherent cherenkov radiation
  from a cylindrical dielectric-lined waveguide}.
\newblock {\em Physical Review Letters}, 103(9):1--4, 2009.

\bibitem{Tokura_2006}
Y~Tokura.
\newblock Critical features of colossal magnetoresistive manganites.
\newblock {\em Reports on Progress in Physics}, 69(3):797--851, feb 2006.

\bibitem{Fu2015}
Feichao Fu, Rui Wang, Pengfei Zhu, Lingrong Zhao, Tao Jiang, Chao Lu,
  Shengguang Liu, Libin Shi, Lixin Yan, Haixiao Deng, Chao Feng, Qiang Gu,
  Dazhang Huang, Bo~Liu, Dong Wang, Xingtao Wang, Meng Zhang, Zhentang Zhao,
  Gennady Stupakov, Dao Xiang, and Jie Zhang.
\newblock Demonstration of nonlinear-energy-spread compensation in relativistic
  electron bunches with corrugated structures.
\newblock {\em Phys. Rev. Lett.}, 114:114801, Mar 2015.

\bibitem{Deng2014}
Haixiao Deng, Meng Zhang, Chao Feng, Tong Zhang, Xingtao Wang, Taihe Lan, Lie
  Feng, Wenyan Zhang, Xiaoqing Liu, Haifeng Yao, Lei Shen, Bin Li, Junqiang
  Zhang, Xuan Li, Wencheng Fang, Dan Wang, Marie-emmanuelle Couprie, Guoqiang
  Lin, Bo~Liu, Qiang Gu, Dong Wang, and Zhentang Zhao.
\newblock Experimental demonstration of longitudinal beam phase-space
  linearizer in a free-electron laser facility by corrugated structures.
\newblock {\em Phys. Rev. Lett.}, 113:254802, Dec 2014.

\bibitem{Inoue1492}
Ichiro Inoue, Yuichi Inubushi, Takahiro Sato, Kensuke Tono, Tetsuo Katayama,
  Takashi Kameshima, Kanade Ogawa, Tadashi Togashi, Shigeki Owada, Yoshiyuki
  Amemiya, Takashi Tanaka, Toru Hara, and Makina Yabashi.
\newblock Observation of femtosecond x-ray interactions with matter using an
  x-ray{\textendash}x-ray pump{\textendash}probe scheme.
\newblock {\em Proceedings of the National Academy of Sciences},
  113(6):1492--1497, 2016.

\bibitem{GlauberCoherence}
Roy~J. Glauber.
\newblock The quantum theory of optical coherence.
\newblock {\em Phys. Rev.}, 130:2529--2539, Jun 1963.

\bibitem{Gorobtsov}
Oleg~Yu. Gorobtsov, Giuseppe Mercurio, Flavio Capotondi, Petr Skopintsev,
  Sergey Lazarev, Ivan~A. Zaluzhnyy, Miltcho~B. Danailov, Martina Dell'Angela,
  Michele Manfredda, Emanuele Pedersoli, Luca Giannessi, Maya Kiskinova,
  Kevin~C. Prince, Wilfried Wurth, and Ivan~A. Vartanyants.
\newblock Seeded x-ray free-electron laser generating radiation with laser
  statistical properties.
\newblock {\em Nature Communications}, 9(1):4498, 2018.

\bibitem{ModeLockXFEL}
N.~R. Thompson and B.~W.~J. McNeil.
\newblock Mode locking in a free-electron laser amplifier.
\newblock {\em Phys. Rev. Lett.}, 100:203901, May 2008.

\bibitem{StimXray}
Thomas Kroll, Clemens Weninger, Roberto Alonso-Mori, Dimosthenis Sokaras,
  Diling Zhu, Laurent Mercadier, Vinay~P. Majety, Agostino Marinelli, Alberto
  Lutman, Marc~W. Guetg, Franz-Josef Decker, S\'ebastien Boutet, Andy Aquila,
  Jason Koglin, Jake Koralek, Daniel~P. DePonte, Jan Kern, Franklin~D. Fuller,
  Ernest Pastor, Thomas Fransson, Yu~Zhang, Junko Yano, Vittal~K. Yachandra,
  Nina Rohringer, and Uwe Bergmann.
\newblock Stimulated x-ray emission spectroscopy in transition metal complexes.
\newblock {\em Phys. Rev. Lett.}, 120:133203, Mar 2018.

\bibitem{HalavanauXLO}
A.~{Halavanau}, A.~{Benediktovitch}, A.~A. {Lutman}, D.~{DePonte}, D.~{Cocco},
  N.~{Rohringer}, U.~{Bergmann}, and C.~{Pellegrini}.
\newblock {Design and Characteristics of an X-ray Laser Oscillator}.
\newblock {\em arXiv e-prints}, page arXiv:1912.03554, Dec 2019.

\bibitem{Scheinker2019}
Alexander Scheinker, Dorian Bohler, Sergey Tomin, Raimund Kammering, Igor
  Zagorodnov, Holger Schlarb, Matthias Scholz, Bolko Beutner, and Winfried
  Decking.
\newblock Model-independent tuning for maximizing free electron laser pulse
  energy.
\newblock {\em Phys. Rev. Accel. Beams}, 22:082802, Aug 2019.

\bibitem{Durissimo}
Joseph Duris, Dylan Kennedy, Adi Hanuka, Jane Shtalenkova, Auralee Edelen,
  P~Baxevanis, Adam Egger, T~Cope, M~McIntire, S~Ermon, et~al.
\newblock Bayesian optimization of a free-electron laser.
\newblock {\em Physical Review Letters}, 124(12):124801, 2020.

\bibitem{FELML}
A.~Sanchez-Gonzalez, P.~Micaelli, C.~Olivier, T.~R. Barillot, M.~Ilchen, A.~A.
  Lutman, A.~Marinelli, T.~Maxwell, A.~Achner, M.~Ag{\aa}ker, N.~Berrah,
  C.~Bostedt, J.~D. Bozek, J.~Buck, P.~H. Bucksbaum, S.~Carron Montero,
  B.~Cooper, J.~P. Cryan, M.~Dong, R.~Feifel, L.~J. Frasinski, H.~Fukuzawa,
  A.~Galler, G.~Hartmann, N.~Hartmann, W.~Helml, A.~S. Johnson, A.~Knie, A.~O.
  Lindahl, J.~Liu, K.~Motomura, M.~Mucke, C.~O'Grady, J-E Rubensson, E.~R.
  Simpson, R.~J. Squibb, C.~S{\aa}the, K.~Ueda, M.~Vacher, D.~J. Walke,
  V.~Zhaunerchyk, R.~N. Coffee, and J.~P. Marangos.
\newblock Accurate prediction of x-ray pulse properties from a free-electron
  laser using machine learning.
\newblock {\em Nature Communications}, 8(1):15461, 2017.

\bibitem{EmmaML2018}
C.~Emma, A.~Edelen, M.~J. Hogan, B.~O'Shea, G.~White, and V.~Yakimenko.
\newblock Machine learning-based longitudinal phase space prediction of
  particle accelerators.
\newblock {\em Phys. Rev. Accel. Beams}, 21:112802, Nov 2018.

\bibitem{FemtosecImaging}
M.~P. Minitti, J.~M. Budarz, A.~Kirrander, J.~S. Robinson, D.~Ratner, T.~J.
  Lane, D.~Zhu, J.~M. Glownia, M.~Kozina, H.~T. Lemke, M.~Sikorski, Y.~Feng,
  S.~Nelson, K.~Saita, B.~Stankus, T.~Northey, J.~B. Hastings, and P.~M. Weber.
\newblock Imaging molecular motion: Femtosecond x-ray scattering of an
  electrocyclic chemical reaction.
\newblock {\em Phys. Rev. Lett.}, 114:255501, Jun 2015.

\bibitem{Erk2013}
B.~Erk, D.~Rolles, L.~Foucar, B.~Rudek, S.~W. Epp, M.~Cryle, C.~Bostedt,
  S.~Schorb, J.~Bozek, A.~Rouzee, A.~Hundertmark, T.~Marchenko, M.~Simon,
  F.~Filsinger, L.~Christensen, S.~De, S.~Trippel, J.~K{\"{u}}pper,
  H.~Stapelfeldt, S.~Wada, K.~Ueda, M.~Swiggers, M.~Messerschmidt, C.~D.
  Schr{\"{o}}ter, R.~Moshammer, I.~Schlichting, J.~Ullrich, and A.~Rudenko.
\newblock {Ultrafast Charge Rearrangement and Nuclear Dynamics upon Inner-Shell
  Multiple Ionization of Small Polyatomic Molecules}.
\newblock {\em Physical Review Letters}, 110(5):1--5, 2013.

\bibitem{McFarland2014}
B.~K. McFarland, J.~P. Farrell, S.~Miyabe, F.~Tarantelli, A.~Aguilar,
  N.~Berrah, C.~Bostedt, J.~D. Bozek, P.~H. Bucksbaum, J.~C. Castagna, R.~N.
  Coffee, J.~P. Cryan, L.~Fang, R.~Feifel, K.~J. Gaffney, J.~M. Glownia, T.~J.
  Martinez, M.~Mucke, B.~Murphy, A.~Natan, T.~Osipov, V.~S. Petrovi{\'{c}},
  S.~Schorb, Th~Schultz, L.~S. Spector, M.~Swiggers, I.~Tenney, S.~Wang, J.~L.
  White, W.~White, and M.~G{\"{u}}hr.
\newblock {Ultrafast X-ray Auger probing of photoexcited molecular dynamics}.
\newblock {\em Nature Communications}, 5(May):1--7, 2014.

\bibitem{Zhang2014}
Wenkai Zhang, Roberto Alonso-Mori, Uwe Bergmann, Christian Bressler, Matthieu
  Chollet, Andreas Galler, Wojciech Gawelda, Ryan~G. Hadt, Robert~W. Hartsock,
  Thomas Kroll, Kasper~S. Kj{\ae}r, Katharina Kubiek, Henrik~T. Lemke,
  Huiyang~W. Liang, Drew~A. Meyer, Martin~M. Nielsen, Carola Purser, Joseph~S.
  Robinson, Edward~I. Solomon, Zheng Sun, Dimosthenis Sokaras, Tim~B. {Van
  Driel}, Gy{\"{o}}rgy Vank{\'{o}}, Tsu~Chien Weng, Diling Zhu, and Kelly~J.
  Gaffney.
\newblock {Tracking excited-state charge and spin dynamics in iron coordination
  complexes}.
\newblock {\em Nature}, 509(7500):345--348, 2014.

\bibitem{Siefermann2014}
Katrin~R. Siefermann, Chaitanya~D. Pemmaraju, Stefan Neppl, Andrey Shavorskiy,
  Amy~A. Cordones, Josh Vura-Weis, Daniel~S. Slaughter, Felix~P. Sturm, Fabian
  Weise, Hendrik Bluhm, Matthew~L. Strader, Hana Cho, Ming~Fu Lin, Camila
  Bacellar, Champak Khurmi, Jinghua Guo, Giacomo Coslovich, Joseph~S. Robinson,
  Robert~A. Kaindl, Robert~W. Schoenlein, Ali Belkacem, Daniel~M. Neumark,
  Stephen~R. Leone, Dennis Nordlund, Hirohito Ogasawara, Oleg Krupin, Joshua~J.
  Turner, William~F. Schlotter, Michael~R. Holmes, Marc Messerschmidt,
  Michael~P. Minitti, Sheraz Gul, Jin~Z. Zhang, Nils Huse, David Prendergast,
  and Oliver Gessner.
\newblock {Atomic-scale perspective of ultrafast charge transfer at a
  dye-semiconductor interface}.
\newblock {\em Journal of Physical Chemistry Letters}, 5(15):2753--2759, 2014.

\bibitem{Ismail2020}
Ahmed S.~M. Ismail, Yohei Uemura, Sang~Han Park, Soonnam Kwon, Minseok Kim,
  Hebatalla Elnaggar, Federica Frati, Yasuhiro Niwa, Hiroki Wadati, Yasuyuki
  Hirata, Yujun Zhang, Kohei Yamagami, Susumu Yamamoto, Iwao Matsuda, Ufuk
  Halisdemir, Gertjan Koster, Bert~M. Weckhuysen, and Frank M.~F. de~Groot.
\newblock Direct observation of the electronic states of photoexcited hematite
  with ultrafast 2p3d x-ray absorption spectroscopy and resonant inelastic
  x-ray scattering.
\newblock {\em Phys. Chem. Chem. Phys.}, 22:2685--2692, 2020.

\bibitem{ProteinStructure}
Karol Nass, Anton Meinhart, Thomas R.~M. Barends, Lutz Foucar, Alexander Gorel,
  Andrew Aquila, Sabine Botha, R.~Bruce Doak, Jason Koglin, Mengning Liang,
  Robert~L. Shoeman, Garth Williams, Sebastien Boutet, and Ilme Schlichting.
\newblock {Protein structure determination by single-wavelength anomalous
  diffraction phasing of X-ray free-electron laser data}.
\newblock {\em IUCrJ}, 3(3):180--191, May 2016.

\bibitem{DOEimaging1}
Jacques-Philippe Colletier, Michael~R. Sawaya, Mari Gingery, Jose~A. Rodriguez,
  Duilio Cascio, Aaron~S. Brewster, Tara Michels-Clark, Robert~H. Hice, Nicolas
  Coquelle, S{\'e}bastien Boutet, Garth~J. Williams, Marc Messerschmidt,
  Daniel~P. DePonte, Raymond~G. Sierra, Hartawan Laksmono, Jason~E. Koglin,
  Mark~S. Hunter, Hyun-Woo Park, Monarin Uervirojnangkoorn, Dennis~K. Bideshi,
  Axel~T. Brunger, Brian~A. Federici, Nicholas~K. Sauter, and David~S.
  Eisenberg.
\newblock De novo phasing with x-ray laser reveals mosquito larvicide binab
  structure.
\newblock {\em Nature}, 539(7627):43---47, 2016.

\bibitem{ProteinCrystal}
Michael~R Sawaya, Duilio Cascio, Mari Gingery, Jose Rodriguez, Lukasz
  Goldschmidt, Jacques-Philippe Colletier, Marc~M Messerschmidt, Sébastien
  Boutet, Jason~E Koglin, Garth~J Williams, Aaron~S Brewster, Karol Nass, Johan
  Hattne, Sabine Botha, R~Bruce Doak, Robert~L Shoeman, Daniel~P DePonte,
  Hyun-Woo Park, Brian~A Federici, Nicholas~K Sauter, Ilme Schlichting, and
  David~S Eisenberg.
\newblock Protein crystal structure obtained at 2.9 Å resolution from
  injecting bacterial cells into an x-ray free-electron laser beam.
\newblock {\em Proceedings of the National Academy of Sciences of the United
  States of America}, 111(35):12769—12774, September 2014.

\bibitem{Chapman2011}
Henry~N Chapman, Petra Fromme, Anton Barty, Thomas~A White, Richard~A Kirian,
  Andrew Aquila, Mark~S Hunter, Joachim Schulz, Daniel~P DePonte, Uwe
  Weierstall, R~Bruce Doak, Filipe R N~C Maia, Andrew~V Martin, Ilme
  Schlichting, Lukas Lomb, Nicola Coppola, Robert~L Shoeman, Sascha~W Epp,
  Robert Hartmann, Daniel Rolles, Artem Rudenko, Lutz Foucar, Nils Kimmel,
  Georg Weidenspointner, Peter Holl, Mengning Liang, Miriam Barthelmess, Carl
  Caleman, S{\'{e}}bastien Boutet, Michael~J Bogan, Jacek Krzywinski, Christoph
  Bostedt, Sa{\v{s}}a Bajt, Lars Gumprecht, Benedikt Rudek, Benjamin Erk, Carlo
  Schmidt, Andr{\'{e}} H{\"{o}}mke, Christian Reich, Daniel Pietschner, Lothar
  Str{\"{u}}der, G{\"{u}}nter Hauser, Hubert Gorke, Joachim Ullrich, Sven
  Herrmann, Gerhard Schaller, Florian Schopper, Heike Soltau, Kai-Uwe
  K{\"{u}}hnel, Marc Messerschmidt, John~D Bozek, Stefan~P Hau-Riege, Matthias
  Frank, Christina~Y Hampton, Raymond~G Sierra, Dmitri Starodub, Garth~J
  Williams, Janos Hajdu, Nicusor Timneanu, M~Marvin Seibert, Jakob Andreasson,
  Andrea Rocker, Olof J{\"{o}}nsson, Martin Svenda, Stephan Stern, Karol Nass,
  Robert Andritschke, Claus-Dieter Schr{\"{o}}ter, Faton Krasniqi, Mario Bott,
  Kevin~E Schmidt, Xiaoyu Wang, Ingo Grotjohann, James~M Holton, Thomas R~M
  Barends, Richard Neutze, Stefano Marchesini, Raimund Fromme, Sebastian
  Schorb, Daniela Rupp, Marcus Adolph, Tais Gorkhover, Inger Andersson, Helmut
  Hirsemann, Guillaume Potdevin, Heinz Graafsma, Bj{\"{o}}rn Nilsson, and John
  C~H Spence.
\newblock {Femtosecond X-ray protein nanocrystallography}.
\newblock {\em Nature}, 470(7332):73--77, 2011.

\bibitem{Sawaya2014}
Michael~R. Sawaya, Duilio Cascio, Mari Gingery, Jose Rodriguez, Lukasz
  Goldschmidt, Jacques~Philippe Colletier, Marc~M. Messerschmidt,
  S{\'{e}}bastien Boutet, Jason~E. Koglin, Garth~J. Williams, Aaron~S.
  Brewster, Karol Nass, Johan Hattne, Sabine Botha, R.~Bruce Doak, Robert~L.
  Shoeman, Daniel~P. DePonte, Hyun~Woo Park, Brian~A. Federici, Nicholas~K.
  Sauter, Ilme Schlichting, and David~S. Eisenberg.
\newblock {Protein crystal structure obtained at 2.9 {\AA} resolution from
  injecting bacterial cells into an X-ray free-electron laser beam}.
\newblock {\em Proceedings of the National Academy of Sciences of the United
  States of America}, 111(35):12769--12774, 2014.

\bibitem{Fenalti2015}
Gustavo Fenalti, Nadia~A. Zatsepin, Cecilia Betti, Patrick Giguere, Gye~Won
  Han, Andrii Ishchenko, Wei Liu, Karel Guillemyn, Haitao Zhang, Daniel James,
  Dingjie Wang, Uwe Weierstall, John~C.H. Spence, S{\'{e}}bastien Boutet, Marc
  Messerschmidt, Garth~J. Williams, Cornelius Gati, Oleksandr~M. Yefanov,
  Thomas~A. White, Dominik Oberthuer, Markus Metz, Chun~Hong Yoon, Anton Barty,
  Henry~N. Chapman, Shibom Basu, Jesse Coe, Chelsie~E. Conrad, Raimund Fromme,
  Petra Fromme, Dirk Tourw{\'{e}}, Peter~W. Schiller, Bryan~L. Roth, Steven
  Ballet, Vsevolod Katritch, Raymond~C. Stevens, and Vadim Cherezov.
\newblock {Structural basis for bifunctional peptide recognition at human
  $\delta$-opioid receptor}.
\newblock {\em Nature Structural and Molecular Biology}, 22(3):265--268, 2015.

\bibitem{Zhang2015}
Haitao Zhang, Hamiyet Unal, Cornelius Gati, Gye~Won Han, Wei Liu, Nadia~A.
  Zatsepin, Daniel James, Dingjie Wang, Garrett Nelson, Uwe Weierstall,
  Michael~R. Sawaya, Qingping Xu, Marc Messerschmidt, Garth~J. Williams,
  S{\'{e}}bastien Boutet, Oleksandr~M. Yefanov, Thomas~A. White, Chong Wang,
  Andrii Ishchenko, Kalyan~C. Tirupula, Russell Desnoyer, Jesse Coe, Chelsie~E.
  Conrad, Petra Fromme, Raymond~C. Stevens, Vsevolod Katritch, Sadashiva~S.
  Karnik, and Vadim Cherezov.
\newblock {Structure of the angiotensin receptor revealed by serial femtosecond
  crystallography}.
\newblock {\em Cell}, 161(4):833--844, 2015.

\bibitem{Barends2015}
Thomas~R.M. Barends, Lutz Foucar, Albert Ardevol, Karol Nass, Andrew Aquila,
  Sabine Botha, R.~Bruce Doak, Konstantin Falahati, Elisabeth Hartmann, Mario
  Hilpert, Marcel Heinz, Matthias~C. Hoffmann, J{\"{u}}rgen K{\"{o}}finger,
  Jason~E. Koglin, Gabriela Kovacsova, Mengning Liang, Despina Milathianaki,
  Henrik~T. Lemke, Jochen Reinstein, Christopher~M. Roome, Robert~L. Shoeman,
  Garth~J. Williams, Irene Burghardt, Gerhard Hummer, S{\'{e}}bastien Boutet,
  and Ilme Schlichting.
\newblock {Direct observation of ultrafast collective motions in CO myoglobin
  upon ligand dissociation}.
\newblock {\em Science}, 350(6259):445--450, 2015.

\bibitem{virology2019}
A.~Meents and M.O. Wiedorn.
\newblock Virus structures by x-ray free-electron lasers.
\newblock {\em Annual Review of Virology}, 6(1):161--176, 2019.
\newblock PMID: 31567066.

\bibitem{ProteinMech}
Doeke~R. Hekstra, K.~Ian White, Michael~A. Socolich, Robert~W. Henning, Vukica
  {\v S}rajer, and Rama Ranganathan.
\newblock Electric-field-stimulated protein mechanics.
\newblock {\em Nature}, 540(7633):400--405, 2016.

\bibitem{Miao530}
Jianwei Miao, Tetsuya Ishikawa, Ian~K. Robinson, and Margaret~M. Murnane.
\newblock Beyond crystallography: Diffractive imaging using coherent x-ray
  light sources.
\newblock {\em Science}, 348(6234):530--535, 2015.

\bibitem{Tadesse2019}
Getnet~K. Tadesse, Wilhelm Eschen, Robert Klas, Maxim Tschernajew, Frederik
  Tuitje, Michael Steinert, Matthias Zilk, Vittoria Schuster, Michael
  Z{\"u}rch, Thomas Pertsch, Christian Spielmann, Jens Limpert, and Jan
  Rothhardt.
\newblock Wavelength-scale ptychographic coherent diffractive imaging using a
  high-order harmonic source.
\newblock {\em Scientific Reports}, 9(1):1735, 2019.

\bibitem{Sandberg2007}
Richard~L. Sandberg, Ariel Paul, Daisy~A. Raymondson, Steffen H\"adrich,
  David~M. Gaudiosi, Jim Holtsnider, Ra'anan~I. Tobey, Oren Cohen, Margaret~M.
  Murnane, Henry~C. Kapteyn, Changyong Song, Jianwei Miao, Yanwei Liu, and
  Farhad Salmassi.
\newblock Lensless diffractive imaging using tabletop coherent high-harmonic
  soft-x-ray beams.
\newblock {\em Phys. Rev. Lett.}, 99:098103, Aug 2007.

\bibitem{Ravasio2009}
A.~Ravasio, D.~Gauthier, F.~R. N.~C. Maia, M.~Billon, J-P. Caumes, D.~Garzella,
  M.~G\'el\'eoc, O.~Gobert, J-F. Hergott, A-M. Pena, H.~Perez, B.~Carr\'e,
  E.~Bourhis, J.~Gierak, A.~Madouri, D.~Mailly, B.~Schiedt, M.~Fajardo,
  J.~Gautier, P.~Zeitoun, P.~H. Bucksbaum, J.~Hajdu, and H.~Merdji.
\newblock Single-shot diffractive imaging with a table-top femtosecond soft
  x-ray laser-harmonics source.
\newblock {\em Phys. Rev. Lett.}, 103:028104, Jul 2009.

\bibitem{Gardner2017}
Dennis~F. Gardner, Michael Tanksalvala, Elisabeth~R. Shanblatt, Xiaoshi Zhang,
  Benjamin~R. Galloway, Christina~L. Porter, Robert Karl~Jr, Charles Bevis,
  Daniel~E. Adams, Henry~C. Kapteyn, Margaret~M. Murnane, and Giulia~F.
  Mancini.
\newblock Subwavelength coherent imaging of periodic samples using a 13.5 nm
  tabletop high-harmonic light source.
\newblock {\em Nature Photonics}, 11(4):259--263, 2017.

\bibitem{AttoFermi}
Praveen~Kumar Maroju, Cesare Grazioli, Michele Di~Fraia, Matteo Moioli, Dominik
  Ertel, Hamed Ahmadi, Oksana Plekan, Paola Finetti, Enrico Allaria, Luca
  Giannessi, Giovanni De~Ninno, Carlo Spezzani, Giuseppe Penco, Simone
  Spampinati, Alexander Demidovich, Miltcho~B. Danailov, Roberto Borghes,
  George Kourousias, Carlos~Eduardo Sanches Dos~Reis, Fulvio Bill{\'e},
  Alberto~A. Lutman, Richard~J. Squibb, Raimund Feifel, Paolo Carpeggiani,
  Maurizio Reduzzi, Tommaso Mazza, Michael Meyer, Samuel Bengtsson, Neven
  Ibrakovic, Emma~Rose Simpson, Johan Mauritsson, Tam{\'a}s Csizmadia, Mathieu
  Dumergue, Sergei K{\"u}hn, Harshitha Nandiga~Gopalakrishna, Daehyun You,
  Kiyoshi Ueda, Marie Labeye, Jens~Egebjerg B{\ae}kh{\o}j, Kenneth~J. Schafer,
  Elena~V. Gryzlova, Alexei~N. Grum-Grzhimailo, Kevin~C. Prince, Carlo
  Callegari, and Giuseppe Sansone.
\newblock Attosecond pulse shaping using a seeded free-electron laser.
\newblock {\em Nature}, 578(7795):386--391, 2020.

\bibitem{KrauszRMP}
Ferenc Krausz and Misha Ivanov.
\newblock Attosecond physics.
\newblock {\em Rev. Mod. Phys.}, 81:163--234, Feb 2009.

\bibitem{CorkumAtto}
P.~B. Corkum and Ferenc Krausz.
\newblock Attosecond science.
\newblock {\em Nature Physics}, 3(6):381--387, 2007.

\bibitem{Agostini_2004}
Pierre Agostini and Louis~F DiMauro.
\newblock The physics of attosecond light pulses.
\newblock {\em Reports on Progress in Physics}, 67(6):813--855, may 2004.

\bibitem{rana2019ptychographic}
Arjun Rana, Jianhua Zhang, Minh Pham, Andrew Yuan, Yuan~Hung Lo, Huaidong
  Jiang, Stanley Osher, and Jianwei Miao.
\newblock Ptychographic coherent diffractive imaging for attosecond pulses,
  2019.

\bibitem{Holler2017}
Mirko Holler, Manuel Guizar-Sicairos, Esther H.~R. Tsai, Roberto Dinapoli,
  Elisabeth M{\"u}ller, Oliver Bunk, J{\"o}rg Raabe, and Gabriel Aeppli.
\newblock High-resolution non-destructive three-dimensional imaging of
  integrated circuits.
\newblock {\em Nature}, 543(7645):402--406, 2017.

\bibitem{Renkai2012}
R.~K. Li, K.~G. Roberts, C.~M. Scoby, H.~To, and P.~Musumeci.
\newblock Nanometer emittance ultralow charge beams from rf photoinjectors.
\newblock {\em Phys. Rev. ST Accel. Beams}, 15:090702, Sep 2012.

\bibitem{Li2014}
R~K Li and P~Musumeci.
\newblock {Single-Shot MeV Transmission Electron Microscopy with Picosecond
  Temporal Resolution}.
\newblock {\em Phys. Rev. Applied}, 2(2):24003, aug 2014.

\bibitem{Carlsten2018}
Bruce~E Carlsten.
\newblock {Tutorial on X-Ray Free-Electron Lasers}.
\newblock {\em IEEE Transactions on Plasma Science}, 46(6):1900--1912, 2018.

\bibitem{resonantPWFA}
P.~Manwani, N.~Majernik, and J.~B. Rosenzweig.
\newblock {Resonant excitation of very high gradient plasma wakefield
  accelerators by optical-period bunch trains}.
\newblock {\em Physical Review Accelerators and Beams}, Submitted, 2020.

\bibitem{HuangJC2017}
Jui-Che Huang, Hideo Kitamura, Chin-Kang Yang, Cheng-Hsing Chang, Cheng-Hsiang
  Chang, and Ching-Shiang Hwang.
\newblock Challenges of in-vacuum and cryogenic permanent magnet undulator
  technologies.
\newblock {\em Phys. Rev. Accel. Beams}, 20:064801, Jun 2017.

\bibitem{CryoUndCouprie}
C.~Benabderrahmane, M.~Vall\'eau, A.~Ghaith, P.~Berteaud, L.~Chapuis,
  F.~Marteau, F.~Briquez, O.~Marcouill\'e, J.-L. Marlats, K.~Tavakoli, A.~Mary,
  D.~Zerbib, A.~Lestrade, M.~Louvet, P.~Brunelle, K.~Medjoubi, C.~Herbeaux,
  N.~B\'echu, P.~Rommeluere, A.~Somogyi, O.~Chubar, C.~Kitegi, and M.-E.
  Couprie.
\newblock Development and operation of a
  ${\mathrm{pr}}_{2}{\mathrm{fe}}_{14}\mathrm{B}$ based cryogenic permanent
  magnet undulator for a high spatial resolution x-ray beam line.
\newblock {\em Phys. Rev. Accel. Beams}, 20:033201, Mar 2017.

\bibitem{Mishra17}
G.~Mishra, Mona Gehlot, Geetanjali Sharma, and Frederic Trillaud.
\newblock {Magnetic design and modelling of a 14mm-period prototype
  superconducting undulator}.
\newblock {\em Journal of Synchrotron Radiation}, 24(2):422--428, Mar 2017.

\bibitem{Majernik2019Triangular}
N~Majernik and JB~Rosenzweig.
\newblock Halbach undulators using right triangular magnets.
\newblock {\em Physical Review Accelerators and Beams}, 22(9):092401, 2019.

\bibitem{Peterson2014}
B.~A. Peterson, O.~D. Oniku, W.~C. Patterson, D.~{Le Roy}, A.~Garraud,
  F.~Herrault, N.~M. Dempsey, D.~P. Arnold, and M.~G. Allen.
\newblock Technology development for short-period magnetic undulators.
\newblock {\em Physics Procedia}, 52:36--45, 2014.

\bibitem{Tan_19}
Jun hao Tan, Yi~fei Li, Bao jun Zhu, Chang qing Zhu, Jin guang Wang, Da~zhang
  Li, Xin Lu, Yu~tong Li, and Li~ming Chen.
\newblock Short-period high-strength helical undulator by laser-driven bifilar
  capacitor coil.
\newblock {\em Opt. Express}, 27(21):29676--29684, Oct 2019.

\bibitem{Gadjev2017}
I~Gadjev, R~Candler, C~Emma, J~Harrison, A~Nause, J~Wu, A~Gover, and
  J~Rosenzweig.
\newblock {High-gain, short wavelength FEL in the Raman regime}.
\newblock {\em Nuclear Instruments and Methods in Physics Research Section A:
  Accelerators, Spectrometers, Detectors and Associated Equipment}, 865:20--24,
  2017.

\bibitem{Debus_2019}
Alexander Debus, Klaus Steiniger, Peter Kling, C~Moritz Carmesin, and Roland
  Sauerbrey.
\newblock Realizing quantum free-electron lasers: a critical analysis of
  experimental challenges and theoretical limits.
\newblock {\em Physica Scripta}, 94(7):074001, apr 2019.

\bibitem{twocolorFEL}
\'Angela Sa\'a~Hern\'andez, Eduard Prat, and Sven Reiche.
\newblock Generation of two-color x-ray free-electron-laser pulses from a beam
  with a large energy chirp and a slotted foil.
\newblock {\em Phys. Rev. Accel. Beams}, 22:030702, Mar 2019.

\bibitem{Halavanau2019}
Aliaksei Halavanau, Franz-Josef Decker, Claudio Emma, Jackson Sheppard, and
  Claudio Pellegrini.
\newblock {Very high brightness and power LCLS-II hard X-ray pulses}.
\newblock {\em Journal of Synchrotron Radiation}, 26(3):635--646, May 2019.

\bibitem{Graves:FEL2017-TUB03}
W.S. Graves and et~al.
\newblock Asu compact xfel.
\newblock In {\em Proc. of International Free Electron Laser Conference
  (FEL'17), Santa Fe, NM, USA, August 20-25, 2017}, number~38 in International
  Free Electron Laser Conference, pages 225--228, Geneva, Switzerland, Feb.
  2018. JACoW.
\newblock https://doi.org/10.18429/JACoW-FEL2017-TUB03.

\bibitem{ASUWeb}
{ASU CFEL Workshops}.
\newblock https://biodesign.asu.edu/cxfel/cxfel-workshops, 2020.

\bibitem{UCXFELworkshop}
{UCLA UC-XFEL Workshop website}.
\newblock https://conferences.pa.ucla.edu/free-electron-laser/, 2019.

\bibitem{Reuter1948}
G.~E.~H Reuter and E.~H Sondheimer.
\newblock {The theory of the anomalous skin effect in metals}.
\newblock {\em Proc. R. Soc. A}, 195:336, 1948.

\bibitem{Dingle1953}
R.~B Dingle.
\newblock {The anomalous skin effect and the reflectivity of metals}.
\newblock {\em Physica}, 19:311, 1953.

\bibitem{Kittel2005}
C.~Kittel.
\newblock {\em {Introduction to Solid State Physics}}.
\newblock John Wiley {\&} Sons, Inc, New York, 8 edition, 2005.

\bibitem{Bane2005}
K~Bane and G.~Stupakov.
\newblock {Resistive wall wakefield in the LCLS undulator}.
\newblock {\em Proceedings of the 21st Particle Accelerator Conference,
  Knoxville, TN}, page 3390, 2005.

\bibitem{andersonemit}
S.~G. Anderson, J.~B. Rosenzweig, G.~P. LeSage, and J.~K. Crane.
\newblock Space-charge effects in high brightness electron beam emittance
  measurements.
\newblock {\em Phys. Rev. ST Accel. Beams}, 5:014201, Jan 2002.

\bibitem{marxemit}
D.~Marx, J.~Giner~Navarro, D.~Cesar, J.~Maxson, B.~Marchetti, R.~Assmann, and
  P.~Musumeci.
\newblock Single-shot reconstruction of core 4d phase space of high-brightness
  electron beams using metal grids.
\newblock {\em Phys. Rev. Accel. Beams}, 21:102802, Oct 2018.

\bibitem{murokhyag}
A.~{Murokh}, J.~{Rosenzweig}, I.~{Ben-Zvi}, X.~{Wang}, and V.~{Yakimenko}.
\newblock Limitations on measuring a transverse profile of ultradense electron
  beams with scintillators.
\newblock In {\em PACS2001. Proceedings of the 2001 Particle Accelerator
  Conference (Cat. No.01CH37268)}, volume~2, pages 1333--1335 vol.2, 2001.

\bibitem{KULKE1988241}
B.~Kulke, H.~Shay, F.~Coffield, R.~Frye, and R.~Holmes.
\newblock A compact beam position monitor for use inside a free electron laser
  wiggler.
\newblock {\em Nuclear Instruments and Methods in Physics Research Section A:
  Accelerators, Spectrometers, Detectors and Associated Equipment}, 272(1):241
  -- 246, 1988.

\bibitem{laisievers1}
R.~Lai and A.~J. Sievers.
\newblock Determination of a charged-particle-bunch shape from the coherent far
  infrared spectrum.
\newblock {\em Phys. Rev. E}, 50:R3342--R3344, Nov 1994.

\bibitem{MUROKH1998}
A.~Murokh, J.B. Rosenzweig, M.~Hogan, H.~Suk, G.~Travish, and U.~Happek.
\newblock Bunch length measurement of picosecond electron beams from a
  photoinjector using coherent transition radiation.
\newblock {\em Nuclear Instruments and Methods in Physics Research Section A:
  Accelerators, Spectrometers, Detectors and Associated Equipment}, 410(3):452
  -- 460, 1998.

\bibitem{Andonian:2007}
G~Andonian, M~Dunning, E~Hemsing, and J~Rosenzweig.
\newblock {Observation of Coherent Edge Radiation Emitted by a 100 Femtosecond
  Compressed Electron Beam}.
\newblock {\em International Journal of Modern Physics A}, 22:4101, 2007.

\bibitem{FUKASAWA20142}
A.~Fukasawa, H.~To, S.K. Mahapatra, B.~Baumgartner, A.~Cahill, K.~Fitzmorris,
  R.~Li, P.~Musumeci, J.B. Rosenzweig, B.~Spataro, D.~Alesini, L.~Ficcadenti,
  A.~Valloni, and L.~Palumbo.
\newblock Progress on the hybrid gun project at ucla.
\newblock {\em Physics Procedia}, 52:2 -- 6, 2014.

\bibitem{WuICT}
Yuchi Wu, Dan Han, Bin Zhu, Kegong Dong, Fang Tan, and Yuqiu Gu.
\newblock A new method to calculate the beam charge for an integrating current
  transformer.
\newblock {\em Review of Scientific Instruments}, 83(9):093302, 2012.

\bibitem{Moody2009}
J.~T. Moody, P.~Musumeci, M.~S. Gutierrez, J.~B. Rosenzweig, and C.~M. Scoby.
\newblock Longitudinal phase space characterization of the blow-out regime of
  rf photoinjector operation.
\newblock {\em Phys. Rev. ST Accel. Beams}, 12:070704, Jul 2009.

\bibitem{Anderson2003}
S.~G. Anderson, J.~B. Rosenzweig, P.~Musumeci, and M.~C. Thompson.
\newblock Horizontal phase-space distortions arising from magnetic pulse
  compression of an intense, relativistic electron beam.
\newblock {\em Phys. Rev. Lett.}, 91:074803, Aug 2003.

\bibitem{RTI2012}
J.~Thangaraj, G.~Andonian, R.~Thurman-Keup, J.~Ruan, A.~S. Johnson, A.~Lumpkin,
  J.~Santucci, T.~Maxwell, A.~Murokh, M.~Ruelas, and A.~Ovodenko.
\newblock Demonstration of a real-time interferometer as a bunch-length monitor
  in a high-current electron beam accelerator.
\newblock {\em Review of Scientific Instruments}, 83(4):043302, 2012.

\bibitem{ROSENZWEIG1995}
J.~Rosenzweig, G.~Travish, and A.~Tremaine.
\newblock Coherent transition radiation diagnosis of electron beam
  microbunching.
\newblock {\em Nuclear Instruments and Methods in Physics Research Section A:
  Accelerators, Spectrometers, Detectors and Associated Equipment}, 365(1):255
  -- 259, 1995.

\bibitem{WEIKUM2018369}
M.K. Weikum, G.~Andonian, N.S. Sudar, M.G. Fedurin, M.N. Polyanskiy,
  C.~Swinson, A.~Ovodenko, F.~O’Shea, M.~Harrison, Z.M. Sheng, and R.W.
  Assmann.
\newblock Preliminary measurements for a sub-femtosecond electron bunch length
  diagnostic.
\newblock {\em Nuclear Instruments and Methods in Physics Research Section A:
  Accelerators, Spectrometers, Detectors and Associated Equipment}, 909:369 --
  373, 2018.
\newblock 3rd European Advanced Accelerator Concepts workshop (EAAC2017).

\bibitem{Murokh:2003}
A~Murokh, R~Agustsson, M~Babzien, I~Ben-Zvi, L~Bertolini, K~Van~Bibber, R~Carr,
  M~Cornacchia, P~Frigola, J~Hill, E~Johnson, L~Klaisner, G~Le~Sage, M~Libkind,
  R~Malone, H-D Nuhn, C~Pellegrini, S~Reiche, G~Rakowsky, J~Rosenzweig,
  R~Ruland, J~Skaritka, A~Toor, A~Tremaine, X~Wang, and V~Yakimenko.
\newblock {Properties of the ultrashort gain length, self-amplified spontaneous
  emission free-electron laser in the linear regime and saturation}.
\newblock {\em Physical Review E}, 67(6):5, June 2003.

\bibitem{Andonian:2005}
G~Andonian, A~Murokh, J~B Rosenzweig, R~Agustsson, M~Babzien, I~Ben-Zvi,
  P~Frigola, J~Y Huang, L~Palumbo, C~Pellegrini, S~Reiche, G~Travish,
  C~Vicario, and V~Yakimenko.
\newblock {Observation of Anomalously Large Spectral Bandwidth in a High-Gain
  Self-Amplified Spontaneous Emission Free-Electron Laser}.
\newblock {\em Physical Review Letters}, 95:54801, July 2005.

\bibitem{barber:2020}
S~K Barber, J~H Bin, A~J Gonsalves, F~Isono, J~Van~Tilborg, S~Steinke,
  K~Nakamura, A~Zingale, N~A Czapla, D~Schumacher, C~B Schroeder, C~G~R Geddes,
  W~P Leemans, and E~Esarey.
\newblock {A compact, high resolution energy, and emittance diagnostic for
  electron beams using active plasma lenses}.
\newblock {\em Applied Physics Letters}, 116:234108, June 2020.

\bibitem{Barber:thesis}
Samuel Barber.
\newblock {\em Plasma Wakefield Experiments in the Quasi Nonlinear Regime}.
\newblock PhD thesis, UCLA, 2015.

\bibitem{WALSTON20071}
Sean Walston, Stewart Boogert, Carl Chung, Pete Fitsos, Joe Frisch, Jeff
  Gronberg, Hitoshi Hayano, Yosuke Honda, Yury Kolomensky, Alexey Lyapin,
  Stephen Malton, Justin May, Douglas McCormick, Robert Meller, David Miller,
  Toyoko Orimoto, Marc Ross, Mark Slater, Steve Smith, Tonee Smith, Nobuhiro
  Terunuma, Mark Thomson, Junji Urakawa, Vladimir Vogel, David Ward, and Glen
  White.
\newblock Performance of a high resolution cavity beam position monitor system.
\newblock {\em Nuclear Instruments and Methods in Physics Research Section A:
  Accelerators, Spectrometers, Detectors and Associated Equipment}, 578(1):1 --
  22, 2007.

\end{thebibliography}

\end{document}